\documentclass[useAMS,usenatbib]{mn2e}

\usepackage[dvips]{graphicx}
\def\kms{{\rm km}\;{\rm s}^{-1}}
\title[Galaxy Groups in the Local Universe]
{Clusters and Groups of Galaxies in the Simulated Local Universe}
\author[Casagrande and Diaferio]{Luca Casagrande$^{1,2}$\thanks{E-mail: 
luccas@utu.fi (LC), 
diaferio@ph.unito.it (AD)} and Antonaldo Diaferio$^3$\\
$^1$ Tuorla Astronomical Observatory, Piikkio, Finland\\
$^2$ University of Turku, Finland\\
$^3$Dipartimento di Fisica Generale ``Amedeo Avogadro'', Universit\`a degli 
Studi di Torino, Italy}

\begin{document}

\maketitle

\begin{abstract}

We compare the properties of galaxy groups extracted from the Updated Zwicky 
Catalogue (UZC) with those of groups extracted from $N$-body simulations 
of the local Universe, in a $\Lambda$CDM and a $\tau$CDM cosmology.
In the simulations, the initial conditions of the dark matter density field are 
set to reproduce the present time distribution of the galaxies 
within $80 h^{-1}$ Mpc from the Milky Way.
These initial conditions minimize the uncertainty originated by
cosmic variance, which has affected previous analyses of this small volume
of the Universe. The simulations also model the 
evolution of the photometric properties of the galaxy population 
with semi-analytic prescriptions. The models yield a galaxy
luminosity function sensibly different from that of the UZC and  
are unable to reproduce the distribution of groups and their
luminosity content. The discrepancy
between the model and the UZC reduces substantially, 
if we redistribute the luminosity among the galaxies in the simulation 
according to the UZC luminosity function while preserving the galaxy 
luminosity rank.
The modified $\Lambda$CDM model provides the best match to the UZC:  
the abundances of groups by harmonic
radius, velocity dispersion, mass and luminosity are
consistent with observations. We find that this 
model also reproduces the halo occupation number of groups and clusters. 
However, the large-scale distribution of
groups is marginally consistent with the UZC and the redshift-space correlation
function of galaxies on scales larger than $6 h^{-1}$ Mpc is still 
more than 3-$\sigma$ smaller than observed.
We conclude that reproducing the properties of the observed groups
certainly requires a more sophisticated treatment of galaxy formation,
and possibly an improvement of the dark matter model. 

\end{abstract}


\section{Introduction}\label{intro}

The formation and evolution of galaxies is one of the major challenges of 
cosmology.
Within the framework of the hierarchical
structure formation by gravitational instability, galaxies properties are 
strongly
affected by their environment. Most galaxies reside in groups 
and much work on compact (\citealt{hickson97} and references therein;  
\citealt{barton03}; \citealt{mendes03}; \citealt{lee04}; 
\citealt{kelm04}; \citealt*{tovmassian06}) and loose
groups (\citealt{postman84}; \citealt{allington93}; 
\citealt{tran01}; \citealt{martinez02a},b; 
\citealt{tanvuia03}; \citealt{girardi03}; 
\citealt{balogh04}; \citealt{weinmann05}; \citealt*{zandivarez06};
\citealt{martinez06}; \citealt{braugh06}) 
shows that these systems can be more effective than 
clusters at shaping the properties of galaxies,
namely morphology, luminosity, color, star formation rate and 
cold gas content (see also \citealt{kodama01};  
\citealt{ellingson04}; \citealt{verheijen04}; 
\citealt{tanaka05}).     

Galaxy groups trace the large-scale distribution of galaxies (e.g. \citealt{berlind06}), 
and thus describe the distribution of the optical light from sub-megaparsec scales to
the largest scale probed. Models of galaxy formation should be able to reproduce
the properties of galaxy groups. For an appropriate comparison of models with observations,
however, the surveyed volume of the real groups 
must be sufficiently large to be representative of the Universe. 
Over the last years, the compilation of deep galaxy redshift surveys has provided 
galaxy group catalogues with an increasing number of systems (\citealt{ramella99}; 
\citealt{eke04}; \citealt{berlind06} and references therein).
These catalogues are deep enough to average out the large-scale structure fluctuations
and obtain a robust estimate of the light distribution.

In current models, large-scale structure forms by the gravitational
collapse of dark matter, which dominates the dynamics of galaxy systems. Comparing 
models with data therefore requires a valuable treatment of the dynamics of the baryonic matter
within the dark matter halos. \citet{diaferio1999} were the first to attempt to compare 
real redshift surveys with mock surveys where galaxies were formed and evolved
with semi-analytic prescriptions applied to merger trees of dark matter halos extracted from 
$N$-body simulations.  They compared the model with the northern
sector of the Updated Zwicky Catalog (UZC) 
redshift survey \citep{falco1999}. 
This comparison resulted in a general agreement between the now standard Cold Dark Matter
model with a non-zero cosmological constant ($\Lambda$CDM) and observations, although
the mean luminosity of the simulated groups 
was smaller than observed. A further  topological analysis of the mock redshift surveys 
and the UZC 
quantified how the large-scale structure was much fuzzier in the mock catalogues 
than in the UZC \citep{schmalzing2000}. 
Both disagreements were tentatively attributed to cosmic variance, 
responsible for the lack, in the mock surveys,
of a coherent structure as thin and wide as the Great Wall. 
The UZC is only $150 h^{-1}$ Mpc deep and indeed its volume is not large enough to 
suppress the fluctuations of the large-scale structure. Therefore 
a fair comparison of the model with the UZC requires a not-random choice of the
volume in the simulations.  

\citet{mathis2002} perform $N$-body simulations where the initial conditions 
are constrained
to reproduce the large-scale distribution of galaxies observed in the 
IRAS 1.2 Jy survey \citep{fisher1995} 
within $80 h^{-1}$ Mpc of the Milky Way. 
In their simulations, \citet{mathis2002} also include semi-analytic
recipes to form and evolve galaxies. 
Thus, these simulations seem appropriate to test whether the
current galaxy formation models yield the observed clustering and luminosity 
properties of groups once the large scale distribution of galaxies 
faithfully mirror the real one. This paper is devoted to this test.

Sect. 2 summarizes the properties of the constrained simulations; 
sects. 3 and 4 describe our real and mock
galaxy redshift surveys and our group catalogues in redshift space. 
In sect. 5 we investigate the large-scale distribution of groups and
in sect. 6 we discuss the distribution of light among and within groups. 
We conclude in sect. 7.

\section{The simulation of the local Universe}\label{simulations}

The constrained simulations we use here combine the $N$-body technique with the
semi-analytic approach for the formation of galaxies to
predict the evolution of both the photometric properties of galaxies and their
phase space coordinates, following a strategy pionereed by \citet{roukema97}
and \citet{kauffmann1999}.

The simulations were run by \citet{mathis2002} who provide extensive 
details in their original paper.
They investigate two variants of a flat cold 
dark matter (CDM) universe: a $\Lambda$CDM model, with cosmological density 
parameter $\Omega_{0}=0.3$, cosmological constant $\Lambda = 0.7$, and Hubble 
constant $H_{0}=70\; \textrm{km}\; \textrm{s}^{-1}\textrm{Mpc}^{-1}$, and a 
$\tau$CDM model, with $\Omega_{0}=1$ and $H_{0}=50\; \textrm{km}\; 
\textrm{s}^{-1}\textrm{Mpc}^{-1}$. 
For both models, the normalization of the power spectrum of the initial 
density perturbations is imposed by the abundance of galaxy clusters at the 
present time: $\sigma_8=0.9$ and 0.6 for the $\Lambda$CDM and $\tau$CDM 
models, respectively. 

The initial conditions at $z=50$ are set to reproduce the observed 
present-day galaxy density field, provided by  the IRAS 1.2 Jy survey
(\citealt{fisher1994}; 1995), 
within a sphere of radius $80\;h^{-1}$ Mpc centered on the Milky Way and 
smoothed with a Gaussian filter with one-dimensional dispersion 5 $h^{-1}$ Mpc.
The constrained initial displacement field for the dark matter particles
for all wavenumbers in the range $k_0=2\pi /L$ and $64 k_0$, on
a simulation cube of side $L=240 h^{-1}$ Mpc, is generated with the 
Hoffman-Ribak algorithm (\citealt{hoffman91}, 1992; \citealt{ganon93}).
The high-frequency field in the wavenumber range $64k_0 - 343 k_0$ is 
unconstrained. Therefore,
the distribution of matter on scales smaller than  5 $h^{-1}$ Mpc is unrelated 
to the real Universe. The simulation contains two populations of particles: 
the 
high-resolution particles, which will end up within $80\;h^{-1}$ Mpc from the 
center of the simulation box (the location of the Milky Way) by $z=0$, and
more massive low-resolution particles, which will remain outside this region.
The simulation box contains $\sim 5 \times 10^7 $ high-resolution particles 
with individual mass $0.36\times 10^{10} h^{-1} M_\odot$ and 
$1.2\times 10^{10} h^{-1} 
M_\odot$ in the $\Lambda$CDM and $\tau$CDM model, respectively. 

The equations of motion of the dark matter particles are integrated with the 
tree-code GADGET \citep*{springel2001}. The halo merger trees are then 
extracted from the simulation and provided as input to the semi-analytic code 
used to form and evolve galaxies within the dark matter halos. Apart from a 
few 
differences, \citet{mathis2002} use the same recipes that \citet{kauffmann1999}
applied to the GIF simulations. The simulations have enough mass and spatial 
resolution to follow the formation of all galaxies brighter than the Large 
Magellanic Cloud that will end up, by $z=0$, within the nearly spherical 
high-resolution region of radius 80 $h^{-1}$ Mpc centered on the Milky Way. 
The luminosity resolution of the simulations are $M_B=-15.48+5\log h$
and $-16.94+5\log h$ for the $\Lambda$CDM and $\tau$CDM models, respectively.
Note that the limits mentioned
in \citet{mathis2002} differ from those provided on their web site.
Here we adopt the latter.
To these limits, the $\Lambda$CDM and $\tau$CDM models contain 189122 and 
194243 galaxies in the high-resolution volume.

\section{Mock and real redshift surveys}\label{mock}

\begin{figure*}
\begin{center}
\includegraphics[scale=0.5]{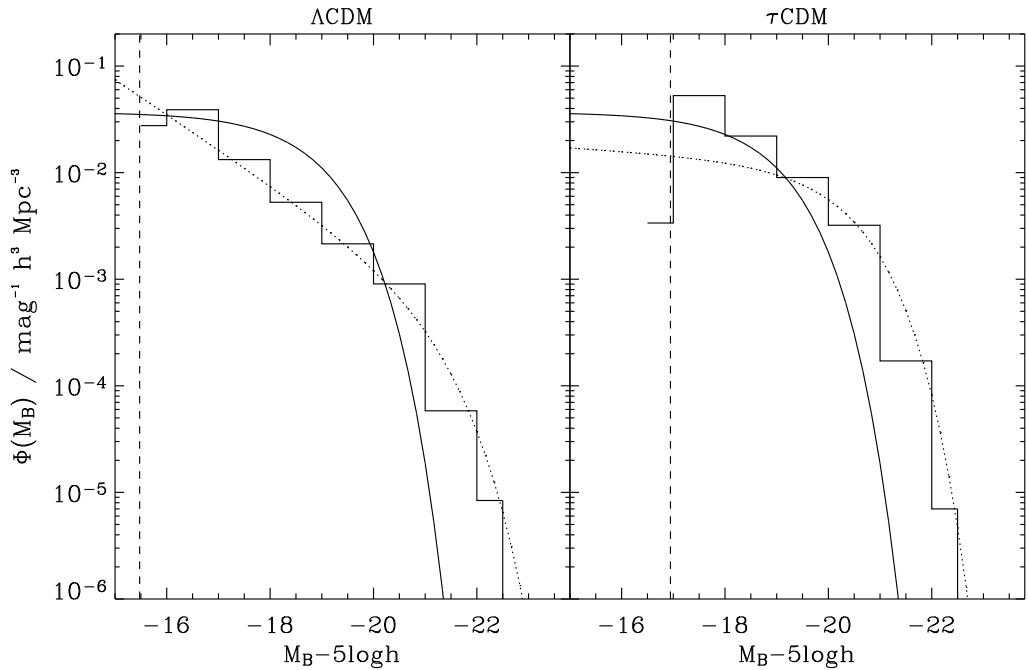}
\caption{The $B$-band luminosity functions in the simulations. The vertical lines 
show the luminosity resolution. The dotted lines are the fits to the Schechter function 
(Table \ref{tab_lum_fun}). The solid lines are the 
$M_{\rm Zw}$-band luminosity function of the UZC \citep*{marzke94}.}
\label{mathi}
\end{center}
\end{figure*}

We extract mock redshift surveys from the simulations to perform a detailed 
comparison with the UZC \citep{falco1999}. 
This catalogue
is currently the magnitude-limited redshift survey of the local universe with 
the widest sky coverage in the optical band. The UZC has been extensively 
investigated:  \citet{schmalzing2000} analyze its topology, 
\citet{padmanabhan2001} compute its redshift space power spectrum, \citet{hoyle02}
investigate the void properties, and much work is dedicated to galaxy
groups and pairs (\citealt{padilla01}; \citealt{ramella2002};
\citealt{focardi02}; 
\citealt{pisani03}; 
\citealt{plionis04}; \citealt{ceccarelli05}; \citealt{berrier06}; \citealt{focardi06}). 
Moreover, \citet{diaferio1999} performed a detailed comparison of the GIF 
simulations \citep{kauffmann1999} with the CfA2N, the subsample of the UZC 
galaxies in the northern galactic emisphere.

Here, we consider the region limited by $-2.5^{\circ} \le \delta_{1950} \le 
50^{\circ}$ and $8^{h} \le \alpha_{1950} \le 17^{h}$ in the North Galactic 
Cap (hereafter NGC) and by $20^{h} \le \alpha_{1950} \le 4^{h}$ in the 
South Galactic Cap (hereafter SGC). This survey region is 98 per cent complete 
down to the Zwicky magnitude $m_{\rm Zw}=15.5$ (Falco et al. 1999). We also apply a
Galactic cut $-13^{\circ} \le b \le 13^{\circ}$ and further discard the
southern Galactic region with $\alpha_{1950} \ge 3^{h}$ to reduce extinction
problems (Padmanabhan et al. 2001; Ramella et al. 2002). Finally, we only 
consider galaxies with redshift $cz < 8000\;\kms$, because the simulated 
high-resolution region is limited to $80 h^{-1}$ Mpc from the Milky Way. 
The redshift surveys we extract cover roughly a quarter of the sky and contain 
more than 7000 galaxies. 

To extract mock redshift surveys from the simulation box, we place an observer 
at the center of the box where the Milky Way is located and assign celestial 
coordinates to the simulated galaxies. Each galaxy in the simulations has 
radial velocity $cz={\bmath v} \cdot {\bmath r}/\Vert {\bmath r}
\Vert +\Vert{\bmath r}\Vert$, where ${\bmath v}$ is the galaxy peculiar 
velocity and ${\bmath r}$ is the galaxy position vector, in units of $\kms$.
Further details on the construction of the samples are given below.

\begin{table}
\centering
\caption{Luminosity function parameters. For the UZC $M^*$ is in the
Zwicky system.}\label{tab_lum_fun}
\begin{tabular}{lccc}
\hline
  & $\alpha$  & $M_B^*-5 \log h$ & $\phi^*/h^3 {\rm Mpc}^{-3} {\rm mag}^{-1}$\\
\hline
UZC  & $-1.00$ & $-18.80$        & $4.00\times10^{-2}$\\
$\Lambda$CDM  & $-1.82$ & $-21.08$     & $8.39\times10^{-4}$\\
$\tau$CDM & $-1.08$ & $-20.29$     & $1.28\times10^{-2}$\\
\hline
\end{tabular}
\end{table}

\subsection{The SALF catalogues}\label{constSALF}

\subsubsection{Mock catalogues}

The UZC is magnitude limited to the Zwicky magnitude 
$m_{\rm Zw}=15.5$, whereas the mock catalogues contain galaxies with $B$-band 
apparent magnitudes $m_B=M_B+25+5\log(r/h^{-1}\textrm{Mpc})$. Therefore, 
we need to transform the Zwicky magnitude limit into the simulated $B$-band.
Detailed calibrations of observed samples seem to imply $m_{\rm Zw}\sim m_B$,
but with a rather large 1-$\sigma$ scatter of $\sim 0.3$ mag  
(\citealt{huchra1976}; \citealt{bothun1990}; \citealt*{marzke94}).
Moreover the luminosity functions of the galaxies in the simulations
and in the UZC are sensibly different, as we discuss below,
and assuming naively $m_{\rm Zw}= m_B$ would not provide reliable
results. Therefore, following \citet{diaferio1999}
and \citet{mathis2002}, we adjust the magnitude cut-off of the simulated 
surveys to obtain a number of galaxies in the mock surveys comparable to
the number in the UZC. 

This constraint imposes the magnitude limits 
\begin{itemize}
\item[(i)]  $m_{B}=m_{\rm Zw} +0.7 = 16.2$ in NGC 
\item[(ii)] $m_{B}=m_{\rm Zw} +0.4 = 15.9 $ in SGC
\end{itemize}
for the the $\Lambda$CDM model, and 
\begin{itemize}
\item[(i)]  $m_{B}=m_{\rm Zw} -0.6 = 14.9$ in NGC 
\item[(ii)] $m_{B}=m_{\rm Zw} -0.9 = 14.6$ in SGC
\end{itemize} 
for the $\tau$CDM model. Unlike \citet{mathis2002}, we therefore use different 
transformations for NGC and SGC. These different limits are consistent with the 0.3
mag scatter mentioned  above and could be due to differences in the photometric 
zero-point across the different regions of the sky. A few of the many claimed 
systematic errors in the Zwicky catalogue are 
supported by clean observational evidence and a unique 
interpretation of this discrepancy between NGC and SGC is still lacking \citep*{marzke94}.
We note that the transformations for NGC are similar to those adopted by 
\citet{diaferio1999} in the GIF simulations which contained semi-analytic 
prescriptions similar to those of \citet{mathis2002}.
In the UZC and mock catalogues we also exclude galaxies with redshift 
$cz< 500\;\kms$ and $m_{\rm Zw}<10$ to avoid bright objects close to the Milky Way.

\subsubsection{The UZC}

In order to make a proper comparison with the simulations, in the UZC we have 
to exclude galaxies that are fainter than the luminosity resolution of
the corresponding simulation. This is especially important for 
NGC, where the nearby Virgo cluster has many faint galaxies that cannot be
resolved by the simulations. 

Therefore we compile two UZC catalogues, one where 
we exclude all galaxies fainter than the $\Lambda$CDM luminosity resolution 
($M_{B}=-15.48+5 \log h $: UZC-$\Lambda$-SALF) and another where we exclude all galaxies 
fainter than the $\tau$CDM luminosity resolution ($M_{B}=-16.94+5 \log h$:
UZC-$\tau$-SALF). 

In principle, the luminosity resolution cut
should be done in the Zwicky and not in the $B$-band magnitude. As we have discussed 
above, the transformation between the two systems is uncertain and possible 
differences in the zero-points of NGC and SGC make the transformation more complicated, 
because different cuts would be required for different regions of the sky. 
Fortunately, 
the luminosity resolution cut only concerns the nearby and faintest 
galaxies and a change of $\pm 0.3$ mag in the luminosity resolution limit 
reflects into a difference of $\pm 170$ galaxies, at most, 
in the final UZC number of galaxies.
Furthermore, in the UZC the luminosity resolution cut applies in 
redshift space, whereas the luminosity resolution limit of the simulation is 
in real space. When building mock redshift surveys, the effect of 
peculiar velocities blurs such a limit.
Given such uncertainties, it is reasonable to use the
simulated $B$-band limits to cut the UZC galaxies in the Zwicky magnitude.

\subsection{The UZCLF catalogues}\label{constUZCLF}

The properties of the galaxy distribution in a magnitude-limited redshift survey depend both on 
the distribution of galaxies in real space and on their luminosity function.
Figure \ref{mathi} shows the galaxy luminosity function in the two simulations,
with their best fits to a Schechter function (Table \ref{tab_lum_fun}). The 
model functions are sensibly different from the luminosity function of the UZC 
derived by \citet{marzke94}.  
In order to separate the role of the large-scale distribution of galaxies from 
their luminosity function in our comparison between the models and the UZC,
we replace the luminosities of galaxies predicted by the semi-analytic 
procedure with luminosities extracted from the UZC luminosity function
\citep{marzke94}. 

For each model we find the new 
luminosity resolution limit $L_{\rm lim}$ which satisfies the relation 
\begin{equation}\label{newlumres}
V\int_{L_{\rm lim}}^{\infty}\phi_{\rm UZC}(L)\textrm{d}L = N
\end{equation}
where $\phi_{\rm UZC}(L)$ is the UZC luminosity function (see Table 
\ref{tab_lum_fun}), $N$ is the number of simulated galaxies more luminous than 
the semi-analytical luminosity resolution and $V$ is the high-resolution 
volume of the simulation box. 
We then randomly sample $N$ luminosities in the range $[L_{\rm lim},+\infty]$ 
which distribute accordingly to $\phi_{\rm UZC}(L)$ and assign these new 
luminosities to the simulated galaxies while preserving the luminosity rank 
predicted by the semi-analytic prescriptions.
The new luminosity resolutions $L_{\rm lim}$ are $M_{\rm Zw}=-15.85+5\log h$ 
and $M_{\rm Zw}=-15.78+5\log h$ for the $\Lambda$CDM and $\tau$CDM models, 
respectively. The luminosity function of the simulated galaxies now perfectly 
matches $\phi_{\rm UZC}(L)$ and the only galaxy property imposed by the model 
is the galaxy luminosity rank. 

Even if the luminosities are now constrained to match the UZC luminosity 
function, a magnitude limit $m_{\rm lim}=15.5$ for the mock catalogues does 
not necessarly return the same number of galaxies as observed. In fact, both 
the aforementioned uncertainties in the Zwicky magnitudes 
and the large-scale distribution of galaxies in the small UZC volume 
can affect the galaxy number of a magnitude-limited survey. 
Even if the simulations we use are intended to minimize the cosmic variance 
uncertainty by reproducing the gross features of the large-scale structure, 
the total number $N$ of the simulated galaxies still depend 
on the semi-analytic prescriptions. 
Therefore, we adopt the following limiting
magnitudes for the mock catalogues, in order to recover a number of galaxies 
approximatevely equal to that of the UZC : 
\begin{itemize}
\item[(i)]  $m_{\rm lim} = 15.3$ in NGC
\item[(ii)] $m_{\rm lim} = 15.0$ in SGC
\end{itemize}
for the the $\Lambda$CDM model, and 
\begin{itemize}
\item[(i)]  $m_{\rm lim} = 15.5$ in NGC
\item[(ii)] $m_{\rm lim} = 15.1$ in SGC
\end{itemize} 
for the $\tau$CDM model. As for the SALF catalogues, the difference between 
NGC and SGC is $\sim 0.3$ mag. All the apparent magnitudes of the galaxies are 
then increased by the difference between the appropriate $m_{\rm lim}$ and 15.5 in
order to have the same zero-point of the photometric system in NGC and SGC.
We now have mock catalogues with the same number of galaxies and the 
same luminosity function as the UZC, but with the large-scale structure and 
the velocity field predicted by the underlying cosmological model. 

As for the SALF catalogues, we also make different catalogues from the real UZC, where 
we exclude all galaxies fainter than the appropriate luminosity resolutions: 
we thus have a UZC-$\Lambda$-UZCLF and a UZC-$\tau$-UZCLF catalogue.

\begin{figure*}
{\begin{minipage}[t]{0.49\linewidth}
\includegraphics[width=1\textwidth]{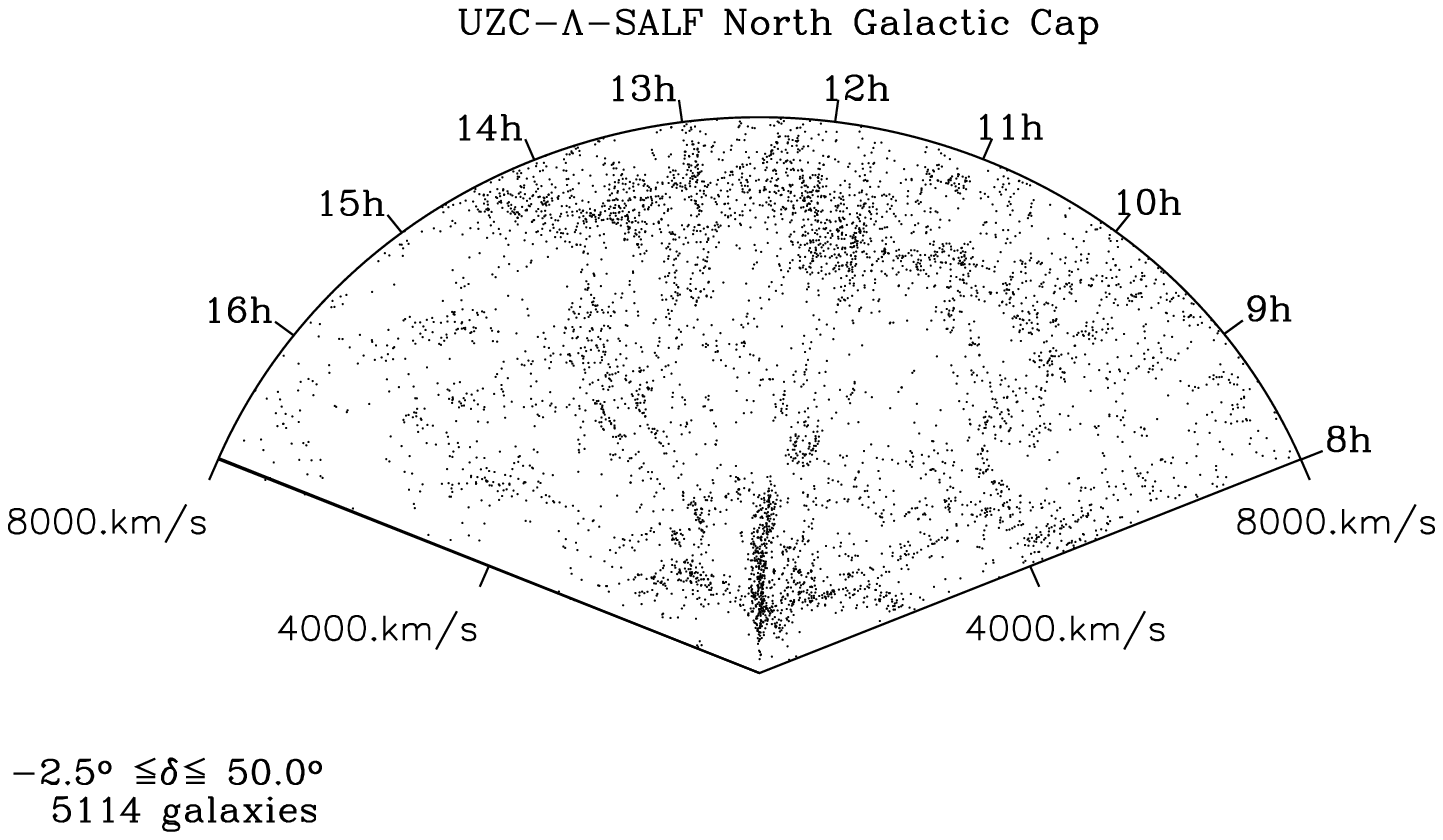}
\end{minipage}
}
\hfill
{\begin{minipage}[t]{0.49\linewidth}
\includegraphics[width=1\textwidth]{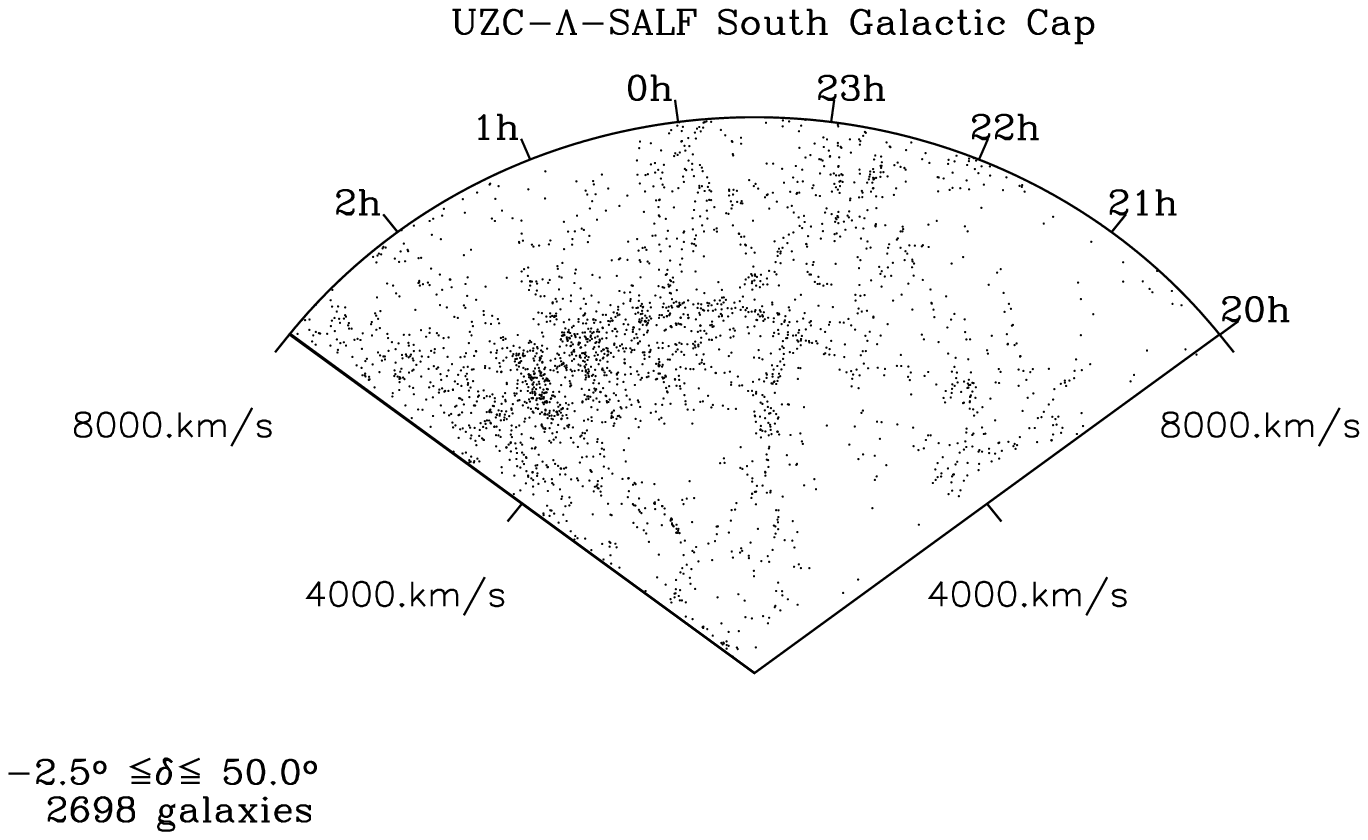}
\end{minipage}
}
{\begin{minipage}[c]{0.49\linewidth}
\includegraphics[width=1\textwidth]{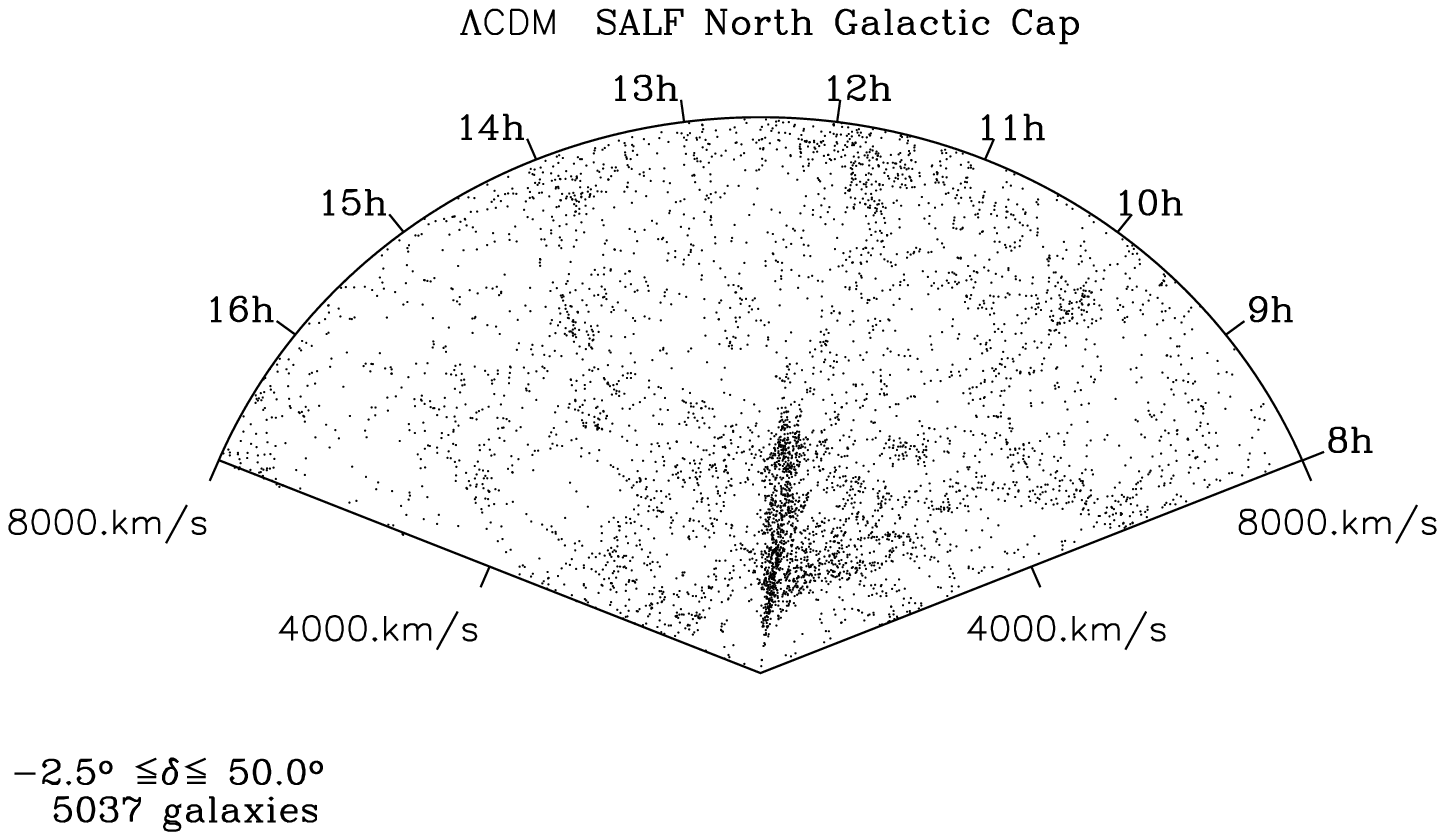}
\end{minipage}
}
\hfill
{\begin{minipage}[c]{0.49\linewidth}
\includegraphics[width=1\textwidth]{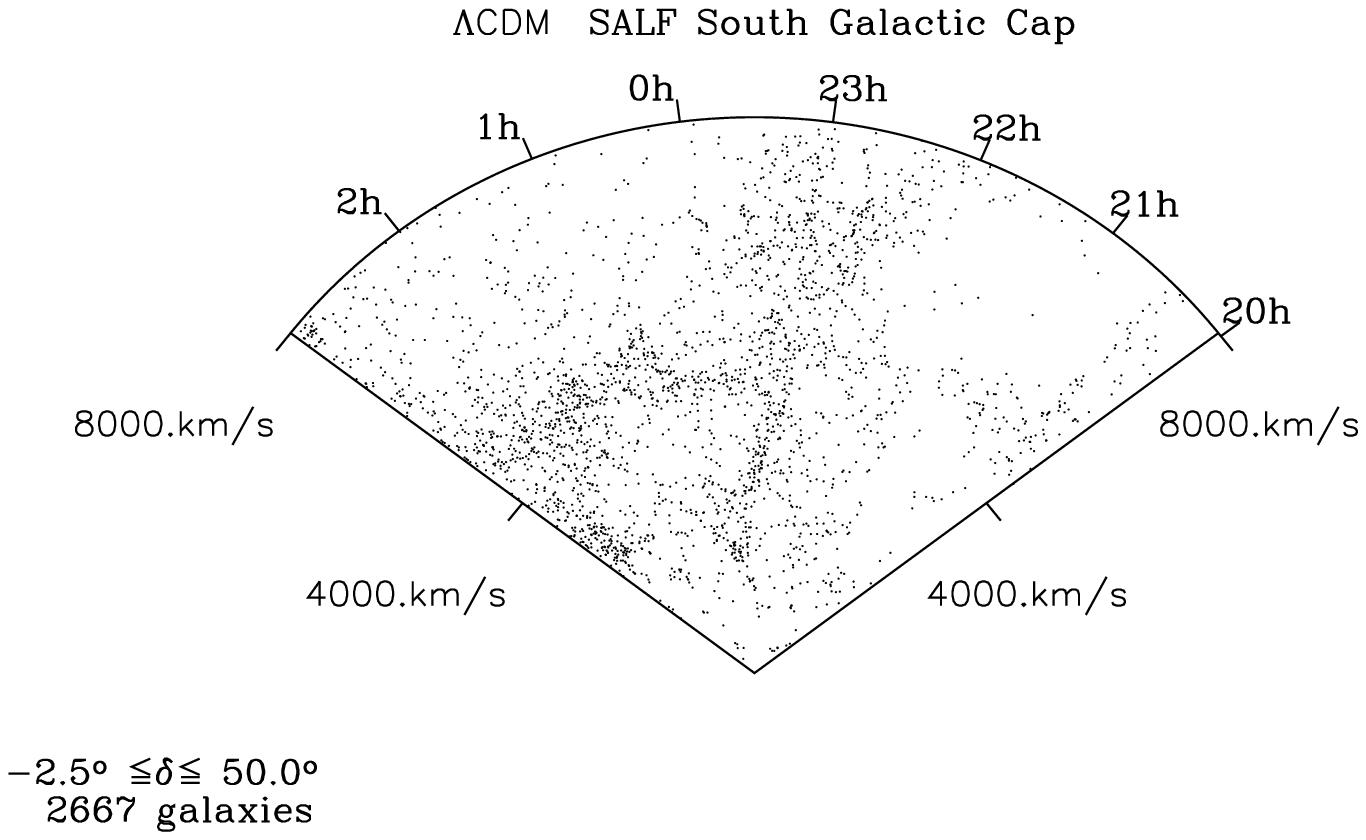}
\end{minipage}
}
\caption{The distribution of the galaxies (dots) in the magnitude-limited 
UZC-$\Lambda$-SALF and in the mock 
$\Lambda$CDM-SALF redshift surveys.}\label{slicesLsalf}
\end{figure*}

\begin{figure*}
{\begin{minipage}[t]{0.49\linewidth}
\includegraphics[width=1\textwidth]{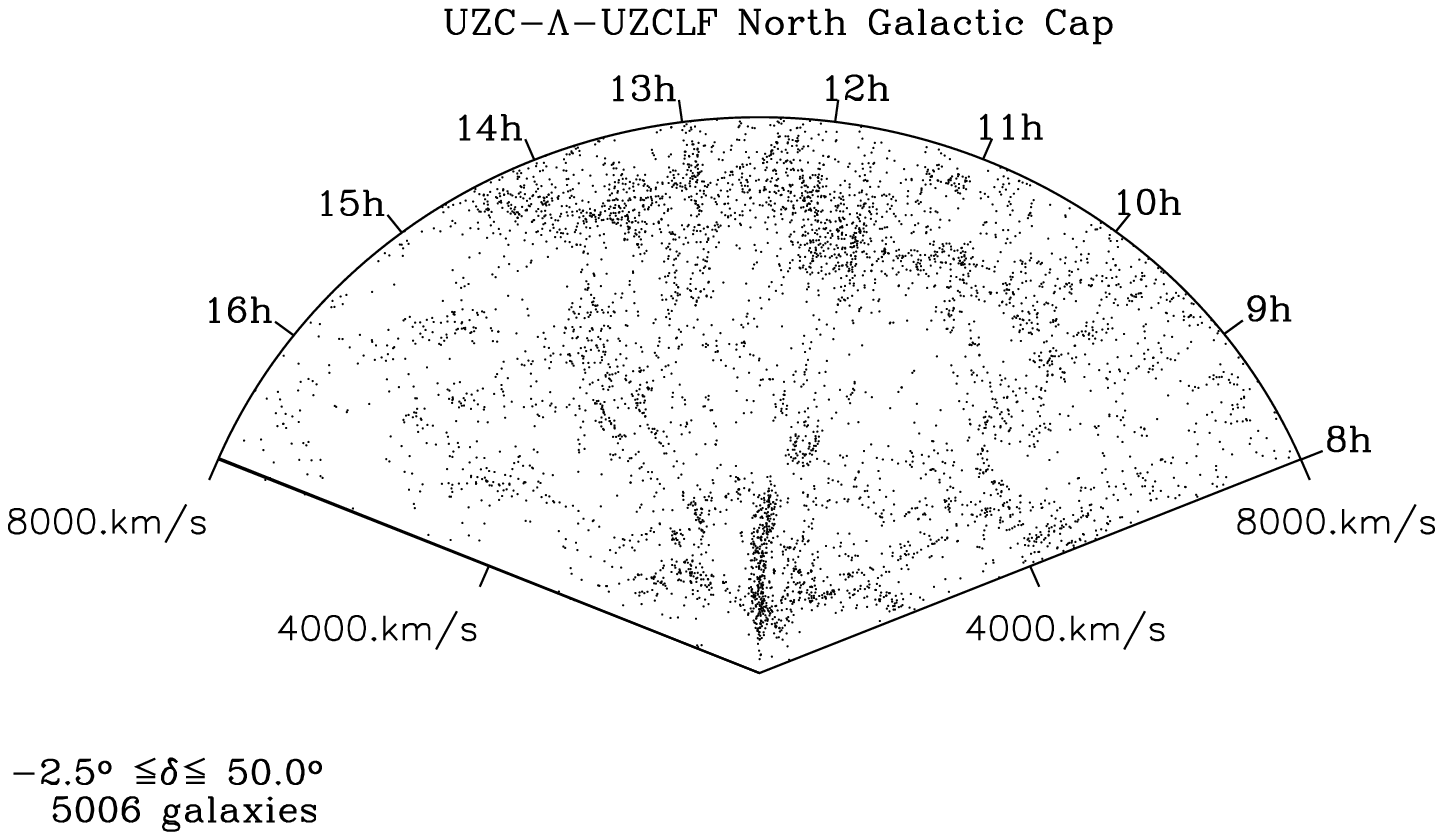}
\end{minipage}
}
\hfill
{\begin{minipage}[t]{0.49\linewidth}
\includegraphics[width=1\textwidth]{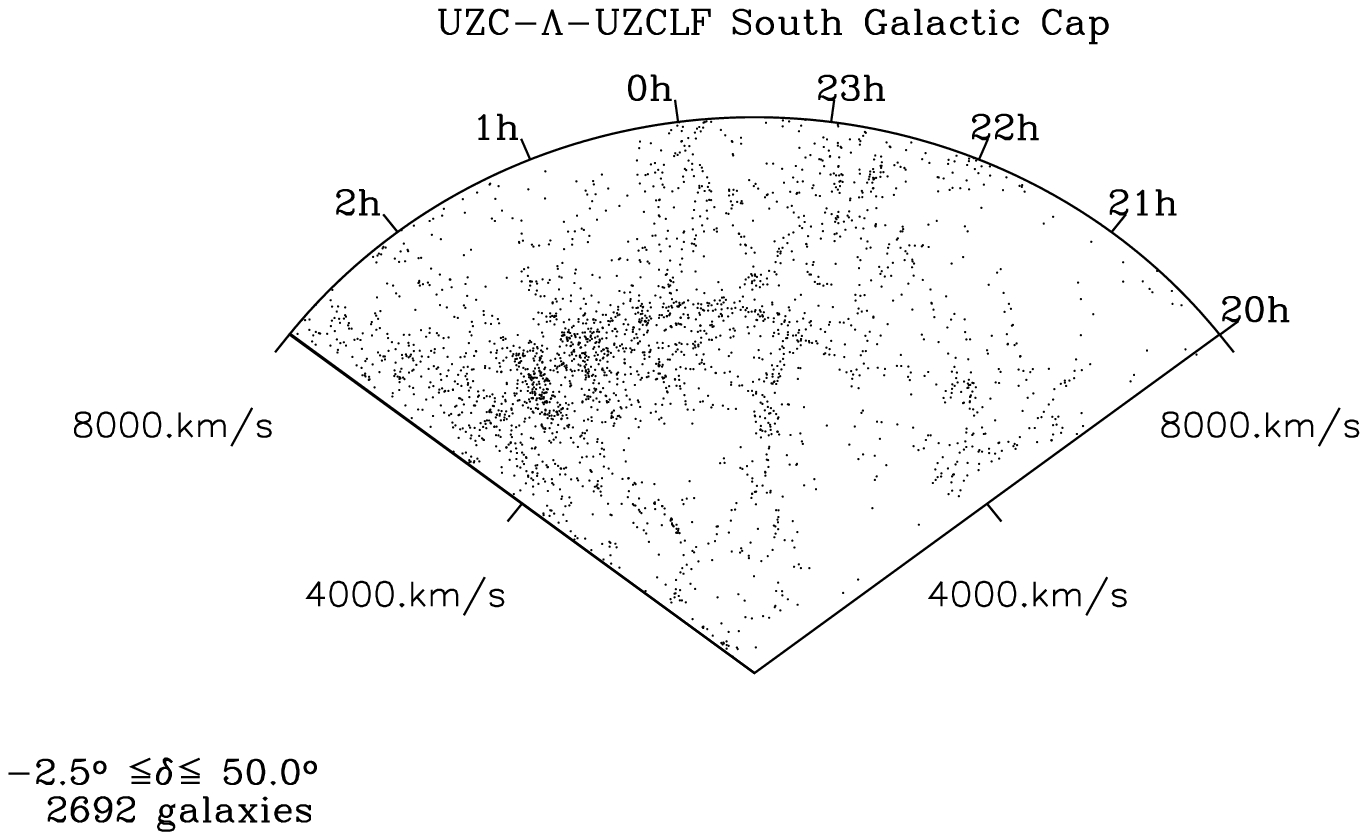}
\end{minipage}
}
{\begin{minipage}[c]{0.49\linewidth}
\includegraphics[width=1\textwidth]{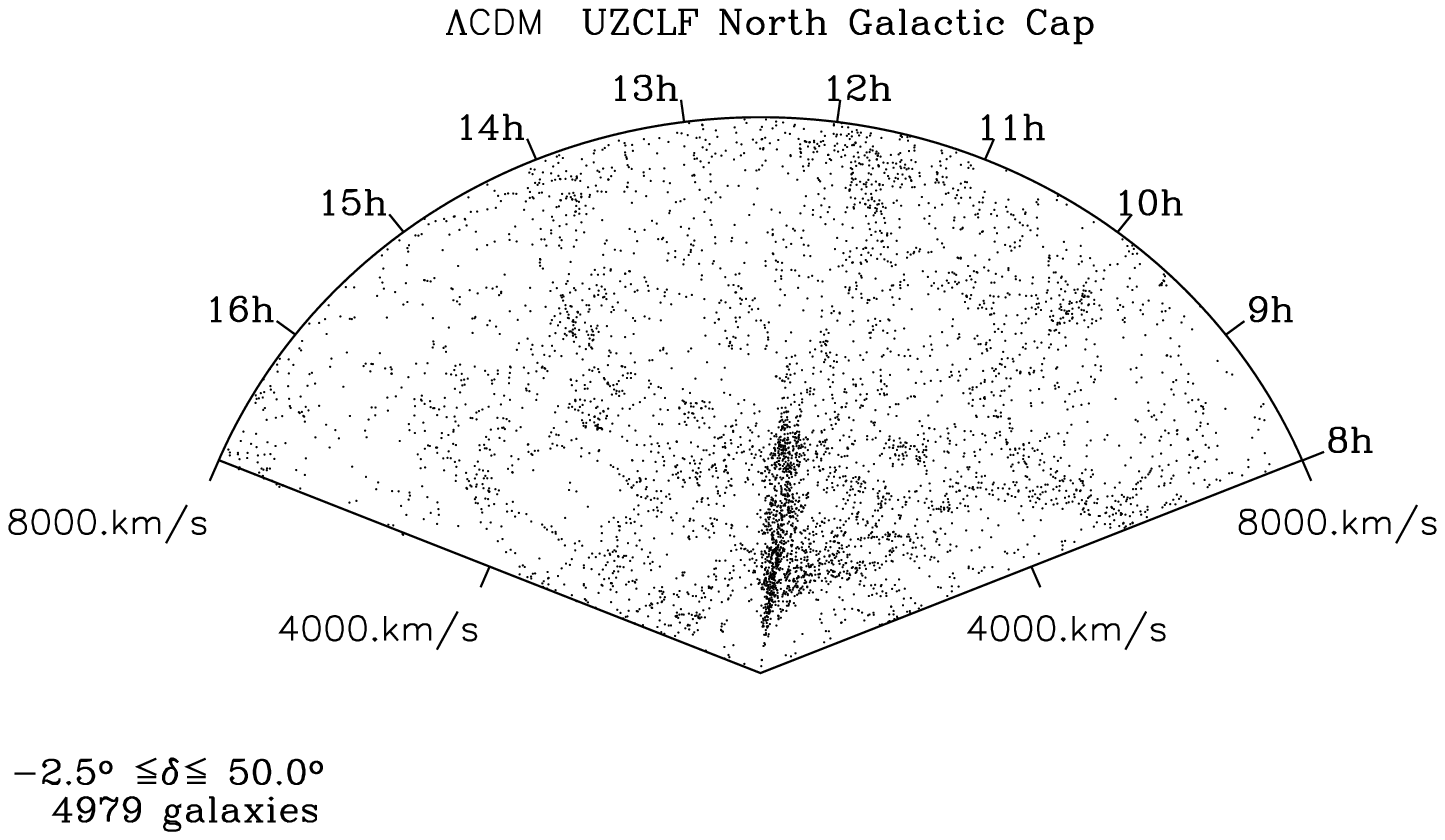}
\end{minipage}
}
\hfill
{\begin{minipage}[c]{0.49\linewidth}
\includegraphics[width=1\textwidth]{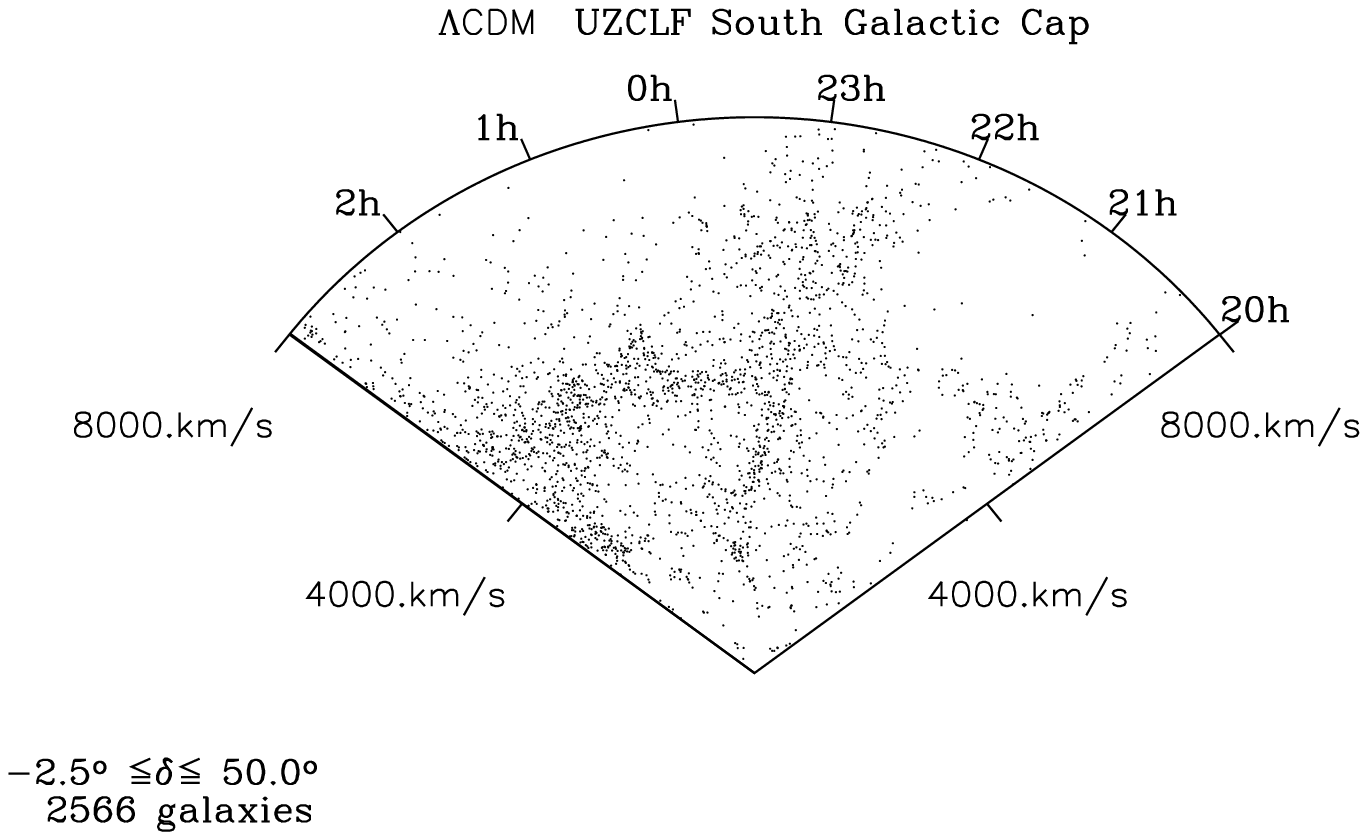}
\end{minipage}
}
\caption{The distribution of the galaxies (dots) in the magnitude-limited
UZC-$\Lambda$-UZCLF and in the mock
$\Lambda$CDM-UZCLF redshift surveys.} 
\label{slicesLuzclf}
\end{figure*}

\begin{figure*}
{\begin{minipage}[t]{0.49\linewidth}
\includegraphics[width=1\textwidth]{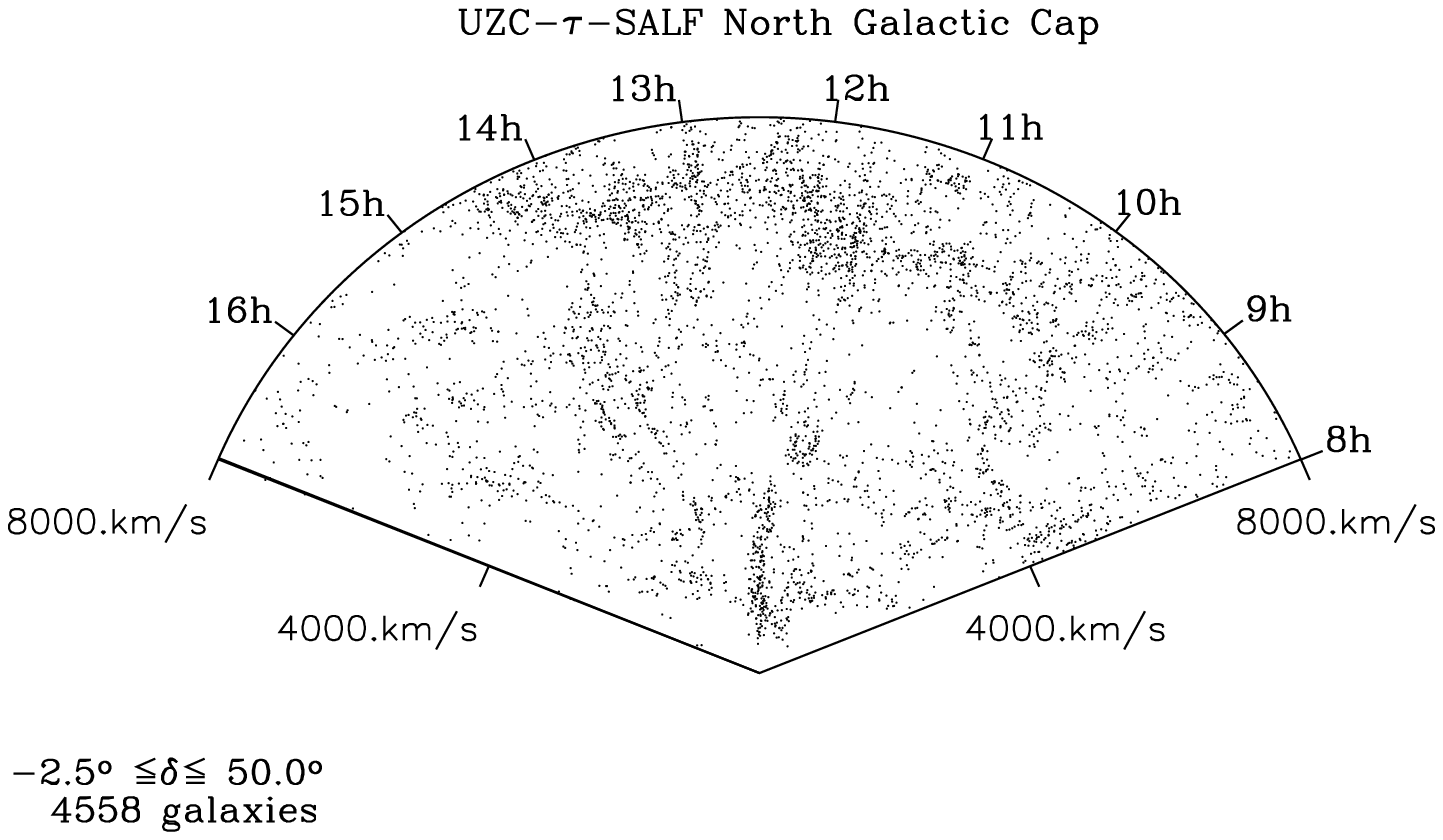}
\end{minipage}
}
\hfill
{\begin{minipage}[t]{0.49\linewidth}
\includegraphics[width=1\textwidth]{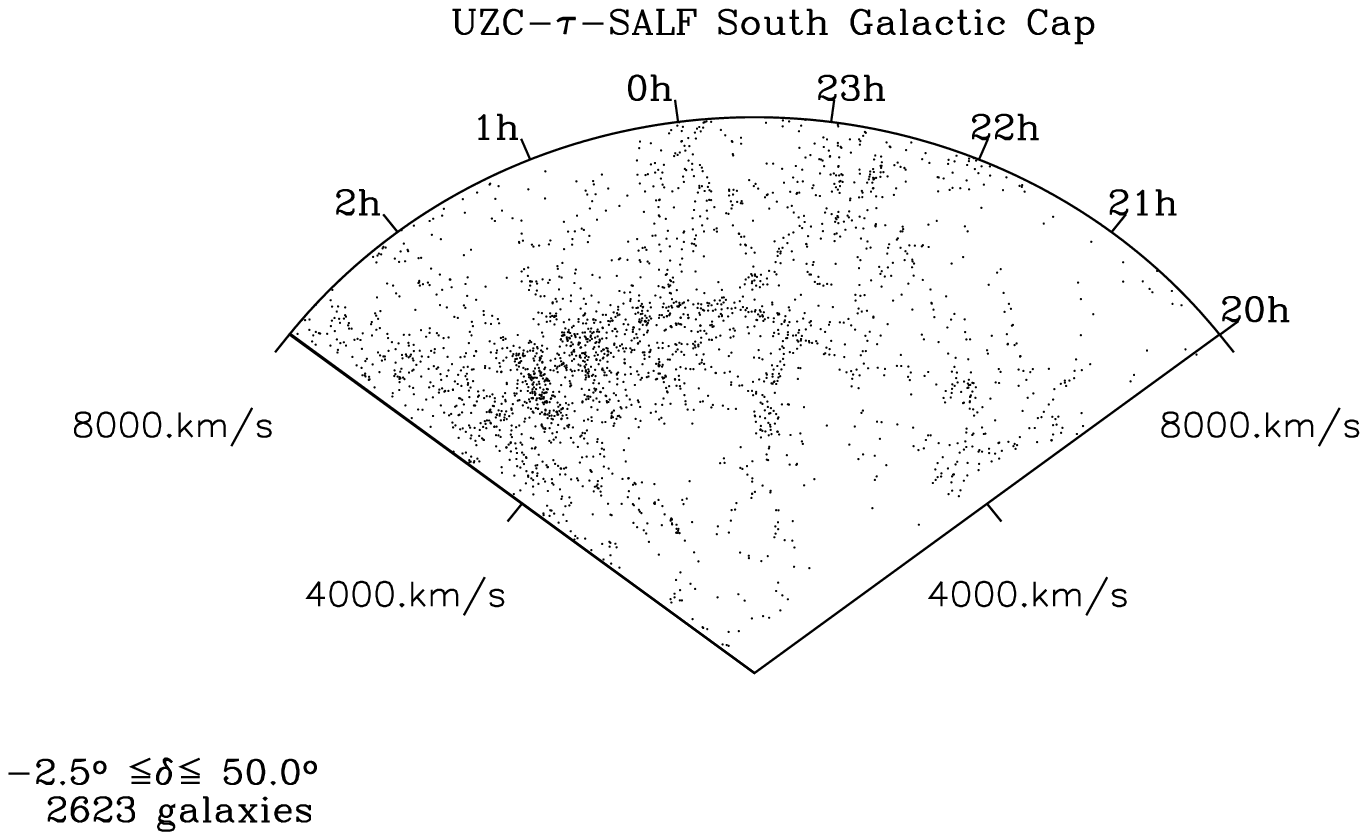}
\end{minipage}
}
{\begin{minipage}[c]{0.49\linewidth}
\includegraphics[width=1\textwidth]{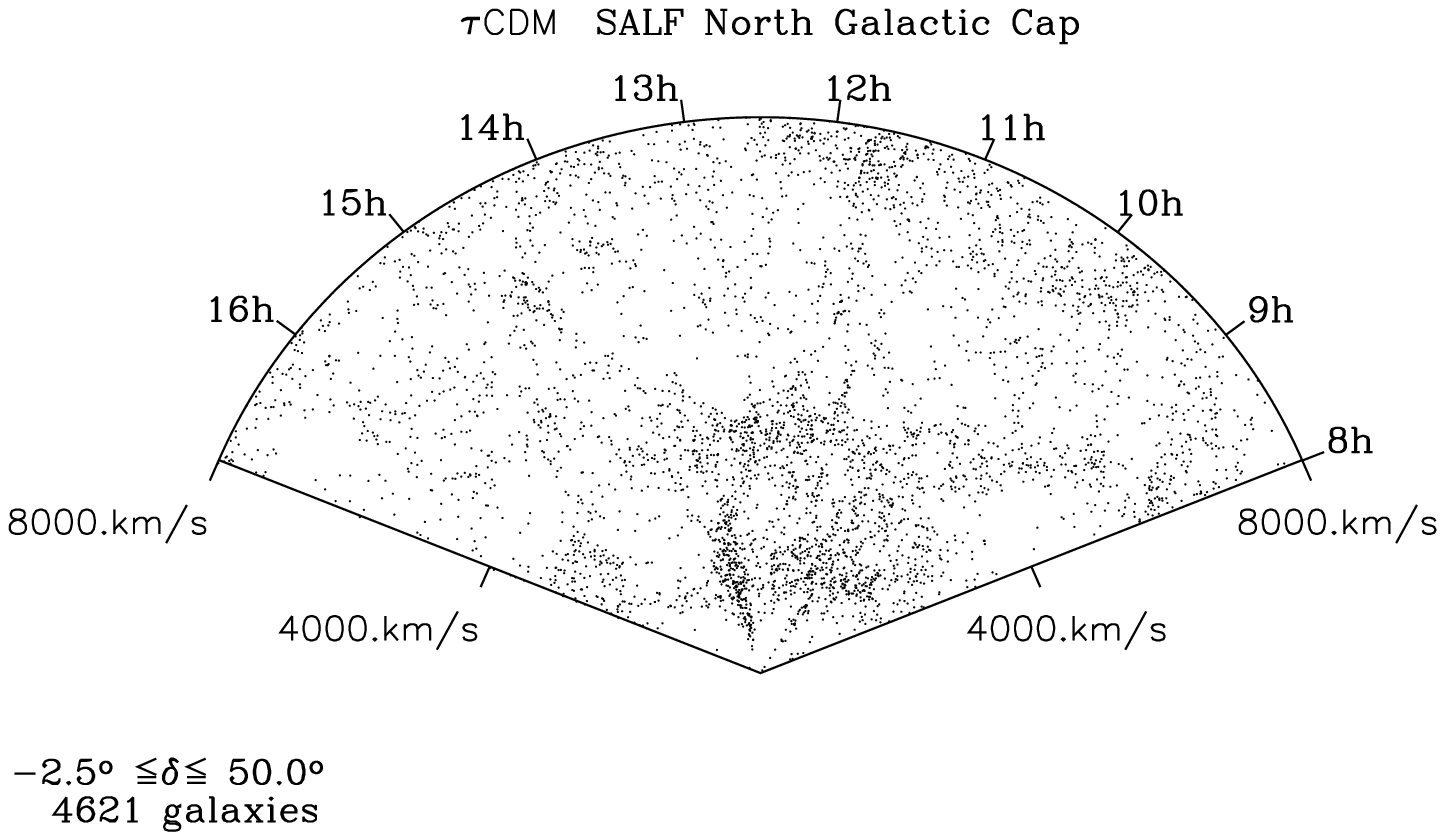}
\end{minipage}
}
\hfill
{\begin{minipage}[c]{0.49\linewidth}
\includegraphics[width=1\textwidth]{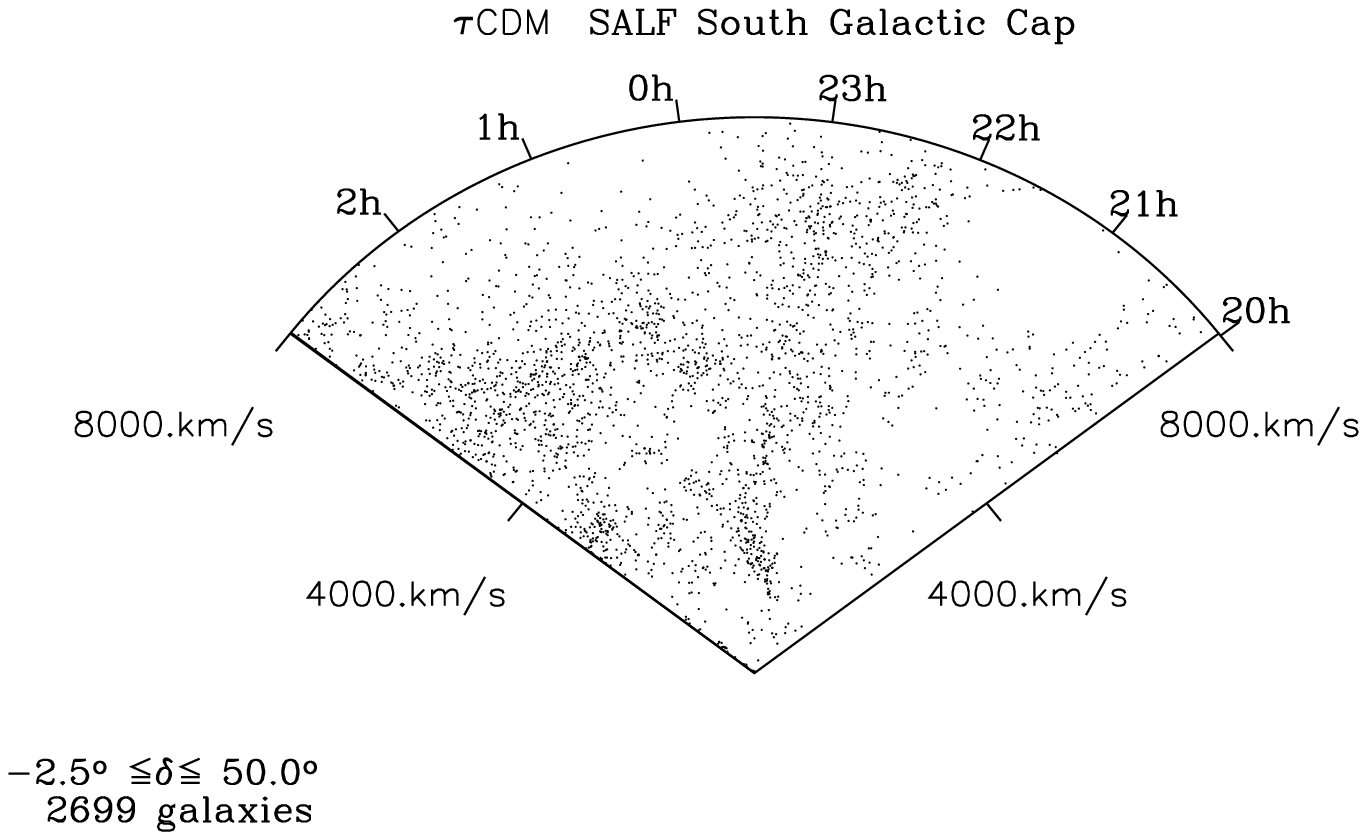}
\end{minipage}
}
\caption{
The distribution of the galaxies (dots) in the magnitude-limited
UZC-$\tau$-SALF and in the mock
$\tau$CDM-SALF redshift surveys.  }\label{slicesTsalf}
\end{figure*}

\begin{figure*}
{\begin{minipage}[t]{0.49\linewidth}
\includegraphics[width=1\textwidth]{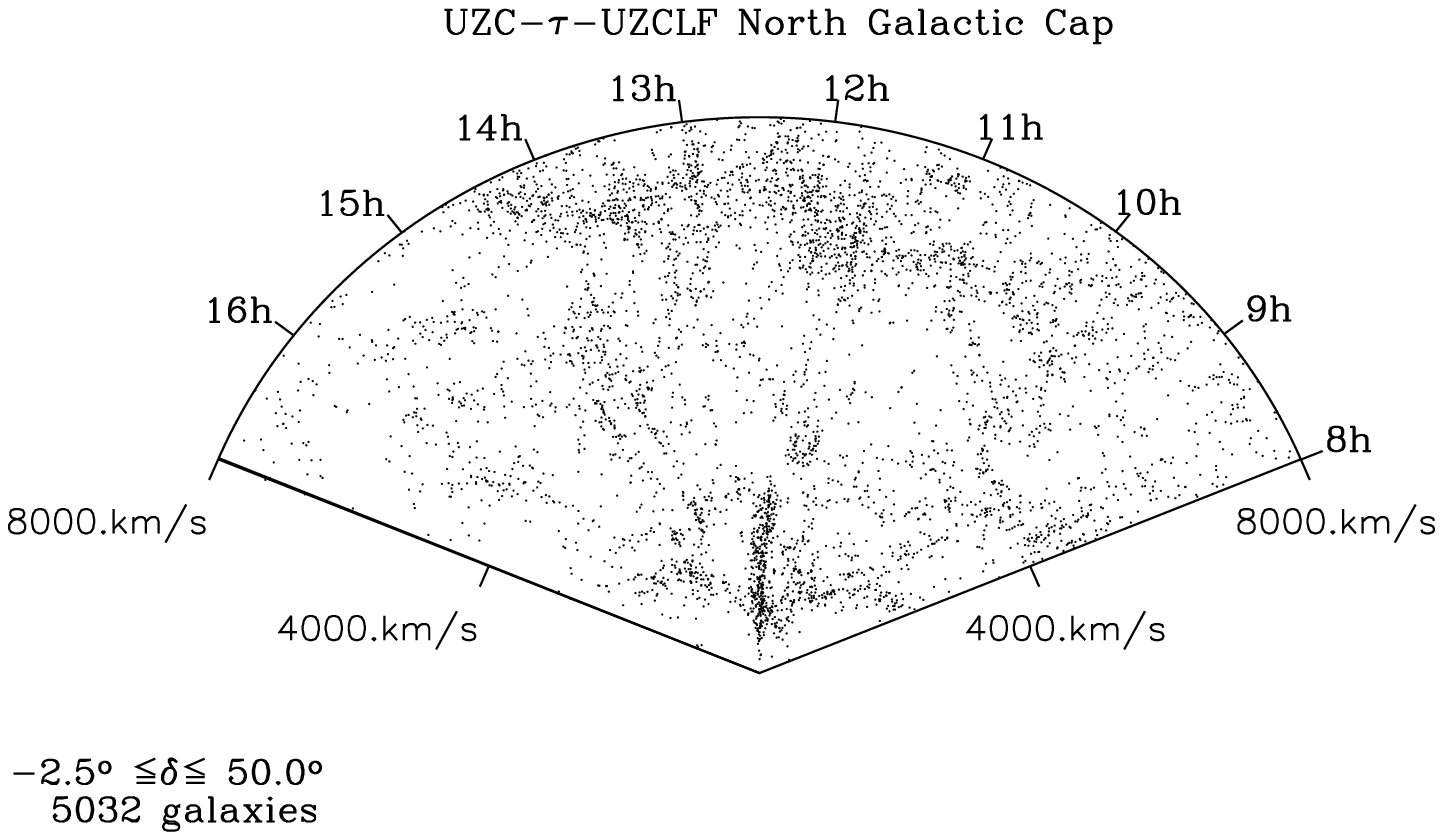}
\end{minipage}
}
\hfill
{\begin{minipage}[t]{0.49\linewidth}
\includegraphics[width=1\textwidth]{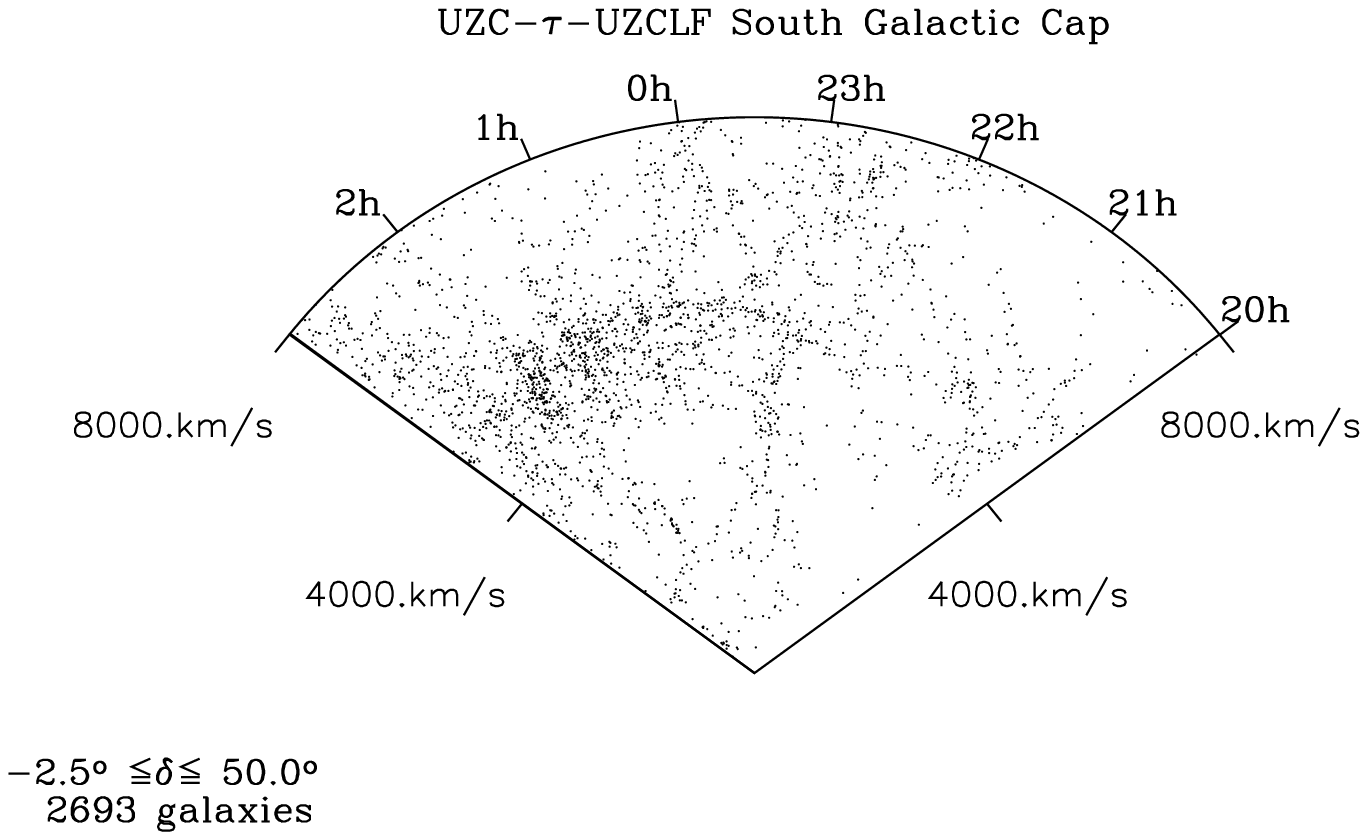}
\end{minipage}
}
{\begin{minipage}[c]{0.49\linewidth}
\includegraphics[width=1\textwidth]{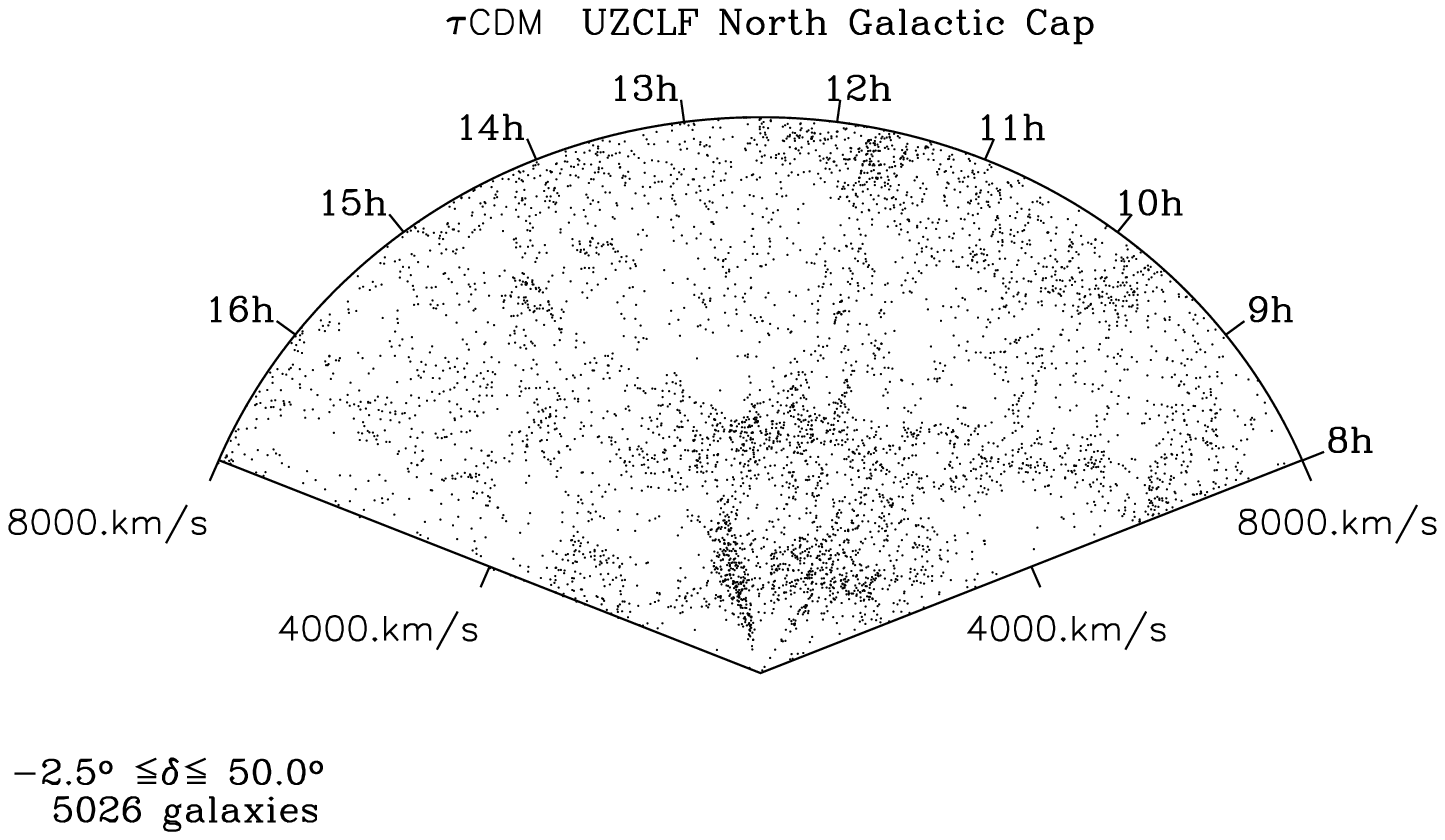}
\end{minipage}
}
\hfill
{\begin{minipage}[c]{0.49\linewidth}
\includegraphics[width=1\textwidth]{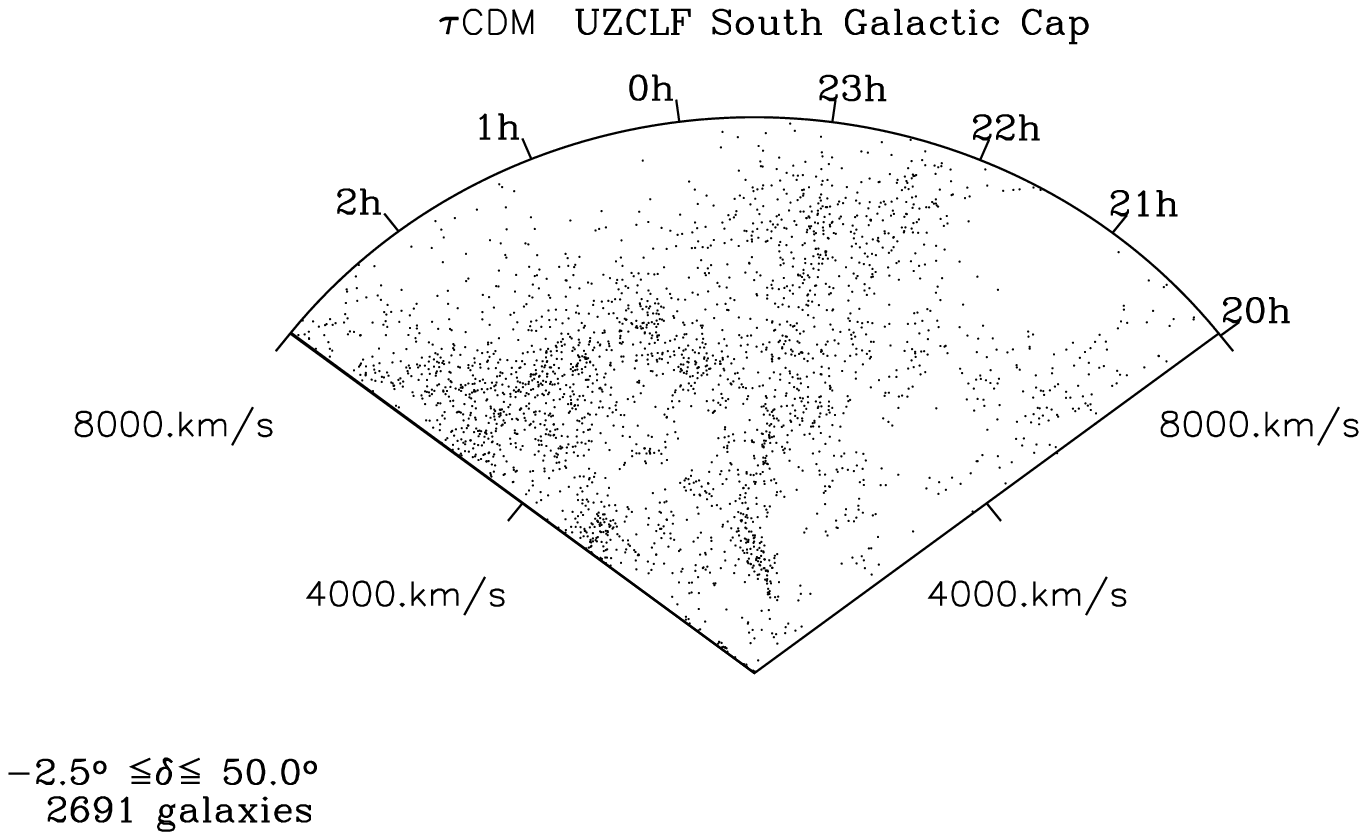}
\end{minipage}
}
\caption{The distribution of the galaxies (dots) in the magnitude-limited
UZC-$\tau$-UZCLF and in the mock
$\tau$CDM-UZCLF redshift surveys.}
\label{slicesTuzclf}
\end{figure*}

Table \ref{allgal} lists the number of galaxies in each catalogue.
Figures \ref{slicesLsalf} -- \ref{slicesTuzclf} show the UZC and the mock 
redshift surveys.

\begin{table}
\centering
\caption{Total number of galaxies in the real and mock redshift surveys.}\label{allgal}
\begin{tabular}{lcccc}
\hline
          &  \multicolumn{2}{c}{SALF}    & \multicolumn{2}{c}{UZCLF}     \\
\hline
\hline
          &  UZC      &  $\Lambda$CDM    &  UZC     &   $\Lambda$CDM     \\
NGC       &  5114     &  5037            &  5006    &   4979             \\
SGC       &  2698     &  2667            &  2692    &   2566             \\
\hline
          &  UZC      &  $\tau$CDM       &  UZC     &   $\tau$CDM        \\
NGC       &  4558     &  4621            &  5032    &    5026            \\
SGC       &  2623     &  2699            &  2693    &    2691            \\
\hline
\end{tabular}
\end{table}

\section{The group identification}\label{groups_red_space}

We identify groups in redshift space in the UZC and in our mock catalogues 
with the friends-of-friends (FOF) algorithm described in \citet*{ramella1997}.
We use the linking parameter $V_{0}= 350\;\kms$ and the number density 
contrast $\delta n/n=80$ at the fiducial velocity $V_{F}=1000\;\kms$.
These parameters minimize the fraction of interlopers and
provide the best estimates of the properties of groups identified in 
real space \citep{diaferio1999}. 
Nevertheless, these linking parameters still yield group catalogues where 
$\sim 40$ per cent of the triplets and $\sim 20$ per cent of the groups 
with four or more members are accidental superpositions of galaxies
(spurious groups). 
Therefore, following \citet{mahdavi2000} and \citet{ramella2002}, we restrict our
catalogues to groups with $N\ge 5$ members when we derive the physical properties of groups.
However, to limit the shot-noise, we include all groups with $N\ge 3$ members 
when we compute the two-point correlation 
function of groups.

\begin{figure*}
{\begin{minipage}[t]{0.49\linewidth}
\includegraphics[width=1\textwidth]{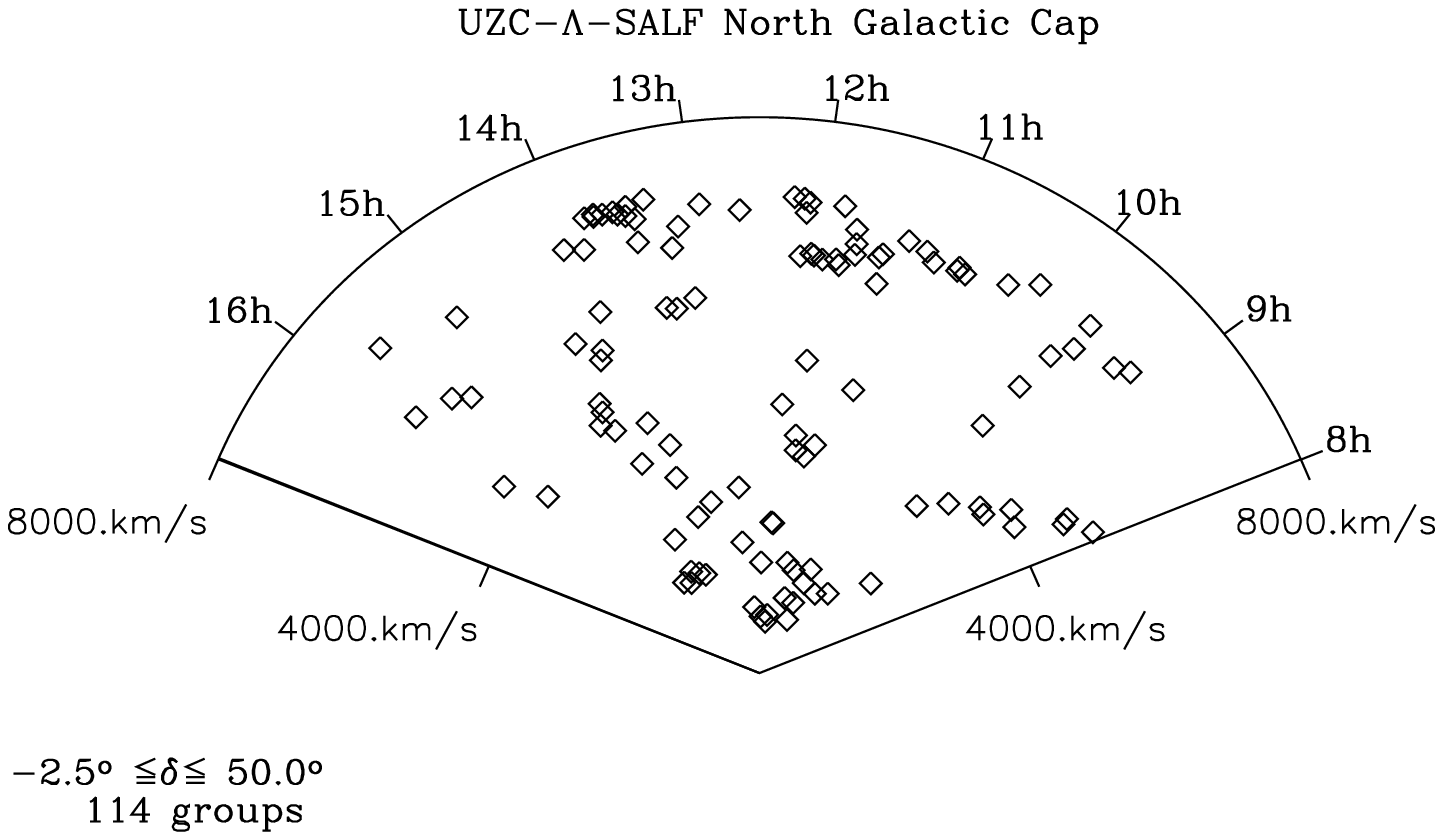}
\end{minipage}
}
\hfill
{\begin{minipage}[t]{0.49\linewidth}
\includegraphics[width=1\textwidth]{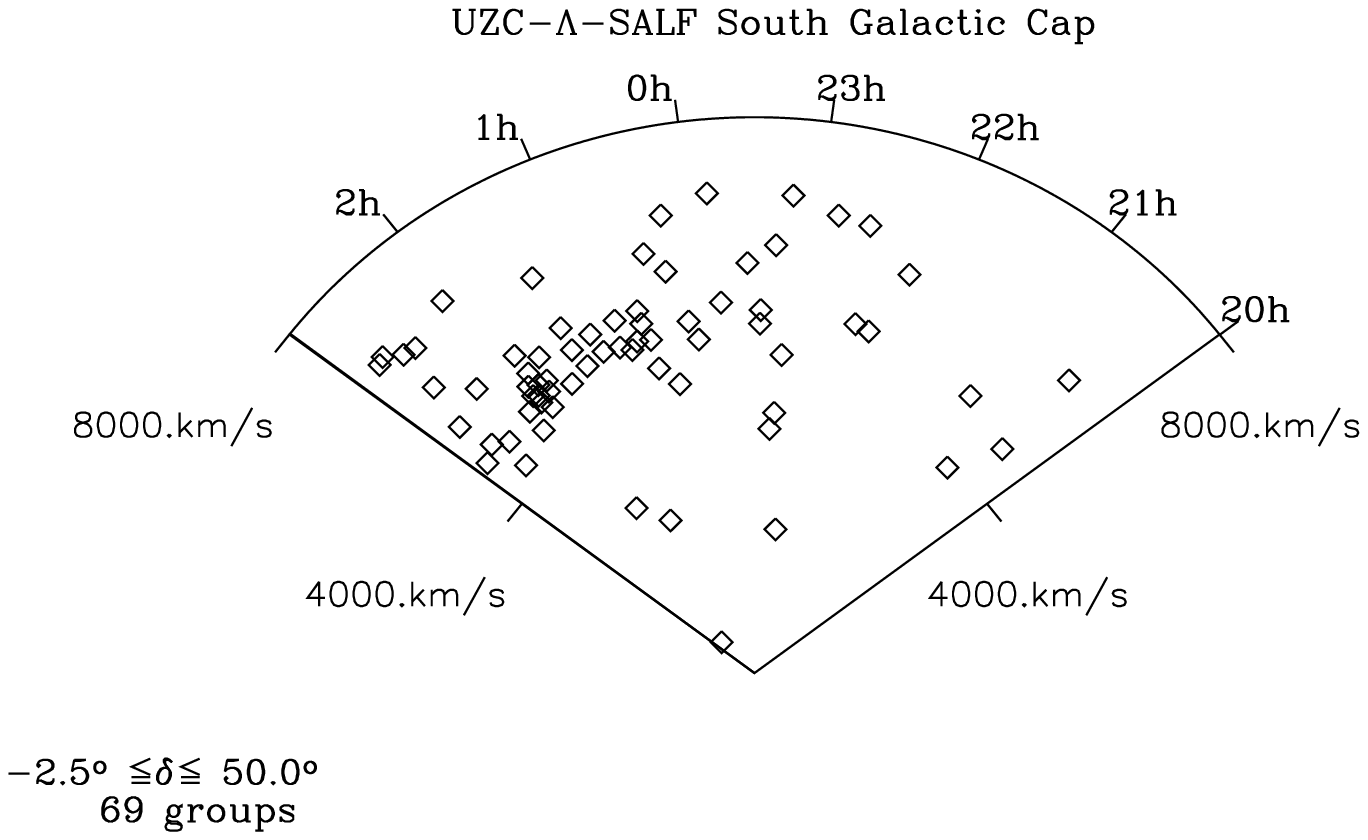}
\end{minipage}
}
{\begin{minipage}[c]{0.49\linewidth}
\includegraphics[width=1\textwidth]{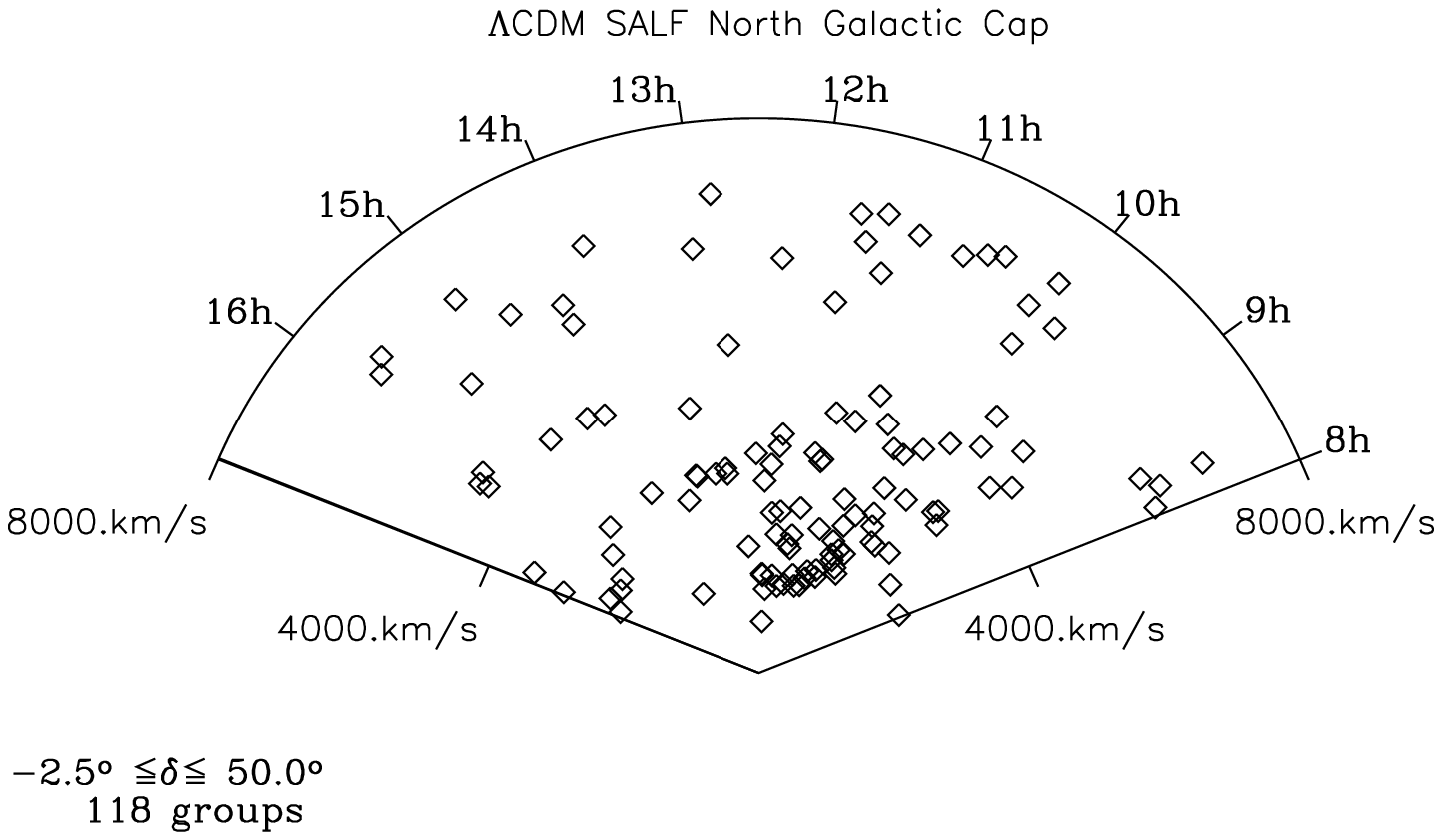}
\end{minipage}
}
\hfill
{\begin{minipage}[c]{0.49\linewidth}
\includegraphics[width=1\textwidth]{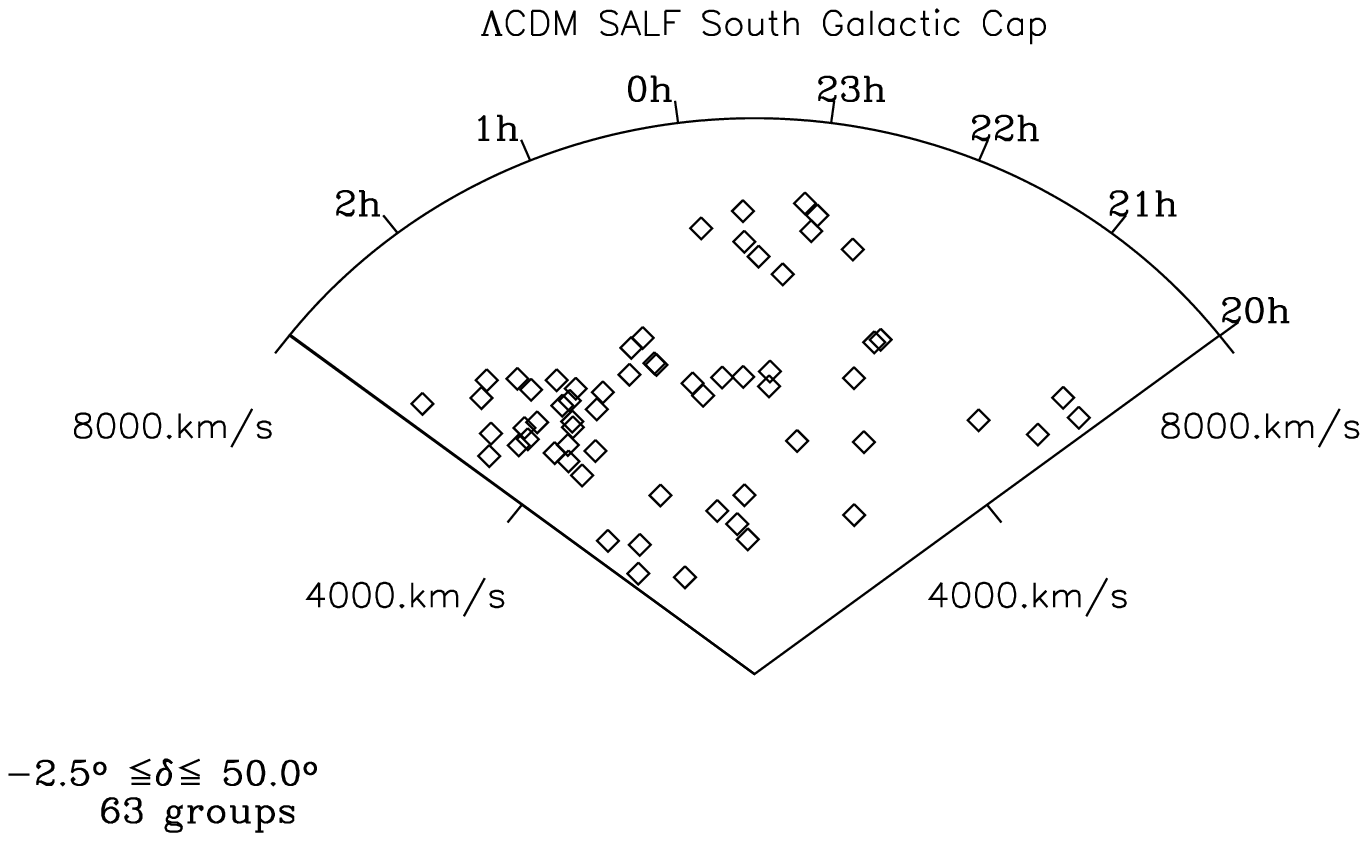}
\end{minipage}
}
\caption{The distribution of groups (diamonds) in the UZC-$\Lambda$-SALF 
and in the mock $\Lambda$CDM-SALF redshift surveys.}
\label{gr_l_salf}
\end{figure*}

\begin{figure*}
{\begin{minipage}[t]{0.49\linewidth}
\includegraphics[width=1\textwidth]{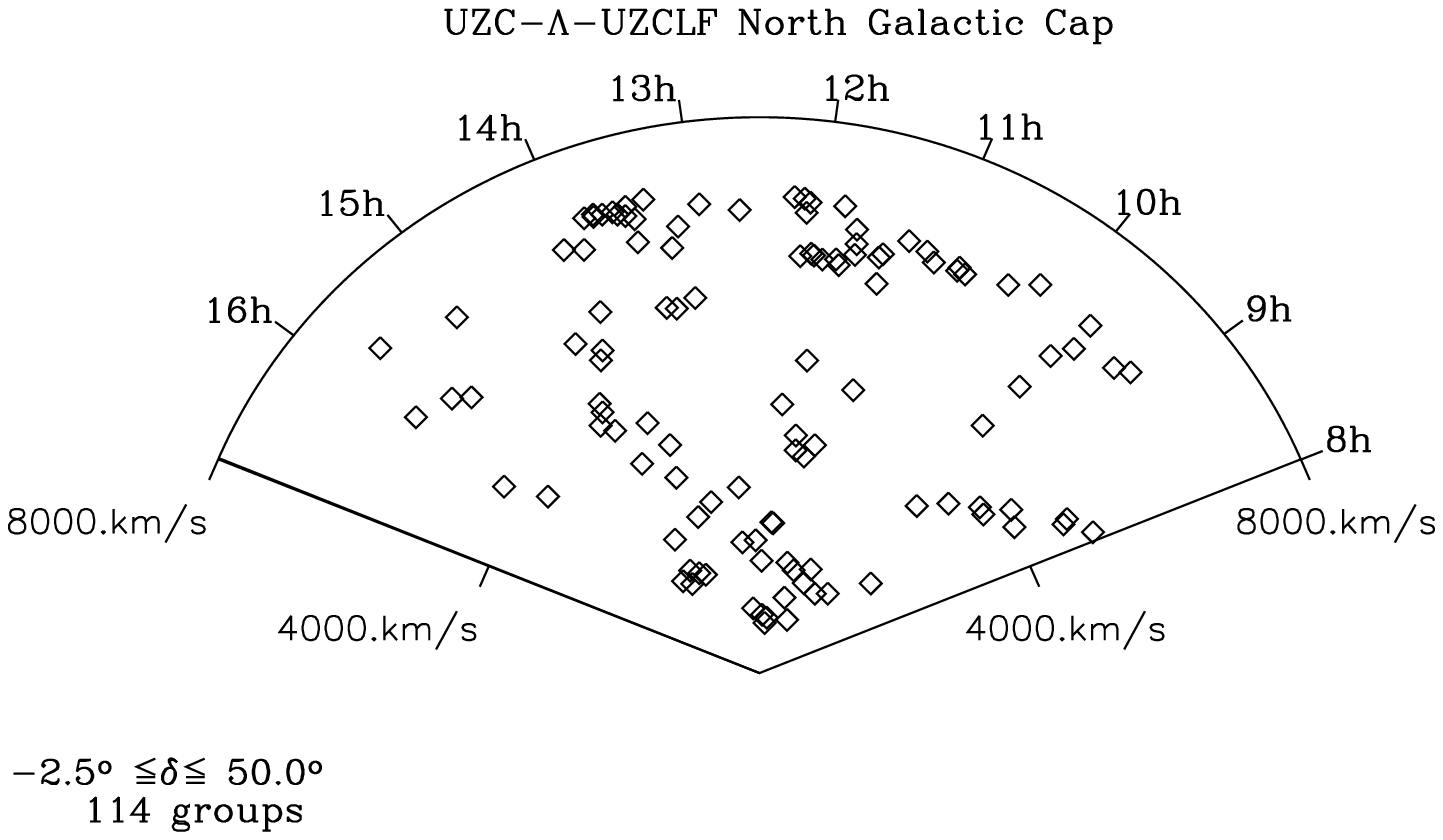}
\end{minipage}
}
\hfill
{\begin{minipage}[t]{0.49\linewidth}
\includegraphics[width=1\textwidth]{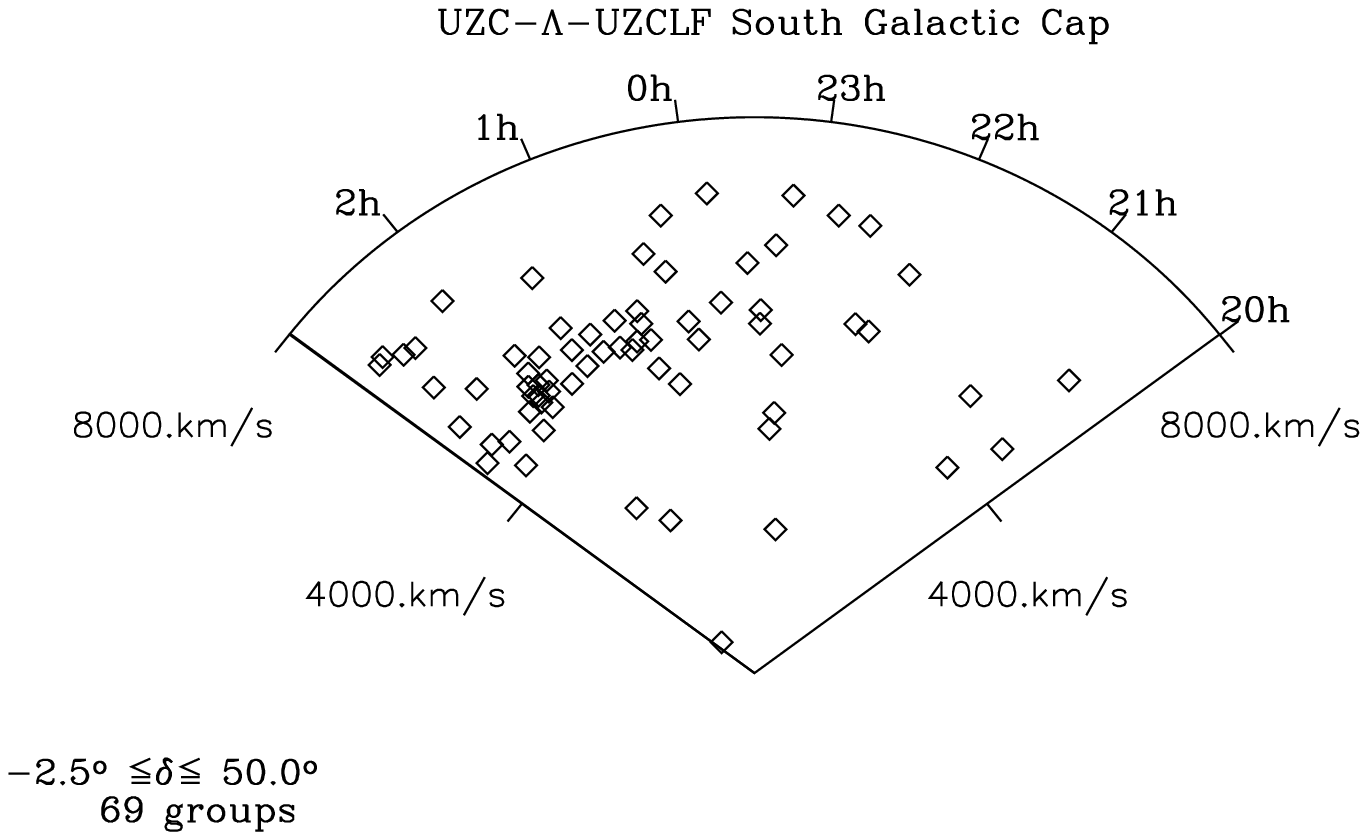}
\end{minipage}
}
{\begin{minipage}[c]{0.49\linewidth}
\includegraphics[width=1\textwidth]{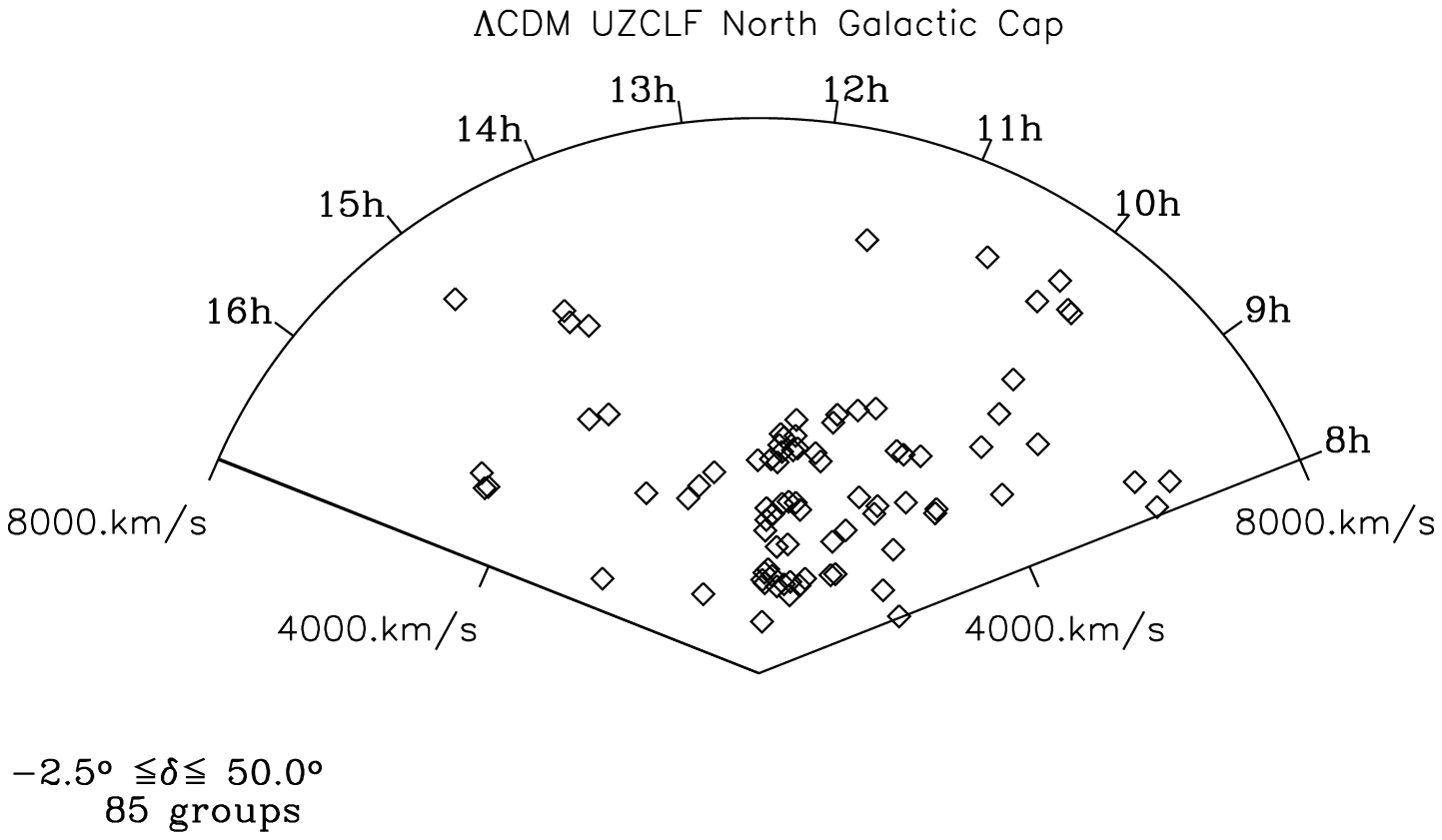}
\end{minipage}
}
\hfill
{\begin{minipage}[c]{0.49\linewidth}
\includegraphics[width=1\textwidth]{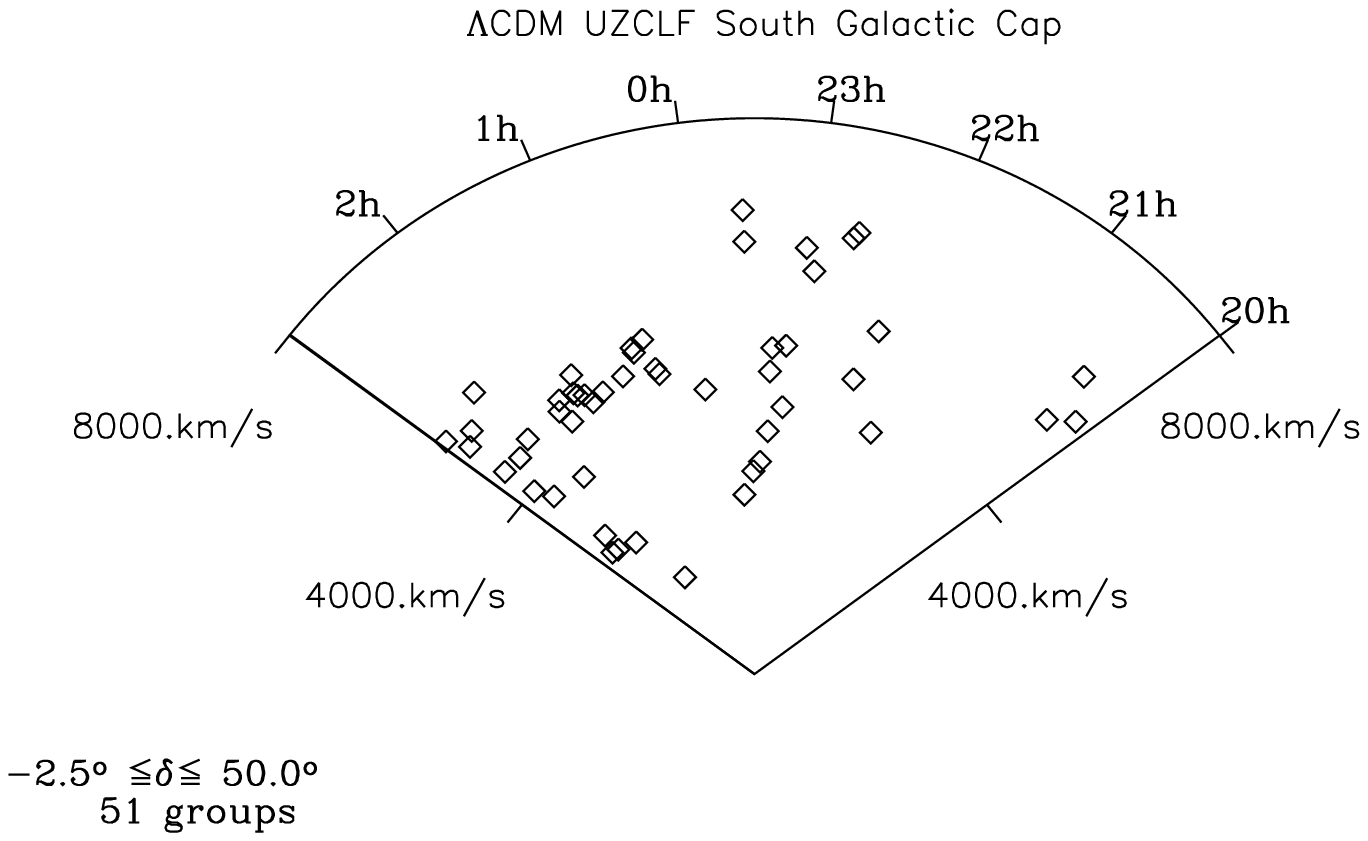}
\end{minipage}
}
\caption{The distribution of groups (diamonds) in the UZC-$\Lambda$-UZCLF
and in the mock $\Lambda$CDM-UZCLF redshift surveys.}
\label{gr_l_uzclf}
\end{figure*}

\begin{figure*}
{\begin{minipage}[t]{0.49\linewidth}
\includegraphics[width=1\textwidth]{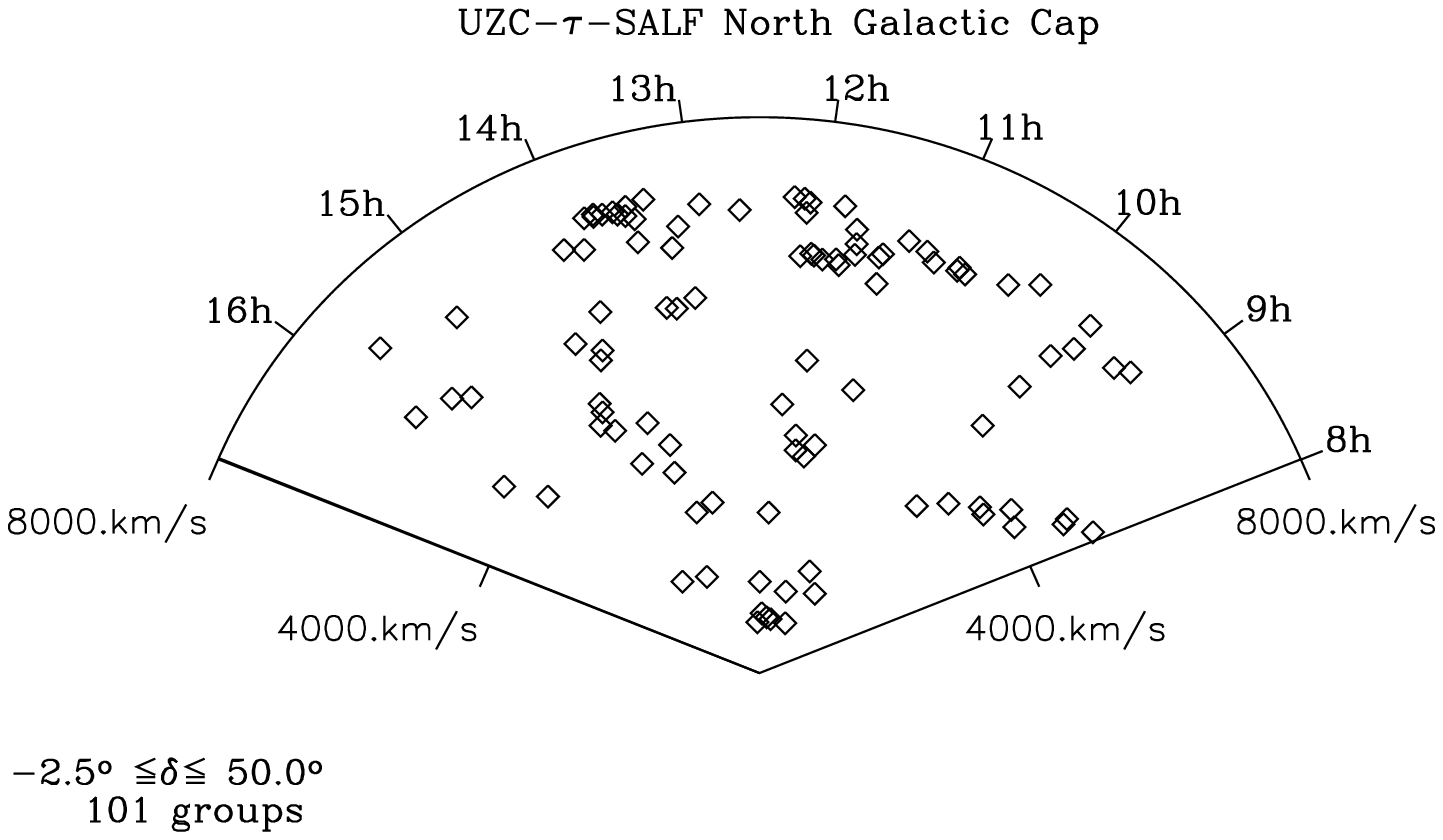}
\end{minipage}
}
\hfill
{\begin{minipage}[t]{0.49\linewidth}
\includegraphics[width=1\textwidth]{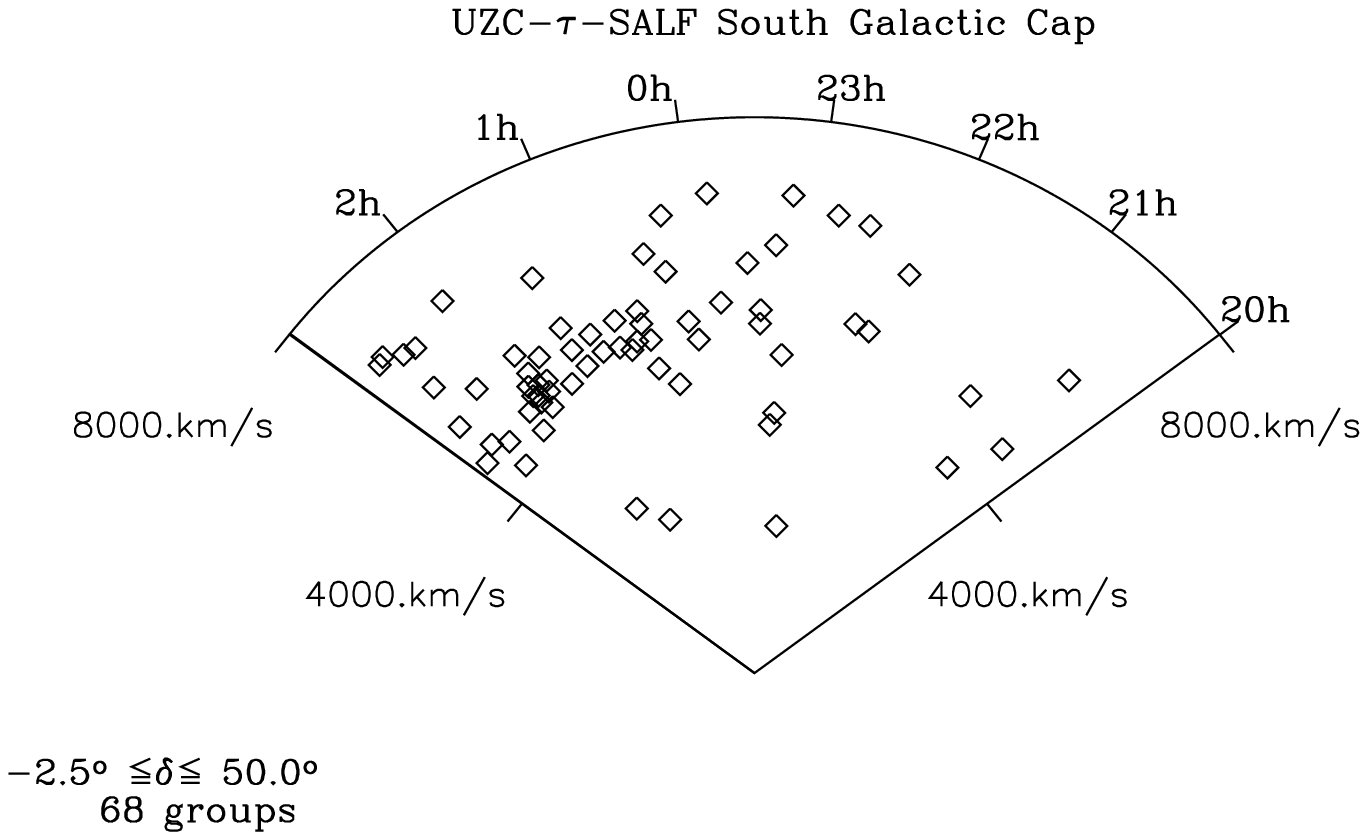}
\end{minipage}
}
{\begin{minipage}[c]{0.49\linewidth}
\includegraphics[width=1\textwidth]{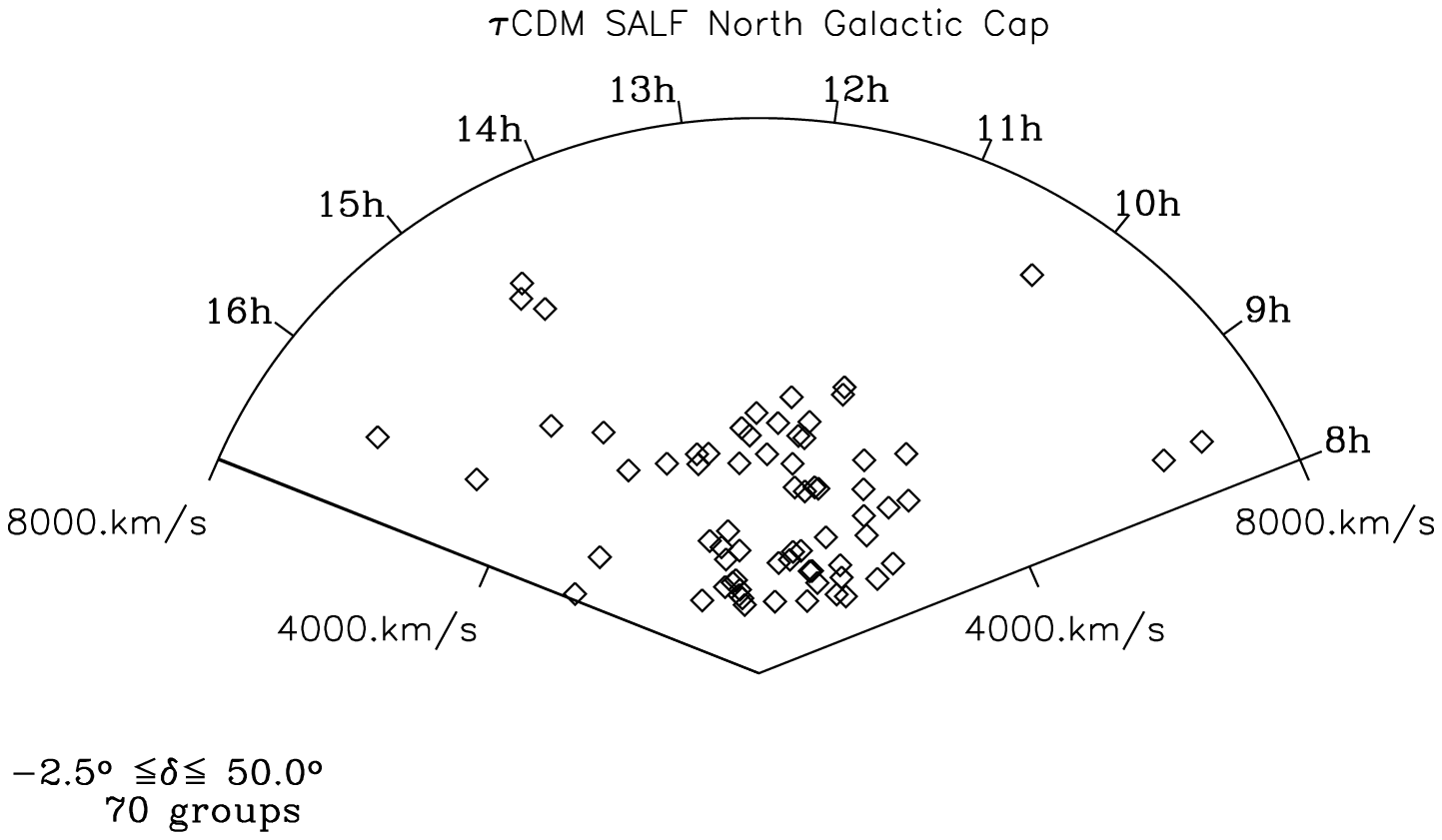}
\end{minipage}
}
\hfill
{\begin{minipage}[c]{0.49\linewidth}
\includegraphics[width=1\textwidth]{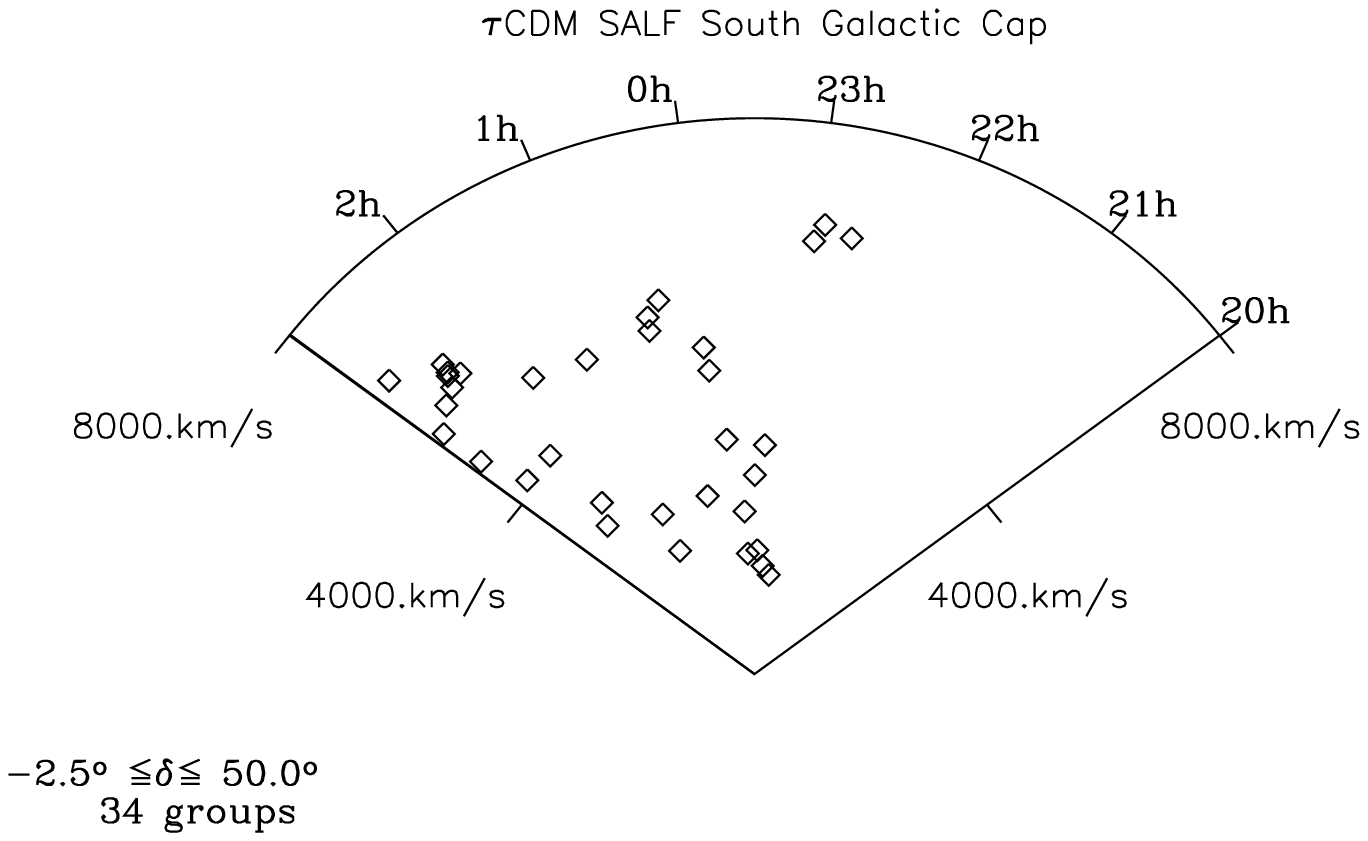}
\end{minipage}
}
\caption{The distribution of groups (diamonds) in the UZC-$\tau$-SALF
and in the mock $\tau$CDM-SALF redshift surveys.}
\label{gr_t_salf}
\end{figure*}

\begin{figure*}
{\begin{minipage}[t]{0.49\linewidth}
\includegraphics[width=1\textwidth]{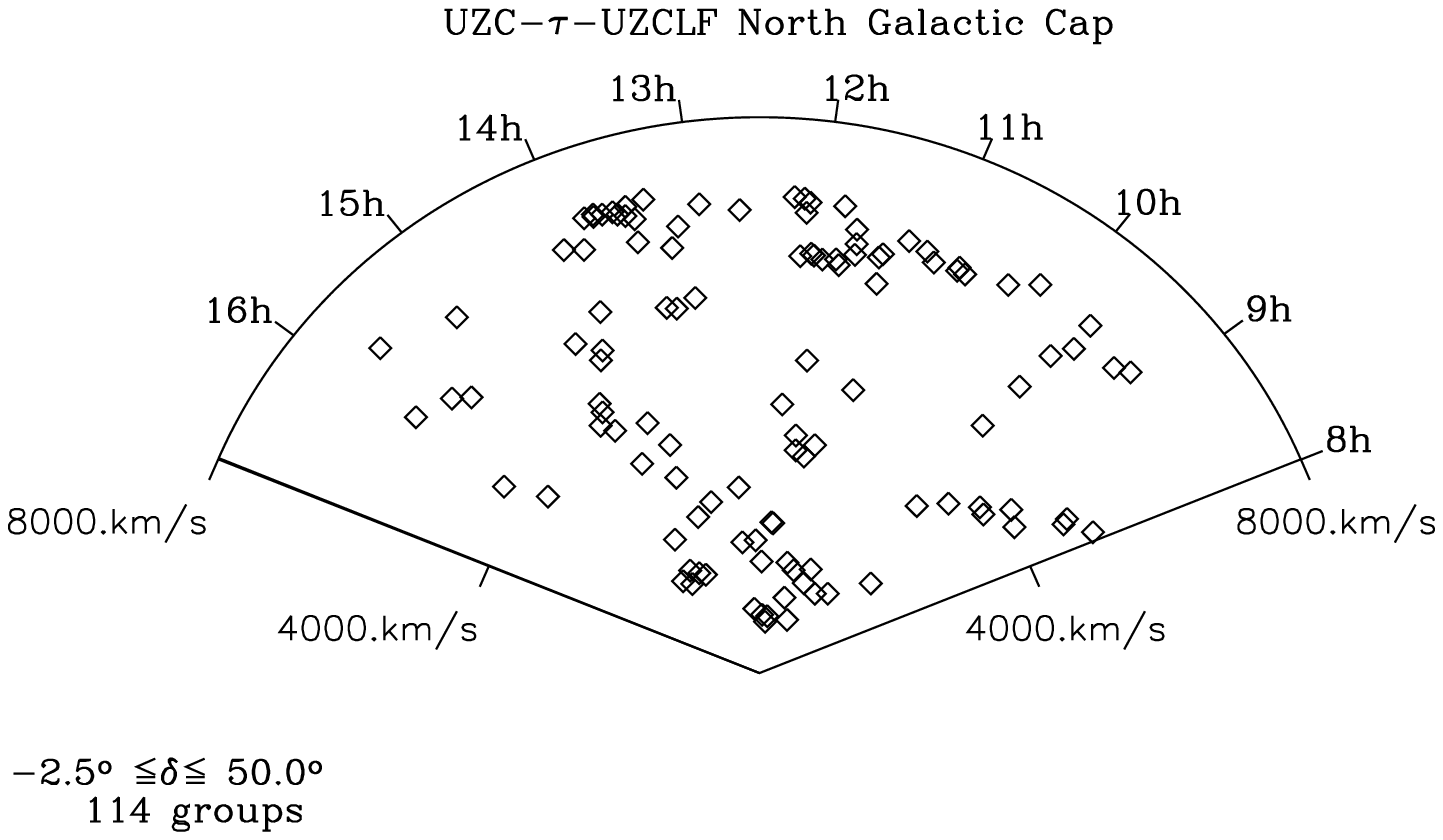}
\end{minipage}
}
\hfill
{\begin{minipage}[t]{0.49\linewidth}
\includegraphics[width=1\textwidth]{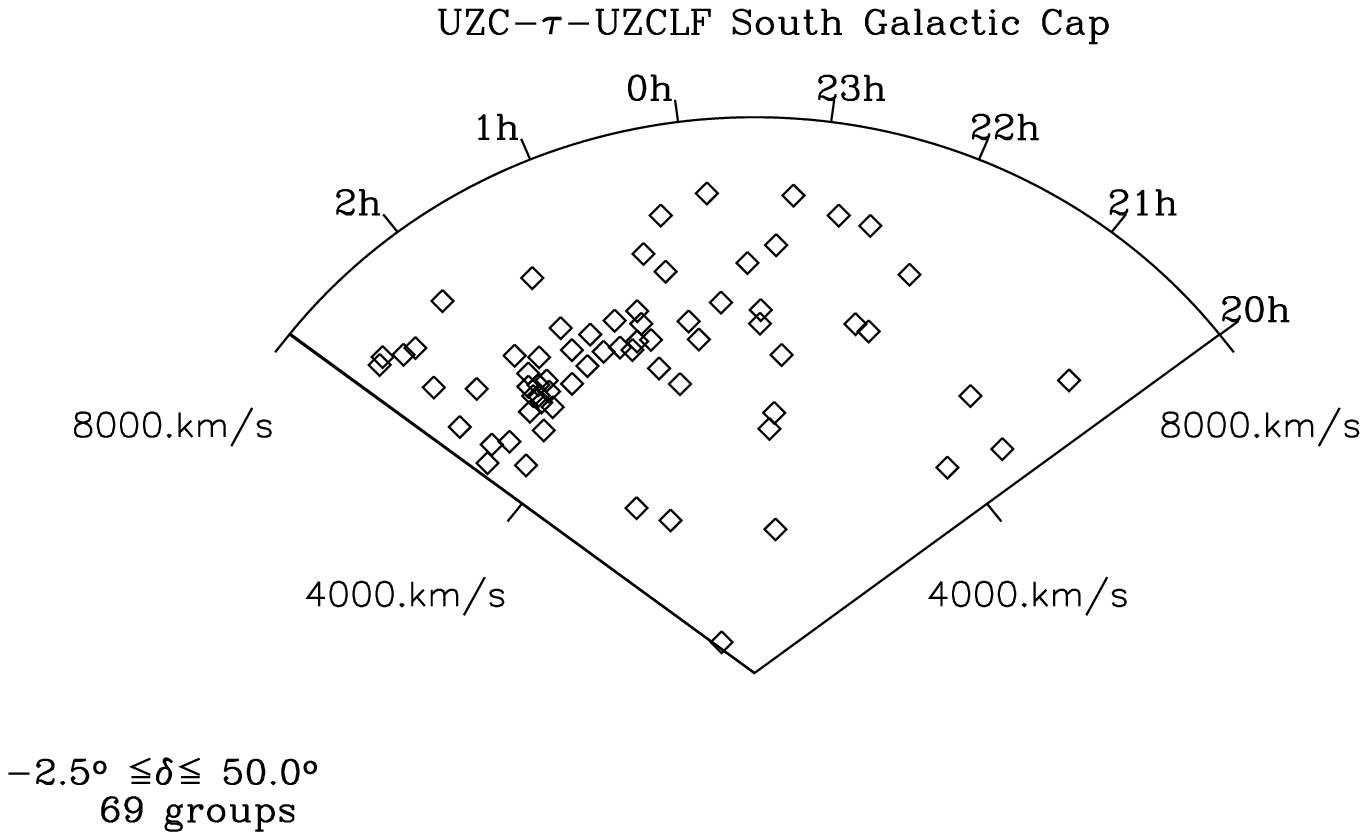}
\end{minipage}
}
{\begin{minipage}[c]{0.49\linewidth}
\includegraphics[width=1\textwidth]{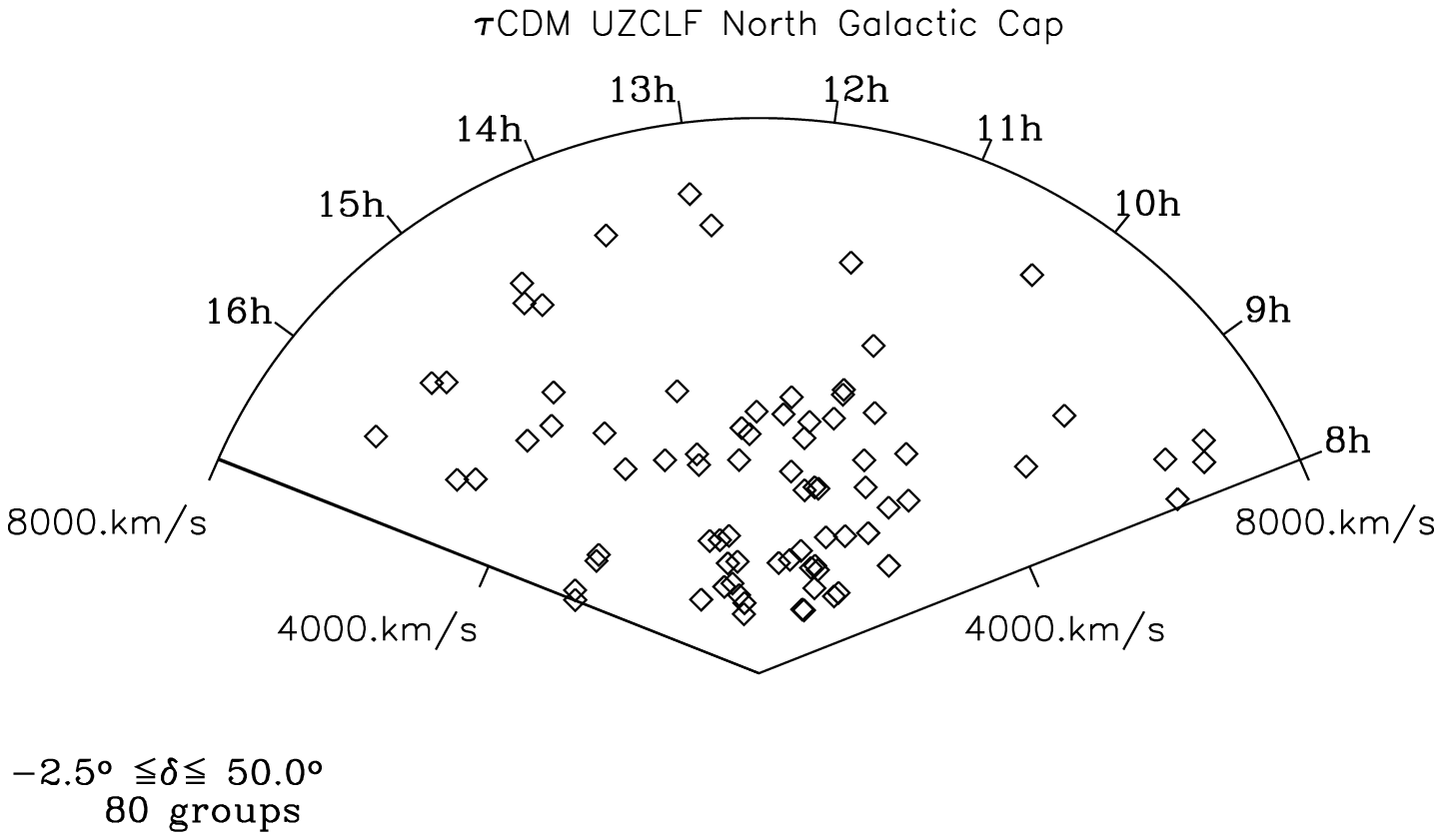}
\end{minipage}
}
\hfill
{\begin{minipage}[c]{0.49\linewidth}
\includegraphics[width=1\textwidth]{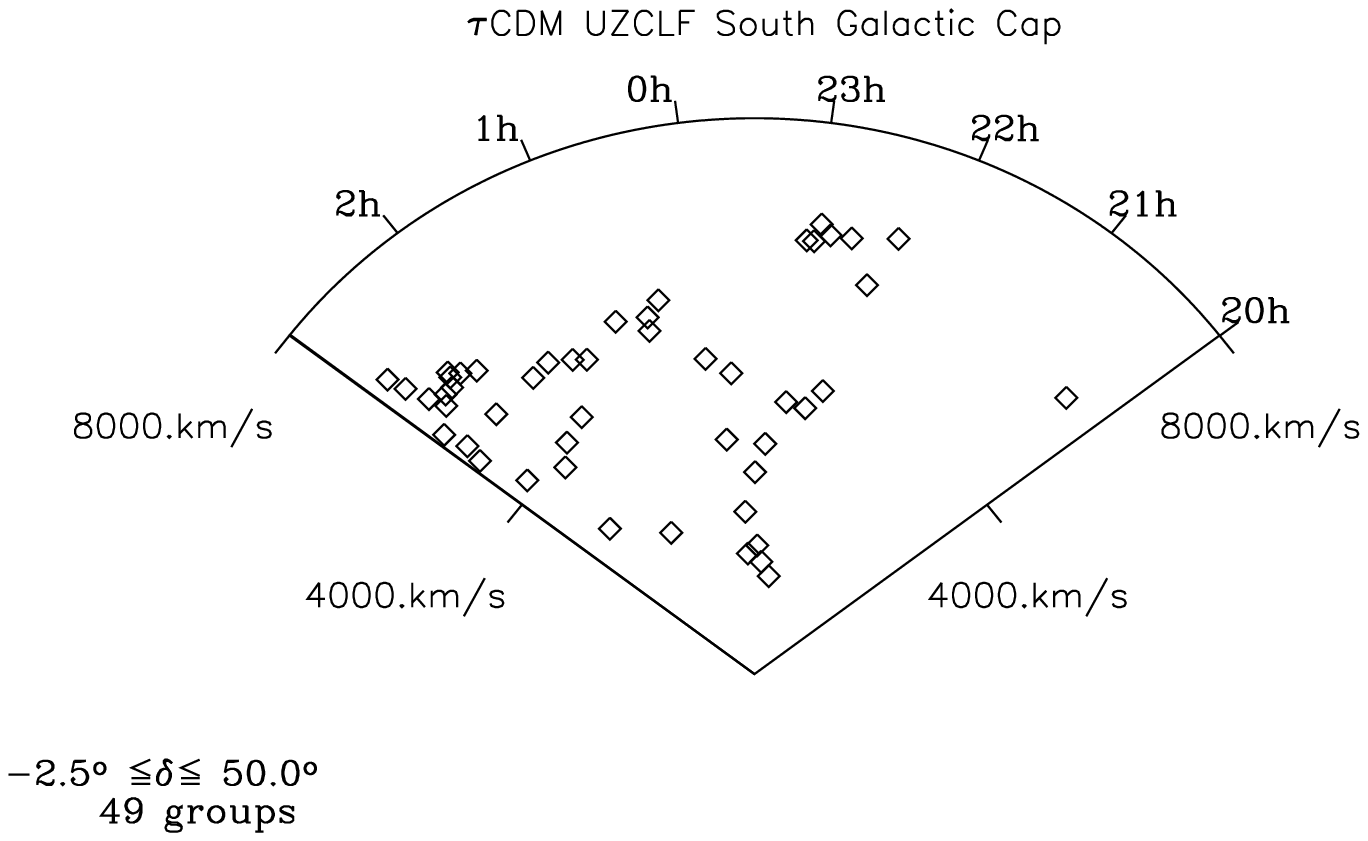}
\end{minipage}
}
\caption{The distribution of groups (diamonds) in the UZC-$\tau$-UZCLF
and in the mock $\tau$CDM-UZCLF redshift surveys.}
\label{gr_t_uzclf}
\end{figure*}

We consider groups with mean velocities in the range $500 < cz <7000\;\kms$. 
The lower limit avoids groups close to the Milky Way and the upper limit 
excludes groups which are too close to the edge of the mock survey.
Figures \ref{gr_l_salf}--\ref{gr_t_uzclf} show the location of the groups
in the UZC and in the model surveys.

\begin{table}
\centering
\caption{Number of groups and fraction of galaxies in groups.}\label{groups}
\begin{tabular}{lcccc}
\hline
          &  \multicolumn{2}{c}{SALF}    & \multicolumn{2}{c}{UZCLF}    \\
\hline
\hline
          &  UZC      &  $\Lambda$CDM    &   UZC       &   $\Lambda$CDM \\
NGC       &  114/0.27 &  118/0.35        &   114/0.26  &   85/0.24      \\
SGC       &  69/0.28  &  63/0.34         &   69/0.28   &   51/0.18      \\
\hline
          &  UZC      &  $\tau$CDM       &   UZC       &   $\tau$CDM    \\
NGC       &  101/0.23 &  70/0.16         &   114/0.27  &   80/0.13      \\
SGC       &  68/0.28  &  34/0.11         &   69/0.28   &   49/0.15      \\
\hline
\end{tabular}
\end{table}

\section{Groups as tracers of the large-scale structure}\label{visual}

Figures \ref{slicesLsalf} -- \ref{slicesTuzclf} show that both the 
$\Lambda$CDM and the $\tau$CDM models reproduce the gross features of the 
local Universe, as imposed by the initial conditions.  
However, neither model has structures as sharply defined as 
in the real Universe. In particular, the $\tau$CDM model has much looser
structures than the $\Lambda$CDM model and the UZC, 
and this reflects into much lower fractions of galaxies in groups 
(see Table \ref{groups}).

\citet{schmalzing2000} quantified this disagreement in mock redshift surveys 
extracted from the GIF simulations. They used the Minkowski functionals to 
show that the UZC has a substantially larger  degree of planarity than the 
models. They concluded that cosmic variance can be responsible for the 
inability of the models at reproducing large two-dimensional structures like 
the Great Wall, which is the main large-scale structure feature in the UZC. 
Unfortunately, the Great Wall, which is at $\sim 7000\;\kms$ from the Milky Way, 
is just on the border of our mock 
catalogues extracted from the constrained simulations. The mock catalogues do 
indeed show a concentration of galaxies at the location of the Great Wall, 
but this concentration appears substantially looser. 

We compute below the two-point correlation functions
of groups. However, this function is not exhaustive
of the large-scale structure description when the density field is not
Gaussian, as in the case of the small UZC volume. 
Therefore, we provide here an alternative simple test of clustering. 

Consider the space ${\cal S}=(\alpha,cz)$, the two-dimensional projection of the
redshift space $(\alpha,\delta,cz)$.
One can consider the galaxies as a random sample of an underlying density 
distribution $N_{2D}$ in this space. If galaxy groups are fair tracers of the 
large-scale distribution of galaxies, we expect a positive Kolmogorov-Smirnov 
(KS) test between the 
distributions of galaxies and groups in ${\cal S}$. Table \ref{grp_lss} shows 
that this is indeed the case in the UZC, at a significance level
larger than 17\%. A large significance level is also provided by 
the $\Lambda$CDM model, whereas the $\tau$CDM groups trace poorly the large-scale 
distribution of galaxies in most cases. For both models the situation
moderately improves when the UZC luminosity function is used, except for the 
NGC of the $\Lambda$CDM model. 
This disagreement is not due to the difficulty of reproducing the Great Wall, 
because cutting the catalogues at 6000 $\kms$ does not substantially improve the agreement. 

Now, if $N_{2D}$ in the models is comparable to $N_{2D}$ in the UZC, 
we also expect a positive KS test between the models and the observations. 
The KS test applied to the galaxy $N_{2D}$ fails because the initial 
conditions of the simulation are not set to reproduce the observed galaxy
distribution on scales smaller than $5h^{-1}$ Mpc. 
However, groups trace the large scale 
distribution of galaxies fairly enough, and the comparison between the $N_{2D}$ of
the groups in the UZC and in the models is appropriate.
Table \ref{grp_grp} shows that the KS tests fail when we consider the entire 
survey region; however, the significance levels increase when 
we limit the region to 6000 $\kms$, therefore excluding the
Great Wall.
 
\begin{table}
\centering
\caption{Groups of galaxies as tracers of the large-scale structure.}\label{grp_lss}
\begin{tabular}{lcccc}
\hline
          &  \multicolumn{2}{c}{SALF}    & \multicolumn{2}{c}{UZCLF}    \\
\hline
\hline
          &  UZC       &  $\Lambda$CDM  &   UZC        &   $\Lambda$CDM  \\
NGC       &  0.17/0.65 &  0.11/0.07     &   0.25/0.54  &   0.03/0.08     \\
SGC       &  0.22/0.18 &  0.32/0.13     &   0.23/0.19  &   0.37/0.20     \\
\hline
          &  UZC       &  $\tau$CDM     &   UZC        &   $\tau$CDM     \\
NGC       &  0.19/0.45 &  0.002/0.01    &   0.23/0.56  &   0.05/0.04     \\
SGC       &  0.29/0.23 &  0.02/0.03     &   0.23/0.18  &   0.21/0.24     \\
\hline
\end{tabular}
\begin{minipage}{0.48\textwidth}
\medskip
{Significance levels of the bidimensional KS test in the 
$\alpha$ -- $cz$ plane between the distributions of galaxies and groups 
in the range $cz \in [500,7000] \; / \; [500,6000] \;\kms$.}
\end{minipage}
\end{table}

\begin{table}
\centering
\caption{Comparison between the large scale distribution of the UZC and 
the simulated groups.}\label{grp_grp}
\begin{tabular}{lcccc}
\hline
          &  \multicolumn{2}{c}{SALF}    & \multicolumn{2}{c}{UZCLF}    \\
\hline
\hline
          &  $\Lambda$CDM  &  $\tau$CDM  &   $\Lambda$CDM  & $\tau$CDM   \\
NGC       &  3e-4/0.06     &  9e-8/0.03  &   2e-5/0.01     &  1e-4/0.01  \\
SGC       &  2e-3/2e-3     &  3e-4/1e-3  &   4e-3/0.01     &  0.04/0.01  \\
\hline
\hline
\end{tabular}
\begin{minipage}{0.48\textwidth}
\medskip
{Bidimensional KS test in the $\alpha$ -- $cz$ plane between 
the distributions of groups in the UZC and the models in the range 
$cz \in [500,7000] \; / \; [500,6000] \;\kms$.}
\end{minipage}
\end{table}

We finally consider groups in volume-limited catalogues (Table \ref{vol_lim}). 
The $\Lambda$CDM model agrees with the UZC better than the
$\tau$CDM model, which yields extremely low numbers of groups 
(Table \ref{vol_lim_gr}). 

All these results suggest that the structures in the $\Lambda$CDM model
are better defined than those in the $\tau$CDM model, although the
agreement with the UZC is not yet satisfactory.
The use of the UZC luminosity function allievates some
problems but it is clearly not the only relevant ingredient.

\begin{table}
\centering
\caption{Number of galaxies brighter than
$-19.02+5 \log h$ in the volume-limited catalogues.}\label{vol_lim}
\begin{tabular}{lcccc}
\hline
          &  \multicolumn{2}{c}{SALF}    & \multicolumn{2}{c}{UZCLF}     \\
\hline
\hline
          &  UZC      &  $\Lambda$CDM    &  UZC     &   $\Lambda$CDM     \\
NGC       &  1788     &  1936            &  1788    &   1431             \\
SGC       &  941      &  1011            &  941     &   479              \\
\hline
          &  UZC      &  $\tau$CDM       &  UZC     &   $\tau$CDM        \\
NGC       &  1788     &  1823            &  1788    &   2004             \\
SGC       &  941      &  875             &  941     &   655              \\
\hline
\end{tabular}
\end{table}

\begin{table}
\centering
\caption{Number of groups and fraction of galaxies in groups for
the volume-limited catalogues.}\label{vol_lim_gr}
\begin{tabular}{lcccc}
\hline
          &  \multicolumn{2}{c}{SALF}    & \multicolumn{2}{c}{UZCLF}     \\
\hline
\hline
          &  UZC      &  $\Lambda$CDM    &  UZC        & $\Lambda$CDM     \\
NGC       &  18/0.08  &   15/0.09        &  18/0.08    &   7/0.06         \\
SGC       &  17/0.15  &   15/0.13        &  17/0.15    &   1/0.01         \\
\hline
          &  UZC      &  $\tau$CDM       &  UZC        &   $\tau$CDM      \\
NGC       &  18/0.08  &  3/0.01          &  18/0.08    &   5/0.02         \\
SGC       &  17/0.15  &  0/0             &  17/0.15    &   1/0.01         \\
\hline
\end{tabular}
\end{table}

\subsection{The correlation function of galaxies and groups}\label{corref}

In the redshift space of the local Universe, 
the vector ${\bf s}_{i}=c z_{i} {\bf r}_{i}$ locates a 
galaxy with redshift $ z_{i} \ll 1$ and celestial coordinates ${\bf r}_{i} =
(\alpha_{i},\delta_{i})$. For small angular separations (smaller than $50^{\circ}$ in
our analysis), the components along the line of sight ($\pi$)
and projected onto the sky ($r_p$) of the pairwise galaxy separation ${\bf
  s}={\bf s}_{i}-{\bf s}_{j}$ are 
\begin{equation}
\pi=\frac{{\bf s \cdot l}}{|{\bf l}|},\phantom{bhuuu!} r^{2}_{p}=s^{2}-\pi^{2},
\end{equation}
where ${\bf l}=({\bf s}_{i}+{\bf s}_{j})/2$. The two-dimensional redshift
space correlation function $\xi(r_{p},\pi)$ measures the excess probability,
compared to a Poisson distribution, that a galaxy pair has separation $(r_{p},
\pi)$. To measure $\xi(r_{p},\pi)$, we compute the distribution of pair
separations in the data and in a Poisson realization of the data with the same
radial and angular selection criteria 
\begin{equation}
\xi(r_{p},\pi)=\frac{DD(r_{p},\pi)}{DR(r_{p},\pi)}\frac{n_R}{n_D}-1 \; ,
\end{equation} 
where DD and DR are the weighted sums (see below) of the data/data and data/random pairs
with separation $r_{p}$ and $\pi$, and $n_D$ and $n_R$ are the mean
densities of the real and random samples, respectively.

The points in the random sample are radially distributed according to the
selection function 
\begin{equation}
\varphi(z)=\frac{\int_{-\infty}^{M(z)}\phi(M)\textrm{d}M}{\int_{-\infty}^
{M(z_{\textrm{\tiny{min}}})}\phi(M)\textrm{d}M} \; , 
\end{equation}
where the luminosity function 
$\phi(M)$ has the parameters, appropriate to the model considered, 
listed in Table \ref{tab_lum_fun}. We consider 
only galaxies and groups in the range $cz=[500,7000]$ $\kms$. 
$M(z_{\textrm{\tiny{min}}})$ is the absolute magnitude correponding to
$m_{\rm lim}$ at the fiducial minimum redshift $cz_{\rm min}=500\,\kms$; 
$M(z)$ is the
absolute magnitude corresponding to $m_{\rm lim}$ at any given redshift $z$. 
We constrain $M(z)$ to be brighter than the luminosity resolution of the
corresponding catalogue. We adopt the usual assumption 
that the group and galaxy selection functions coincide
(\citealt*{ramella90}; \citealt{frederic95}; 
 \citealt{trasarti97}; \citealt{girardi00}).

In magnitude-limited samples, the pair sums DD and DR are weighted 
to correct for the rapid decrease of the galaxy density with distance.
In its simplest form, the weight is given by the
inverse of the selection function $\varphi(z)$. A more appropriate approach,
that we apply here, is the minimum-variance estimate \citep{davis82}. This
approach requires the {\it a-priori} knowledge of the volume integral of the 
real-space correlation function
$\xi(r)$. However, the weights are robust against different choices 
of $\xi(r)$; moreover, because this choice affects each 
catalogue in the same way, the conclusions of our comparison remain unchanged. 
We adopt a power-law correlation
function $\xi(r)=(r/r_0)^{-\gamma}$, with $r_0=5.8 h^{-1}$ Mpc 
and $\gamma=1.8$ for both galaxies and groups. 

\begin{figure*}
\begin{center}
\includegraphics[scale=0.45]{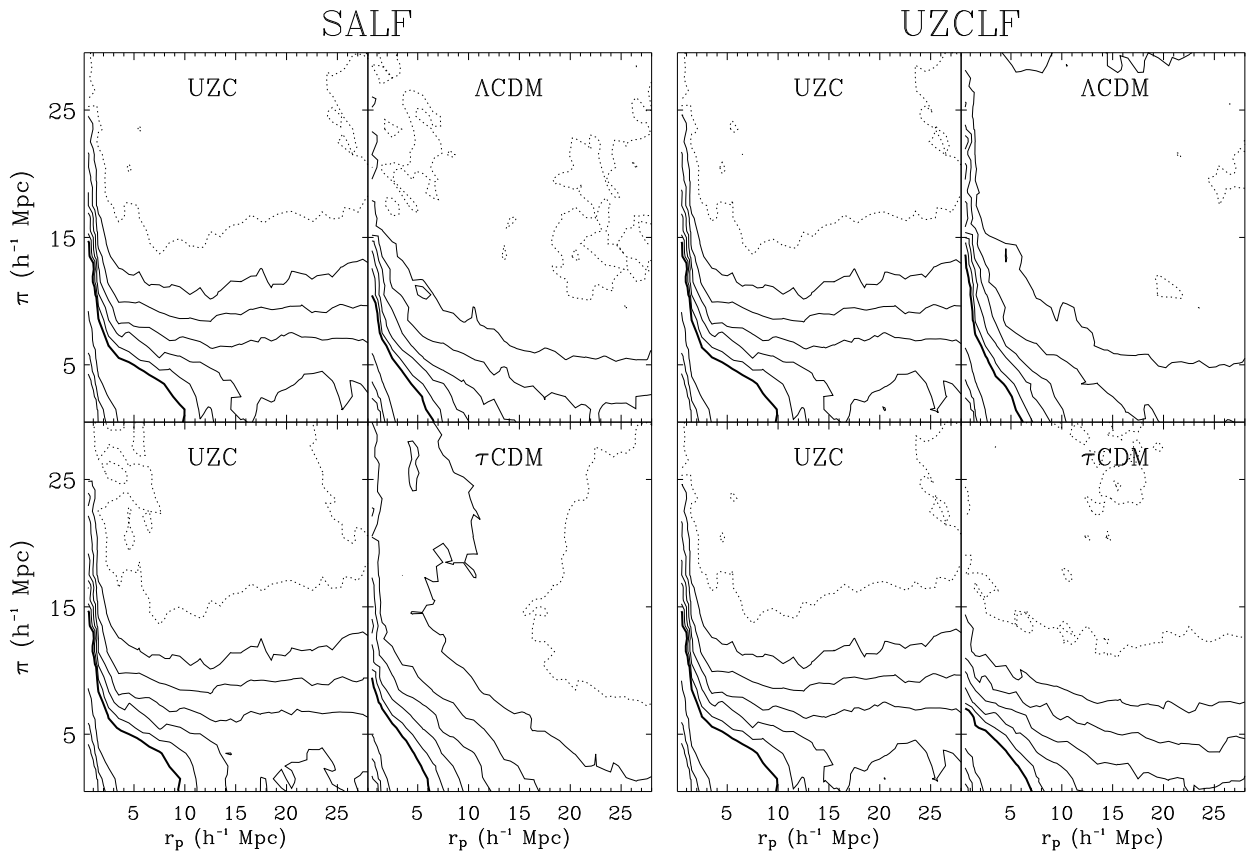}
\caption{The galaxy two-dimensional redshift-space correlation function
  $\xi(r_{p},\pi)$. The bold contour indicates $\xi(r_{p},\pi)=1$. The contour
  levels are separated by logarithmic intervals of 0.2 for $\xi(r_{p},\pi)>1$
and by linear intervals of 0.2 for $\xi(r_{p},\pi)<1$. The dotted contours
  show $\xi(r_{p},\pi)<0$.}\label{rpi}
\end{center}
\end{figure*}

Figure \ref{rpi} shows the $\xi(r_{p},\pi)$ maps of our SALF and UZCLF
galaxy catalogues. These figures clearly show that the models 
have a substantially weaker clustering at large $r_p$ and small $\pi$. This
quantifies the lack of large and coherent structures. The Finger-of-Gods
features (at large $\pi$ and small $r_p$) are instead well reproduced,
a result that indicates that the virial regions of clusters are
correctly represented in the models.
  
We also compute the two-point redshift-space correlation function $\xi(s)$, where the
separation $s$ is 
\begin{equation}
s=\frac{c}{H_0}\sqrt{z_{i}^{2}+z_{j}^{2}-2 z_{i} z_{j} \cos \theta_{ij}}\; ,
\end{equation}
and $\theta_{ij}$ is the angular separation of the two objects (galaxies or
groups) with redshift $z_i$ and $z_j$. $\xi(s)$
is more appropriate to quantify the
clustering of groups, because the small number of objects 
makes $\xi(r_{p},\pi)$ too noisy. To improve 
the statistics, we include systems with three or more members 
when computing the group $\xi(s)$.
\begin{figure*}
\begin{center}
\includegraphics[scale=0.5]{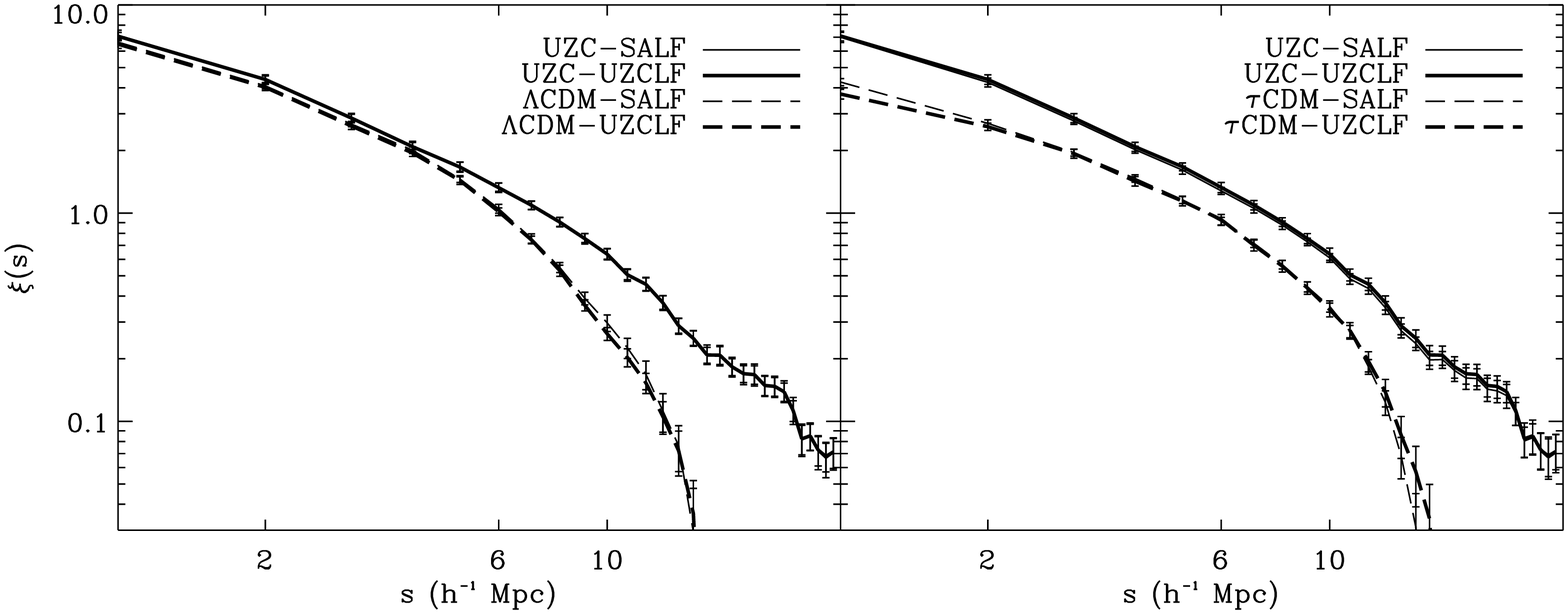}
\caption{The galaxy redshift-space correlation function. The error bars 
  are 1-$\sigma$ bootstrap errors.}\label{xigg}
\end{center}
\end{figure*}

\begin{figure*}
\begin{center}
\includegraphics[scale=0.5]{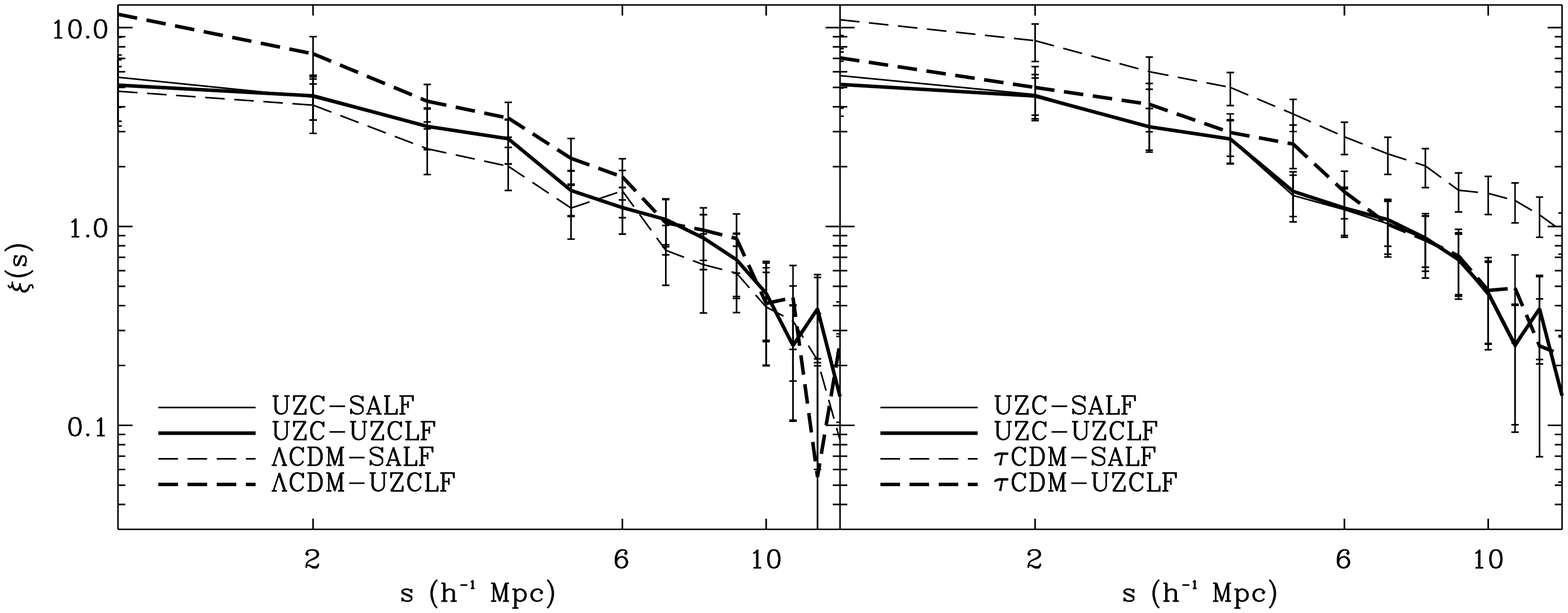}
\caption{The group redshift-space correlation function. The error bars are
  1-$\sigma$ bootstrap errors.}\label{xiGG}
\end{center}
\end{figure*}

To estimate the uncertainties we use a bootstrap procedure with 50 resampled samples in
the case of galaxies and 400 in the case of groups. 

Figures \ref{xigg} and \ref{xiGG} show $\xi(s)$ for galaxies and 
groups. The fits $\xi(s)=(s/s_{0})^{-\gamma}$ in the
interval $2-10h^{-1}$ Mpc for each model are summarized in Tables \ref{fitgg} 
and \ref{fitGG}.
\begin{table}
\centering
\caption{Fit parameters of the redshift-space correlation function of galaxies.}\label{fitgg}
\begin{tabular}{lcc}
\hline
                    & $s_{0}$       & $\gamma$      \\
                    & $h^{-1}$ Mpc  &               \\
\hline
\hline
UZC-$\Lambda$-SALF            & $7.3 \pm 0.3$  & $1.18 \pm 0.01$ \\
$\Lambda$CDM-SALF   & $5.5 \pm 0.1$  & $1.63 \pm 0.01$ \\
\hline
UZC-$\Lambda$-UZCLF           & $7.3 \pm 0.2$  & $1.19 \pm 0.01$ \\
$\Lambda$CDM-UZCLF  & $5.4 \pm 0.1$  & $1.67 \pm 0.01$ \\
\hline
UZC-$\tau$-SALF            & $7.1 \pm 0.2$  & $1.19 \pm 0.01$  \\
$\tau$CDM-SALF      & $5.1 \pm 0.1$  & $1.25 \pm 0.01$  \\
\hline
UZC-$\tau$-UZCLF           & $7.3 \pm 0.3$  & $1.18 \pm 0.01$  \\
$\tau$CDM-UZCLF     & $5.0 \pm 0.2$  & $1.25 \pm 0.01$  \\
\hline
\end{tabular}
\end{table}

\begin{table}
\centering
\caption{Fit parameters of the redshift-space correlation function of groups with 3 or more 
members.}\label{fitGG}
\begin{tabular}{lcc}
\hline
                    & $s_{0}$       & $\gamma$      \\
                    & $h^{-1}$ Mpc  &               \\
\hline
\hline
UZC-$\Lambda$-SALF            & $6.9 \pm 1.0$  & $1.37 \pm 0.03$ \\
$\Lambda$CDM-SALF   & $6.0 \pm 1.0$  & $1.39 \pm 0.04$ \\
\hline
UZC-$\Lambda$-UZCLF           & $6.9 \pm 1.0$  & $1.38 \pm 0.03$ \\
$\Lambda$CDM-UZCLF  & $7.5 \pm 0.8$  & $1.65 \pm 0.03$ \\
\hline
UZC-$\tau$-SALF            & $6.8 \pm 1.0$  & $1.40 \pm 0.03$  \\
$\tau$CDM-SALF      & $14.5 \pm 2.0$ & $1.15 \pm 0.02$  \\
\hline
UZC-$\tau$-UZCLF           & $6.9 \pm 1.0$  & $1.38 \pm 0.03$  \\
$\tau$CDM-UZCLF     & $7.4 \pm 1.1$  & $1.49 \pm 0.04$  \\
\hline
\end{tabular}
\end{table}

Our results confirm the measure of $\xi(s)$ by \citet{mathis2002} 
(our Figure \ref{xigg} is equivalent to their Figure 16): the $\tau$CDM model provides a 
galaxy $\xi(s)$ which is systematically
lower than the UZC $\xi(s)$ by a factor 0.6 to 3. The $\Lambda$CDM model
provides a much better match on scales smaller than $ 5 h^{-1}$ Mpc, whereas its $\xi(s)$
is up to a factor 2 smaller at larger scales.
We remind that the initial conditions are drawn from the IRAS 1.2 Jy survey
rather than from the UZC optically-selected galaxies, and indeed \citet{mathis2002}
show that the $\xi(s)$'s of the model and of the PSCz survey do agree on all scales
in the $\Lambda$CDM model and on scales larger than $5 h^{-1}$ Mpc in the 
$\tau$CDM model.
To compare the models with the PSCz correlation function, however, \citet{mathis2002} assign 
a far-infrared luminosity to each galaxy by matching
the simulated infrared luminosity function to the observed IRAS luminosity function,
a procedure similar to the reassignment of the optical luminosities
we apply to compile the UZCLF catalogues.
Therefore, the fact that this by-hand luminosity assignments can yield satisfactory results in
some bands but not in others suggests that 
the clustering power of the underlying dark matter distribution
is probably correct, whereas the model describing how 
galaxies acquire their luminosity in the various bands has to be 
improved.

Unlike the galaxy $\xi(s)$, the group $\xi(s)$ depends on
the adopted optical luminosity function. 
$\Lambda$CDM-SALF model is in better
agreement with observations than the $\Lambda$CDM-UZCLF, which tends to give a higher
degree of clustering on smaller scales; in any case both models agree 
with the UZC $\xi(s)$ within the errors. 
The $\tau$CDM-UZCLF model also agrees with observations, whereas
the $\tau$CDM-SALF is at least 2-$\sigma$ above the UZC.
The higher degree of clustering of the $\tau$CDM-SALF model 
is a consequence of the fact that most groups reside in the nearby region around Virgo
(see Figure \ref{gr_t_salf}).

Our results show that the amplitude of the UZC group correlation
function is smaller than or equal to 
the amplitude of the galaxy correlation function 
(\citealt{jing88}; \citealt{ramella90}). 
This lower amplitude is confirmed in the much larger group sample extracted from the 2dF
survey \citep{padilla04}. 
Other work suggests instead that groups in small
(\citealt{girardi00};
\citealt{giuricin01}; \citealt{padilla01}) and large samples
\citep*{zandivarez03} are more clustered than galaxies.
Most of this disagreement however is likely to be due to the fact
that many poor groups are spurious and decrease the amplitude of $\xi(s)$ toward
the $\xi(s)$ of field galaxies. Indeed the normalization $s_0$ systematically increases
with the exclusion of groups with fewer and fewer members \citep{padilla04}.
In fact, we also find that, in the UZC, the $\xi(s)$ of groups 
with at least five members has $s_{0}=8.8\pm2.1h^{-1}$Mpc and $\gamma=1.30\pm0.05$,
and is larger than that of galaxies.

\section{The light distribution in the nearby universe}

In this section we investigate the luminosity content of 
individual groups. We study the group luminosity function, the mass-to-light
ratio, the halo occupation number and the light distribution among galaxy 
members. 

\subsection{The luminosity function and abundances of groups}\label{groups_lum_fun}

To estimate the space density of redshift-space groups
we weight each group by $1/\Gamma_{\rm max}$, where
\begin{equation}
\Gamma_{\rm max}=\frac{\omega}{3} \left(\frac{cz_5}{H_0}\right)^{3} 
\left[1-\frac{3cz_5}{2}(1+q_0)\right]
\end{equation} 
is the volume sampled by the group. In fact, $cz_5/H_0$ is the smallest 
distance between our group distance cut off $7000\,\kms$ and the maximum 
distance beyond which the fifth brightest galaxy member would be fainter than 
the apparent magnitude limit $m_{\rm lim}$; moreover, $\omega$ is the solid angle 
of the catalogue, $c$ the light speed, and $q_0$ the deceleration 
parameter which we suppose to be unknown and set equal to 0.5. This choice 
is irrelevant, because, for the small distances considered here, the correct 
value $q_0=-0.55$ for the $\Lambda$CDM model yields a $\Gamma_{\rm max}$ which 
differs by less than 3 per cent from the $\Gamma_{\rm max}$ estimated with 
$q_0=0.5$.

\begin{figure*}
\begin{center}
\includegraphics[scale=0.44]{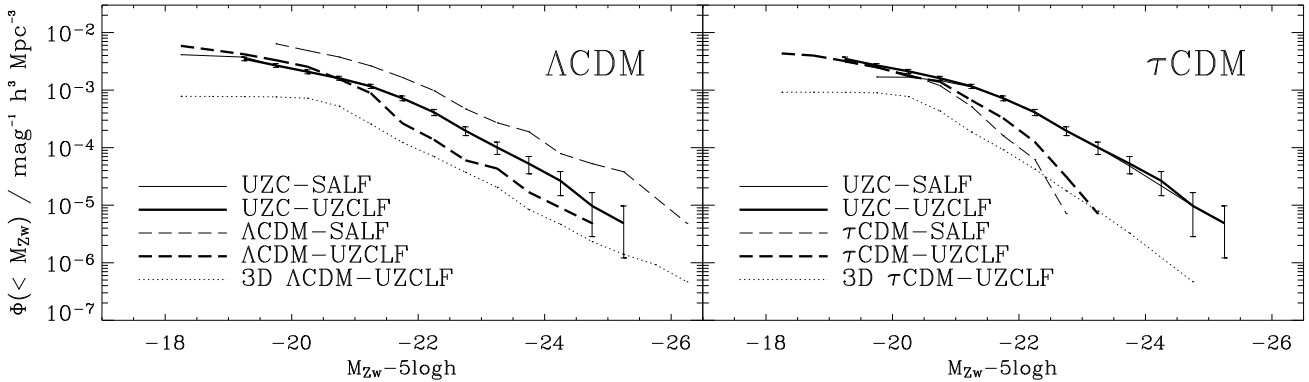}
\caption{Cumulative luminosity function. The error bars are 
1-$\sigma$ Poisson errors and are only shown for the UZC for clarity. The 
error bars for the models have similar size.
}\label{lf}
\end{center}
\end{figure*}

In real space, we consider a group each dark matter halo containing
$N\ge 5$ members brighter than the appropriate luminosity resolution.
We compute the total luminosity $L_{\rm tot}$ 
of a real-space group as the sum of the 
luminosities of the $N$ resolved galaxies within the halo,
$L_{\rm gal}=\sum_iL_i$, and the luminosity of the galaxies fainter than the 
luminosity resolution limit $L_{\rm lim}$   
\begin{equation}\label{lfaint}
L_{\rm faint}=\frac{N}{\int_{L_{\rm lim}}^{\infty}\phi(L)\textrm{d}L}\times 
\int_{0}^{L_{\rm lim}}\phi(L)L\textrm{d}L\; ,
\end{equation}
where $\phi(L)$ is the adopted galaxy luminosity function.

In redshift space, we use the same equations with the only difference that 
the luminosity resolution limit $L_{\rm lim}$ is replaced by the brightest 
luminosity between the luminosity resolution and the absolute 
magnitude $M_{\rm lim}=m_{\rm lim}-25-5 \log (\langle cz \rangle/H_0)$,
where $\langle cz\rangle/H_0$ is the group distance.

Not surprisingly, Figure \ref{lf} shows that neither the $\Lambda$CDM- nor 
the $\tau$CDM-SALF mock catalogues reproduce
the UZC group luminosity function. With the UZCLF catalogues the 
situation improves, but it is still unsatisfactory, because the KS and 
Wilcoxon-Rank-Sum (WRS) tests indicate that the mock distributions are not 
drawn from the same population of the UZC at very high significance levels 
(Tables \ref{ks_all} and
\ref{wil_all}).

Nevertheless the $\Lambda$CDM-UZCLF appears to yield a luminosity function 
closer to the UZC than the $\tau$CDM-UZCLF. In fact, the $\tau$CDM model fails 
at providing 
numerous enough luminous groups, despite the fact that the $\tau$CDM model 
produces a number of bright galaxies larger than both the  
$\Lambda$CDM model and the UZC (Figure \ref{mathi}).
However, the small amplitude of the galaxy two-point correlation function
(Figure \ref{xigg}) 
shows that the galaxies are not enough clustered to yield 
either a sufficiently large fraction of galaxies in groups or
a sufficiently large number of groups.
Consequently, the median luminosities of the $\tau$CDM groups are lower
than in the UZC (Table \ref{magn_lim_results}).

\begin{figure*}
\begin{center}
\includegraphics[scale=0.39]{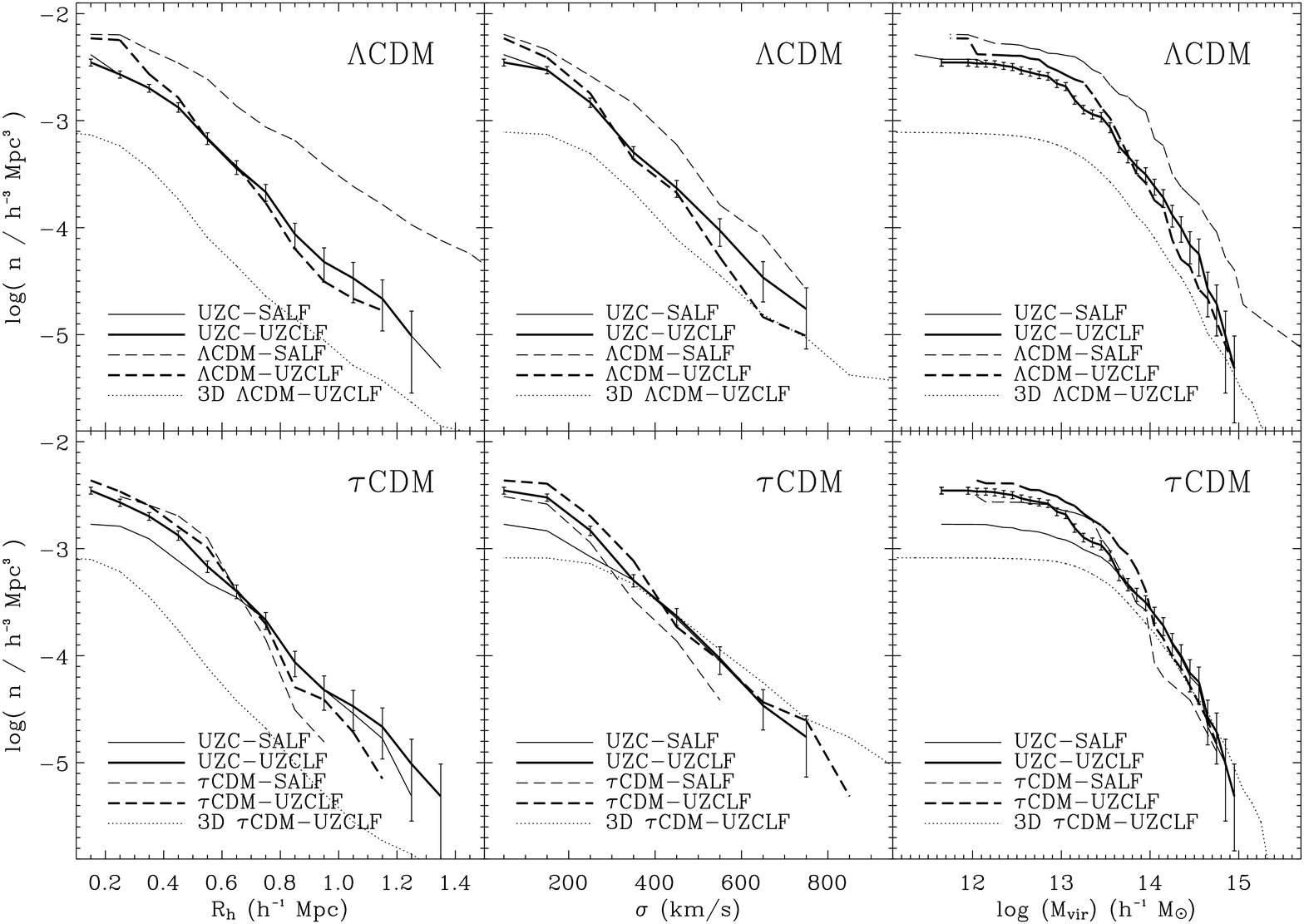}
\caption{Cumulative number density of groups by harmonic radius $R_{h}$, velocity dispersion 
$\sigma$, and virial mass $M_{\rm vir}$. 
The error bars are 1-$\sigma$ Poisson errors and are only shown 
for the UZC for clarity. The error bars for the models have similar size. }\label{cumhisto}
\end{center}
\end{figure*}

Figure \ref{cumhisto} shows the abundances of groups by harmonic radius
\begin{equation}
R_h = {\pi\over 2} {\langle cz\rangle\over H_0} N(N-1) \left[\sum_{i=1}^{N-1}
\sum_{j=i+1}^N {1\over \tan(\theta_{ij}/2)}\right]^{-1} \; ,
\end{equation}
where $N$ is the number of members and $\theta_{ij}$ are the pairwise angular 
separations, velocity dispersion 
\begin{equation}
\sigma=\left[{1\over N-1} \sum_{i=1}^N (cz_i-\langle cz\rangle)^2\right]^{1/2}
\end{equation}
and virial mass 
\begin{equation}
M_{\rm vir} = {6\sigma^2 R_h\over G} \; . 
\end{equation}
To avoid divergences, we set, according to Ramella (private communication), 
$\tan(\theta_{ij}/2)\langle cz\rangle/H_0 = r_{ij}=0.03 h^{-1}$~Mpc when 
$r_{ij}<0.03 h^{-1}$~Mpc.
The SALF catalogues perform very poorly: the medians of these quantities
are sensibly different from the UZC medians (Table \ref{magn_lim_results})
and the KS and WRS tests indicate that the mock groups and the UZC groups
are certainly drawn from different parent populations (Tables \ref{ks_all} 
and \ref{wil_all}).
The only exceptions are the $\sigma$ and $M_{\rm vir}$ distributions
of the $\tau$CDM models and the $M_{\rm vir}/L_{\rm tot}$ distribution of the 
$\Lambda$CDM model.

The UZCLF catalogues perform substantially better: with the exception 
of the $L_{\rm tot}$ and $M_{\rm vir}/L_{\rm tot}$ distributions, both the 
$\Lambda$CDM and the $\tau$CDM models provide a rather
good match to the UZC group abundances, with the $\Lambda$CDM model
yielding the largest significance levels.

Figure \ref{cumhisto} also shows the number density of the groups in real space
(dotted lines). For these groups 
\begin{equation}
R_h = {N(N-1)\over 2} \left[\sum_{i=1}^{N-1}
\sum_{j=i+1}^N {1\over \vert {\bf r}_{ij}\vert }\right]^{-1} \; ,
\end{equation}
where ${\bf r}_{ij}$ are the pairwise separations,
\begin{equation}
\sigma=\left[{1\over 3(N-1)} \sum_{i=1}^N ({\bf v}_i-\langle {\bf v}\rangle)^2\right]^{1/2}
\end{equation}
and $M_{\rm vir} = 6\sigma^2 R_h/ G$. 

In the $\Lambda$CDM model, the number densities of groups estimated in redshift 
space are larger than the real-space number densities, whereas in the 
$\tau$CDM model the largest discrepancy
appears in the number density by harmonic radius. This result agrees with 
tha fact that  $R_h$ can be overestimated by a factor as large as two
because of the presence of interlopers \citep{diaferio1999}. 
In the virial mass estimation, however, this bias is 
less relevant than the bias on $\sigma$, because of the
different powers that $R_h$ and $\sigma$ have in the virial mass relation:
on the scale of clusters, where $\sigma$ is almost unbiased, the recovered 
mass function is in reasonable agreement with the real-space mass function. 

$N$-body simulations, where galaxies
were identified as density peaks (\citealt{nolthenius87}; \citealt*{moore93};
\citealt{frederic95a},b) or formed and evolved with semi-analytic prescriptions
\citep{diaferio1999}, show that 
the FOF algorithm generally returns groups with statistical
properties comparable to those of the real-space groups; 
however, interlopers and spurious groups
can severly affect the estimates of the intrinsic properties of individual
groups and this bias is more dramatic for groups with fewer members
(\citealt{diaferio1999}; \citealt{eke04b}; \citealt{berlind06}). Figure \ref{cumhisto} 
shows that the discrepancy between the average properties 
of real- and redshift-space groups might be larger than usually
assumed. 

Recently, group-finder algorithms alternative to the
FOF have been proposed and might identify real-space groups
more accurately than the FOF algorithm (\citealt{eke04}; \citealt{yang05b}).
However, the FOF algorithm has been traditionally
applied to the UZC and we use it here to make our results
easily comparable to previous work. Moreover, 
we are interested in the comparison between models and
observations when they are both analyzed with the same technique, 
rather than in the determination of the most reliable group finder.
It is indeed a relevant result that the UZCLF catalogues agree with the UZC remarkably well,
despite the fact that the recover of the
average properties of the real-space groups with the FOF algorithm is only partially correct.

\begin{table*}
\centering
\caption{Median values of the group properties.}
\label{magn_lim_results}
\begin{tabular}{lcccc}
\hline
       & \multicolumn{2}{c}{SALF} & \multicolumn{2}{c}{UZCLF} \\
\hline
\hline
                    &  UZC/$\Lambda$CDM/3D   & UZC/$\tau$CDM/3D & UZC/$\Lambda$CDM/3D & UZC/$\tau$CDM/3D \\
$R_{h}$                      &  0.42  / 0.64  / 0.29  & 0.42  / 0.51  / 0.28  & 0.42  / 0.44  / 0.29  & 0.42  / 0.47  / 0.28  \\
$\sigma$                     &  219   / 284   / 230   & 222   / 217   / 320   & 219   / 217   / 230   & 217   / 250   / 320   \\
$\log M_{vir}$               &  13.45 / 13.81 / 13.29 & 13.48 / 13.54 / 13.57 & 13.45 / 13.44 / 13.29 & 13.45 / 13.59 / 13.57 \\
$\log L_{\rm tot}$           &  10.88 / 11.07 / 10.58 & 10.92 / 10.56 / 10.65 & 10.88 / 10.64 / 10.49 & 10.88 / 10.64 / 10.38 \\
$\log (M_{vir}/L_{\rm tot})$ &  2.59  / 2.70  / 2.70  & 2.58  / 2.94  / 2.93  & 2.60  / 2.85  / 2.79  & 2.60  / 2.96  / 3.20  \\
$\Omega_{0}$                 &  0.27  / 0.27  / --    & 0.27  / 0.82  / --    & 0.28  / 0.50  / --    & 0.28  / 0.64  / --    \\
\hline
\end{tabular}
\begin{minipage}{1\textwidth}
Median properties of the galaxy groups 
with more than five members. $R_{h}$, $\sigma$,
$M_{\rm vir}$, $L_{\rm tot}$ and $M_{\rm vir}/L_{\rm tot}$ are in units of $h^{-1}\;{\rm Mpc}$, $\kms$, 
$h^{-1}\rm M_{\odot}$, $h^{-2}\rm L_{\odot}$ and $h\rm M_{\odot}/L_{\odot}$,  
respectively.
\end{minipage}
\end{table*}

\begin{table}
\centering
\caption{Comparison of the UZC and the mock group abundances: KS test.}\label{ks_all}
\begin{tabular}{lcccc}
\hline
                    &  \multicolumn{2}{c}{SALF}  & \multicolumn{2}{c}{UZCLF} \\
\hline
\hline
                    &  $\Lambda$CDM & $\tau$CDM  & $\Lambda$CDM & $\tau$CDM \\ 
  $R_{h}$            &  $10^{-9}$    & $10^{-3}$  & $0.15$       & $0.14$   \\
  $\sigma$           &  $10^{-3}$    & $0.10$     & $0.27$       & $0.08$   \\
  $\log M_{\rm vir}$     &  $10^{-5}$    & $0.11$     & $0.34$       & $0.02$   \\
  $\log L_{\rm tot}$           &  $10^{-5}$    & $10^{-11}$ & $10^{-7}$    & $10^{-5}$\\
  $\log (M_{\rm vir}/L_{\rm tot})$ &  $0.15$       & $10^{-11}$ & $10^{-4}$    & $10^{-9}$\\

\hline
\end{tabular}
\begin{minipage}{0.48\textwidth}
\medskip
{Significance levels of the KS test for the null hypothesis
that the UZC and the mock groups are drawn from the same parent population.} 
\end{minipage}
\end{table}

\begin{table}
\centering
\caption{Comparison of the UZC and the mock group abundances: WRS test.}\label{wil_all}
\begin{tabular}{lcccc}
\hline
                    &  \multicolumn{2}{c}{SALF}  & \multicolumn{2}{c}{UZCLF} \\
\hline
\hline
                    &  $\Lambda$CDM & $\tau$CDM  & $\Lambda$CDM & $\tau$CDM \\ 
  $R_{h}$            &  $0$          & $10^{-3}$  & $0.16$       & $0.04$   \\
  $\sigma$           &  $10^{-3}$    & $0.39$     & $0.41$       & $0.03$   \\
  $\log M_{\rm vir}$     &  $10^{-6}$    & $0.25$     & $0.29$       & $0.01$   \\
  $\log L_{\rm tot}$           &  $10^{-5}$    & $0$        & $10^{-7}$    & $10^{-6}$ \\
  $\log (M_{\rm vir}/L_{\rm tot})$ &  $0.01$       & $0$        & $10^{-6}$    & $0$ \\

\hline
\end{tabular}
\begin{minipage}{0.48\textwidth}
\medskip
{Significance levels of the Wilcoxon Rank-Sum (WRS) test for the null hypothesis
that the UZC and the mock groups are drawn from the same parent population.}
\end{minipage}
\end{table}

\subsection{Mass-to-light ratio vs. mass}\label{3dml}

\begin{figure*}
\begin{center}
\includegraphics[scale=0.5]{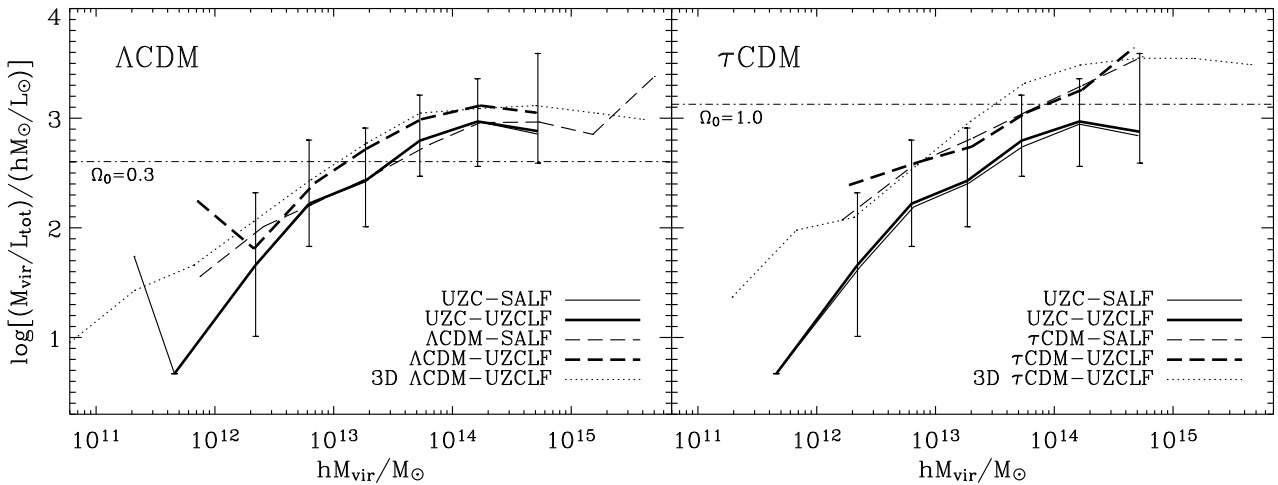}
\caption{Group mass-to-light ratio vs. 
  virial mass. The lines connect the medians of each bin and the 
error bars, only shown for the UZC for clarity, indicate the 
  0.1 and 0.9 percentiles. The error bars for the models
have similar size. The dot-dashed line shows the universal
mass-to-light ratio in the model.}
\label{scaling}
\end{center}
\end{figure*}

Figure \ref{scaling} shows that the mass-to-light ratio $M_{\rm vir}/L_{\rm tot}$ 
of the UZC groups increases with mass, as other samples of groups
have already demonstrated 
(\citealt{marinoni2002}; \citealt{lanzoni2003}; 
\citealt{eke04b}, 2006). Semi-analytic models of galaxy formation show that the
 dependence of $M_{\rm vir}/L_{\rm tot}$ on mass 
naturally arises in hierarchical models (\citealt{kauffmann1999};
\citealt{benson2000}), and extensions of the halo model use this relation 
as a sensitive probe to constrain the efficiency of galaxy formation
and the normalization of the power spectrum of the primordial
density fluctuations $\sigma_8$ (\citealt{vandenbosch03b}; \citealt{tinker05}).

The universal value of the mass-to-light ratio
is $\langle M/L\rangle=\rho_{\rm crit}\Omega_0/j$, where 
$\rho_{\rm crit}$ is the critical density of the Universe and 
$j=\phi^*L^*\Gamma(\alpha+2)$
is the luminosity density derived from the galaxy luminosity function.
In our models, the real-space relations approaches a universal value at sufficiently
large masses; this value does not correspond to the correct $\Omega_0$, 
however, because the virial mass $M_{\rm vir}$ estimated
in real space is still $\sim 30-40\%$ larger then the actual mass of the dark matter halo 
\citep{diaferio1999}. In redshift space the $M/L-M$ relations do not clearly show this flattening, 
but their rather large percentile ranges include the real-space relations.
The $\Lambda$CDM model matches the UZC results rather well, whereas
the $\tau$CDM model provides too large mass-to-light ratios both because
it does not provides luminous groups and because the real Universe seems
to have a low value of $\Omega_0$ and, consequently, a low value of 
$\langle M/L\rangle$.

\subsection{Halo occupation number}

\begin{figure*}
\begin{center}
\includegraphics[scale=0.45]{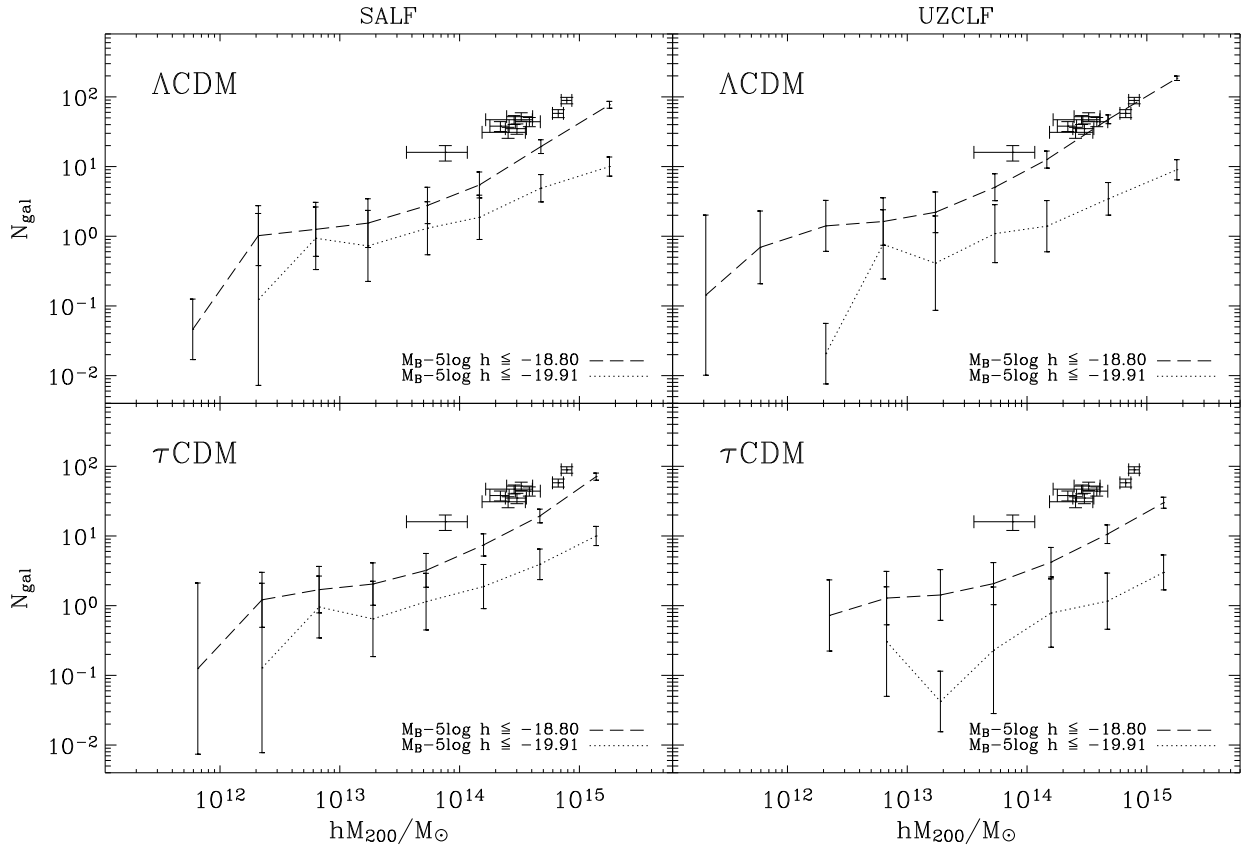}
\caption[]{\footnotesize{The halo occupation number (HON) in real space. 
The top (bottom) panels show the $\Lambda$CDM ($\tau$CDM) model.
In each panel lines are for galaxies brighter than $M_B-5 \log h=
-18.80$, and $-19.91$ (top to bottom) in the $\Lambda$CDM model, and
$-18.80$, and $-19.91$ (top to bottom) in the $\tau$CDM model.
The HON values smaller than one are due to haloes containing no galaxies
brighter than the threshold. The error bars are 1-$\sigma$ Poisson errors.
The points with error bars on both axes show the nine CAIRNS clusters \citep{rines04}.
}}\label{honlt}
\end{center}
\end{figure*}

The halo occupation number (HON) is the mean number of galaxies brighter than a 
luminosity threshold in dark matter halos of a given mass; it quantifies how 
efficient galaxy formation is in halos of different masses 
(\citealt{kauff97}; \citealt{benson2000}; \citealt{scocci01}; \citealt{vandenbosch03a};
\citealt{berlind03}; \citealt{kravtsov04}; \citealt{zheng05}). 
With appropriate extensions, one can use the HON to 
model the dependence of galaxy properties on environment (\citealt{yang05};
\citealt{sheth05}; \citealt{skibba06}; \citealt{sheth06}).

Estimating the HON of real systems requires accurate photometry
and accurate determination of the total mass and 
galaxy membership. To avoid these difficulties, a common approach is 
to assume an analytic form for the HON and adjust its parameters to match the
observed galaxy two-point correlation function (\citealt{jing98}; 
\citealt{maglio2003}; \citealt{aba04}; \citealt{zehavi05}).
Alternatively, one can directly use the galaxy luminosity function to estimate 
the expected number of galaxies in the cluster volume 
(\citealt{marinoni2002}; \citealt{lin04}).

These approaches have provided an indirect comparison of the observed HON 
with $N$-body or semi-analytic simulations. 
A direct comparison requires galaxy counts for a sample of galaxy systems
(\citealt{kochanek03}, \citealt{lin04}, \citealt{popesso06}). 
One of these direct counts was performed by \citet{rines04} for the
nine CAIRNS clusters, thanks to their accurate 2MASS infrared photometry and
spectroscopic uniform sky coverage. 
Figure \ref{honlt} shows the CAIRNS cluster results.

Before comparing the HON of the UZC groups with the model groups,
we check what model, if any, can reproduce the HON of these real clusters.
This comparison is more robust than the comparison with groups, 
because the galaxy membership in groups is substantially more uncertain than in
clusters, and the comparison between models and observations will be obscured by the large
error bars (see Figure \ref{hon-l-mass} below). 

To compare the CAIRNS HON with the HON in the simulations
we need to convert the 2MASS $K_s$-band to the $B$-band. 
According to \citet{jarrett00}, $B-K_s$  
varies between 2.86 and 3.97, depending on the galaxy morphology; these 
colours have rms uncertainties in the range 0.30-0.80.
Therefore, the luminosity limit imposed by \citet{rines04} lies in the 
$B-$band range $[-18.80,-19.91]$. Figure \ref{honlt} shows the HON of
the simulations in real space for both 
luminosity limits. Because early-type galaxies tend to reside in clusters and 
have redder colours, the fainter limit (dashed line) is probably the most appropriate for the 
comparison. 

The SALF catalogues provide a factor of $\sim 3$ fewer galaxies, whereas
only the $\Lambda$CDM-UZCLF model provides a good match to the CAIRNS clusters.
The slight underestimate still present in this case might originate from the fact that, 
because of the difficulties at identifying galaxy members, 
\citet{rines04} include all the galaxies projected onto the cluster.

\begin{figure*}
\begin{center}
\includegraphics[scale=0.45]{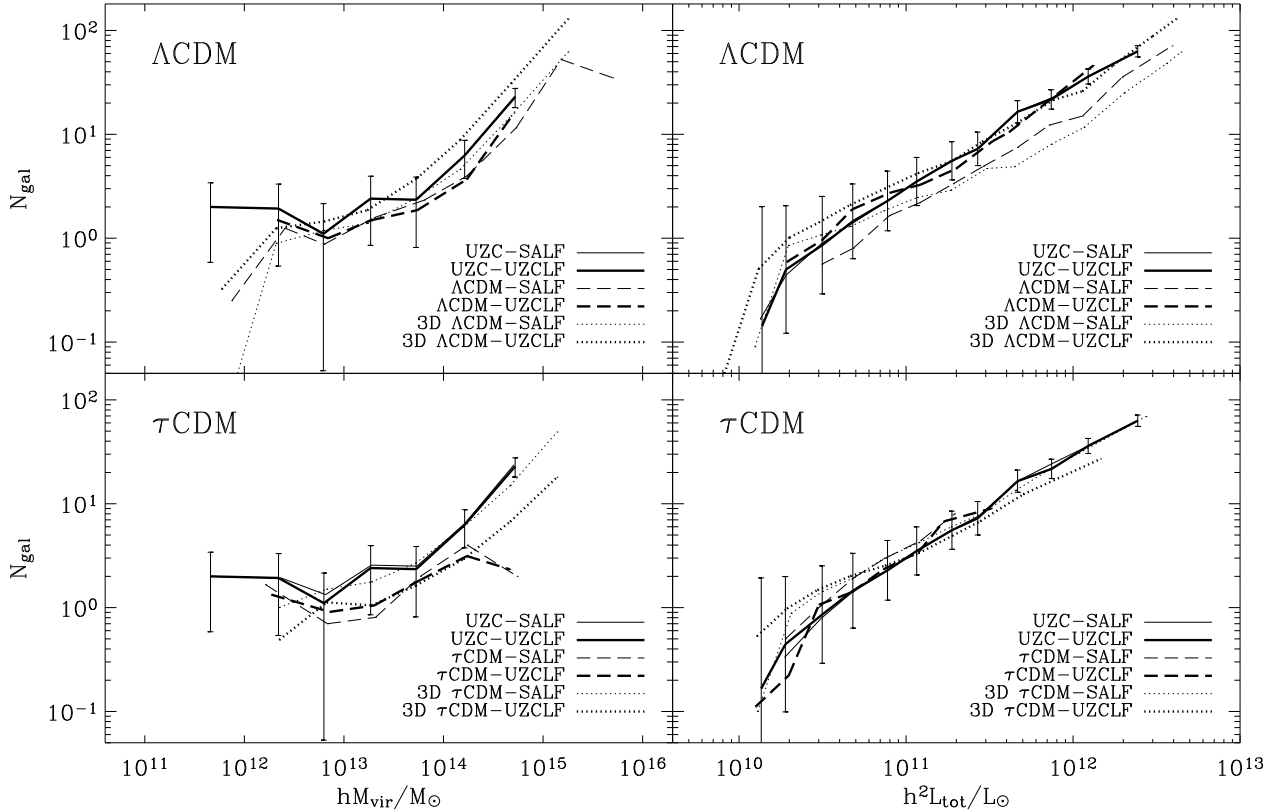}
\caption{ The halo occupation number (HON) in redshift space for groups with
  5 or more members. The top (bottom) 
panels show the $\Lambda$CDM ($\tau$CDM) model. The lines are for galaxies 
brighter than $M_B=-19.02 + 5\log h$, the luminosity of a 
galaxy with apparent magnitude $m_{\rm lim}=15.5$ at the cut off 
distance $8000\,\kms$. The dotted lines are the real space HON at the same 
$M_B=-19.02 + 5\log h$.
The HON is shown as a function of group mass $M_{\rm vir}$ (left panels) and group 
total luminosity $L_{\rm tot}$ (right panels). 
The HON values smaller than one are due to haloes containing no galaxies
brighter than the chosen threshold.
The error bars are 1-$\sigma$ Poisson uncertainties and are only shown for 
the UZC for clarity. The error bars for the models have similar size. 
}
\label{hon-l-mass}
\end{center}
\end{figure*}

When extending the comparison to the scales of groups, 
uncertainties on galaxy memberships and mass estimates become larger.
Therefore, because the luminosity $L_{\rm tot}$ appears to be a more robust
quantity than the virial mass $M_{\rm vir}$ \citep{eke04b}, we measure the HON as a 
function
of both $L_{\rm tot}$ and $M_{\rm vir}$ (Figure \ref{hon-l-mass}). In general
the models underestimate the UZC $N_{\rm gal}-M_{\rm vir}$ relation,
regardless of the luminosity function adopted. On the contrary, the simulated
$N_{\rm gal}-L_{\rm tot}$ relations agree well with the UZC, 
particularly the $\Lambda$CDM-UZCLF model. 
The $\tau$CDM models reproduce the observed slope and normalization, but 
their relations vanish at large $L_{\rm tot}$, because their mock surveys do 
not contain enough luminous groups. 

\subsection{The light distribution within groups}\label{lfred}

In the semi-analytic models analyzed here mergings between galaxies are 
allowed only when one of the two galaxies is the central galaxy of the dark 
matter halo: satellite-satellite merging is not implemented. Moreover
gas cooling is allowed only onto the central galaxy and satellite galaxies 
stop forming stars when their internal reservoir of cold gas is exhausted. 
These two processes yield a luminosity difference $\Delta L_{12}$ between the 
first and the second brightest galaxies in a group that can be substantially 
larger than observed. 

\begin{figure*}
\begin{center}
\includegraphics[scale=0.45]{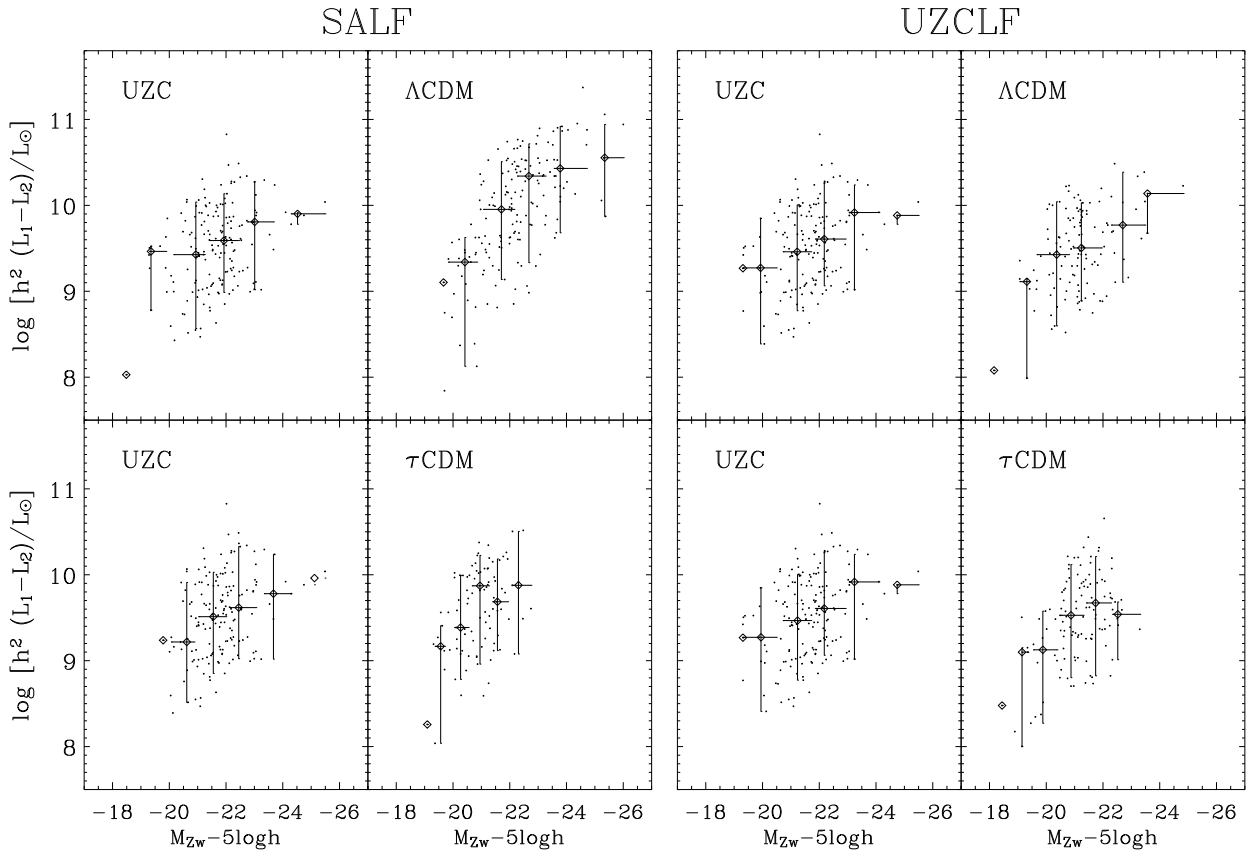}
\caption{Luminosity difference between the first and the second rank galaxies 
in a group as a function of the group total luminosity.}
\label{diffl}
\end{center}
\end{figure*}

In Figure \ref{diffl} we plot $\Delta L_{12}$ as a function of the group total 
luminosity. The median $\Delta L_{12}$ is $3.3\times 10^9 h^{-2} L_\odot$ for 
the UZC and $9.3\times 10^{9} h^{-2} L_\odot$ for the $\Lambda$CDM-SALF
model. The $\Delta L_{12}$ density distribution of the $\tau$CDM-SALF model is
similar to the UZC (its median is $\Delta L_{12}=3.7\times 10^9 h^{-2}
L_\odot$), but the tail of the high-luminosity groups is missing.
To quantify the $L_{\rm tot}-\Delta L_{12}$ relation, 
Table \ref{spearman} lists the Spearman rank-order 
correlation coefficients $r$ and their probability $P$: smaller $P$'s indicate 
stronger correlations. In the $\Lambda$CDM-SALF model, the correlation is 
orders of magnitude stronger than in the UZC. By reassigning the galaxy 
luminosity according to the UZC luminosity function the correlation weakens to 
the observed level.  

\begin{table}
\centering
\caption{Correlation between $\Delta L_{12}$ and the group total luminosity.}
\label{spearman}
\begin{tabular}{lcccc}
\hline
            &  \multicolumn{2}{c}{SALF}  & \multicolumn{2}{c}{UZCLF} \\
\hline
\hline
            &  UZC          & $\Lambda$CDM  & UZC          & $\Lambda$CDM \\ 
$r$         &  0.30         & 0.63          & 0.29         & 0.33         \\
$P$         &  $10^{-4}$    & $10^{-19}$    & $10^{-4}$    & $10^{-4}$    \\
\hline
            &  UZC          & $\tau$CDM     & UZC          & $\tau$CDM    \\ 
$r$         &  0.31         & 0.35          & 0.29         & 0.36         \\
$P$         &  $10^{-4}$    & $10^{-4}$     &$10^{-4}$     & $10^{-4}$    \\
\hline
\end{tabular}
\begin{minipage}{0.48\textwidth}
\medskip
{Spearman rank correlation coefficient ($r$) and the significance of
its deviation from zero ($P$). A smaller value of $P$ indicates a stronger
correlation.}
\end{minipage}
\end{table}

The semi-analytic recipes produce many more bright galaxies than observed in 
the UZC (Figure \ref{mathi}). This fact can boost, in some cases, the
luminosity difference $\Delta L_{12}$, as one can see by comparing the
two $\Lambda$CDM panels in Figure \ref{diffl}. We therefore
also consider the ratio $L_2/L_1$ of the luminosities of the two brightest
galaxies. The medians are 0.69 for the UZC and 0.52 and 0.63 for the 
$\Lambda$CDM-SALF and the $\tau$CDM-SALF models, respectively, and confirm 
that the second brightest galaxy is on average fainter in the models than in 
the UZC. 

A more quantitative comparison is obtained by considering the
density distributions of the luminosity ratios. 
Table \ref{ldiff} lists the significance levels of the KS and WRS tests for
the null hypothesis that the UZC and the model $L_2/L_1$'s
are drawn from the same parent population.
Clearly, the $\Lambda$CDM-SALF
and the UZC groups have different parent populations, whereas
the 1\% significance level of the $\tau$CDM-SALF is a consequence
of the failure of this model to yield groups brighter than $M_{B} - 5 \log h
\sim -22$, as we have clarified above (Figure \ref{diffl}).
The comparison is instead satisfactory when the galaxy luminosities
are reassigned according to the UZC luminosity function (UZCLF catalogues): 
both models now agree with the UZC to a significance level greater
than 38\%.

\begin{table}
\centering
\caption{UZC-model comparison of the luminosity ratio between the two 
brightest galaxies in a group.}\label{ldiff}
\begin{tabular}{lcccc}
\hline
            &  \multicolumn{2}{c}{SALF}  & \multicolumn{2}{c}{UZCLF} \\
\hline
\hline
            &  $\Lambda$CDM & $\tau$CDM  & $\Lambda$CDM & $\tau$CDM \\ 
 KS         &  $10^{-5}$    & 0.01       & 0.78         & 0.67      \\
 WRS        &  $10^{-7}$    & $10^{-3}$  & 0.41         & 0.38      \\

\hline
\end{tabular}
\begin{minipage}{0.48\textwidth}
\medskip
{Significance levels of the KS and WRS tests for the null hypotesis that the
  luminosity ratio $L_{2}/L_{1}$ between the second and first brightest 
galaxies in the UZC and in the simulated groups are drawn from the same parent
population.}
\end{minipage}
\end{table}

\section{Conclusions}\label{conclusion}

We have compared the properties of groups and clusters of galaxies extracted 
from the UZC \citep{falco1999} with those of systems extracted from mock 
redshift surveys. 
We have compiled these mock surveys from $N$-body simulations constrained
to reproduce the large-scale distribution of galaxies in the nearby Universe
\citep{mathis2002}.
In the simulations, galaxies are formed and evolved according to a 
semi-analytic procedure and we are thus able to constrain both the group 
clustering and the group luminosity content. 

By using simulations with constrained initial conditions, we minimize the 
possible role played by cosmic variance, and differences between real and
mock catalogues should mostly originate from the galaxy formation recipes.
Our approach thus differs from previous attempts of 
comparing simulations with real group catalogues, where 
either cosmic variance was an issue \citep{diaferio1999} or the real catalog
was large enough that cosmic variance was naturally suppressed 
(\citealt{eke04}; \citealt{berlind06}).

The gross large-scale distribution of galaxies, including
the location of the major nearby clusters, is very similar in the mock
and real surveys, as mostly imposed by the initial conditions. However,
the simulated large-scale structures are not
as sharply defined as in the UZC, confirming earlier results
of mock surveys extracted from unconstrained simulations 
\citep{schmalzing2000}.
This disagreement amplifies substantially when we consider the properties
and the large-scale distribution of groups. 

The group-finder algorithm strongly depends on the galaxy luminosity function.
The numerical recipes adopted to form and evolve galaxies in
the simulations provide a galaxy luminosity function which is considerably
different from that of the UZC. We test
that this difference is the major responsible for the model inability at 
reproducing the group properties. We assign new 
luminosities to the simulated galaxies according to the luminosity function 
of the UZC, while preserving the luminosity rank predicted by the model.
We thus have new mock catalogues where galaxy luminosities
distribute according to the UZC luminosity function (UZCLF), besides the catalogues 
where galaxy luminosities distribute according to the 
semi-analytic luminosity function (SALF).
 
Unlike the groups extracted from the SALF and the $\tau$CDM-UZCLF
mock catalogues, the groups extracted from the $\Lambda$CDM-UZCLF catalogue 
have statistical properties
in satisfactory agreement with observations: group abundances by 
luminosity, harmonic radius, velocity dispersion and mass are generally within 
3-$\sigma$ errors, or less, from the UZC group abundances. These groups also 
reproduce the UZC relations between the group mass-to-light ratio and mass, 
and the galaxy number and mass, namely the halo occupation number.
Finally, the $\Lambda$CDM-UZCLF groups,
similarly to the UZC groups and unlike the SALF groups, 
show a weak correlation between the luminosity difference between the
two brightest galaxies in a group and the total group luminosity. This result
indicates that the two semi-analytic prescriptions of allowing merging of
satellite galaxies only with the central galaxy of the dark matter halo and gas cooling
only onto this same central galaxy produces a too large
luminosity difference between the two brightest galaxies in a group.

The success in the statistical properties of groups, obtained by adopting the 
observed luminosity function, is not shared by  the large-scale distribution 
of groups. Specifically, the simulated groups in the North Galactic Cap of the 
$\Lambda$CDM-UZCLF catalogue trace the large-scale distribution of galaxies at 
significance levels as much as seven times smaller than those of the UZC. 
This is a consequence of the looser large-scale structures in the models: in 
fact, the redshift-space correlation function of galaxies is more 
than 3-$\sigma$ below the observations on scales larger than $6 h^{-1}$ Mpc 
(Figure \ref{xigg}; \citealt{mathis2002}) and the number of groups is 25\% 
smaller than in the UZC (Table \ref{groups}). 

We therefore conclude that (1) the semi-analytic
recipes used in \citet{mathis2002} have seriuos difficulties at distributing
the luminosity among galaxies correctly; and (2) even the most successful
$\Lambda$CDM model does not yield cosmic structures as coherent as observed. 
Whereas the first problem might be solved, in principle,
with more sophisticated semi-analytic modellings (e.g. \citealt{croton06}; 
\citealt{weinmann06b})
the second problem appears to be more fundamental,
although the disagreement with observation is less dramatic:
in this case, both different dark matter models 
and an appropriate treatment of the 
interplay between gas dynamics and dark matter might be necessary to reconcile
the model with group properties.

\section*{Acknowledgments}

Part of this work served as the undergraduate thesis of LC at the University 
of Torino.
AD acknowledges the hospitality of Chris Flynn and the Tuorla Astronomical 
Observatory where part of this work was carried out.
We thank an anonymous referee for a prompt report and relevant comments, and 
Hughes Mathis and his collaborators for making publicly available
their simulations. The simulations were carried out on the T3E supercomputer at the Computing Centre 
of the Max-Planck Society in Garching, Germany.
The data are available at 
{\tt www.mpa-garching.mpg.de/galform/cr/data.shtml}.

\end{document}